\documentclass[11pt]{article}
\usepackage{theorem,amsfonts,bezier,lscape}

{\theoremstyle{break}
\theorembodyfont{\normalfont}

}
\makeatletter
\@addtoreset{equation}{section}
\makeatother

\def\mylabel#1{\label{#1}}

\def\ignore#1\long{\relax}

\font\tengoth=eufm10 
\font\sevengoth=eufm7 
\font\fivegoth=eufm5
\newfam\gothfam
  \textfont\gothfam=\tengoth 
  \scriptfont\gothfam=\sevengoth
  \scriptscriptfont\gothfam=\fivegoth
\def\goth{\fam\gothfam}  

\font\frak=eufm10 scaled\magstep1
\def\goth#1{\hbox{{\frak#1}}}

\def\be{\begin{equation}}
\def\ee{\end{equation}}
\def\bea{\begin{eqnarray}}
\def\eea{\end{eqnarray}}

\def\Re{{\mathbb R}}
\def\Ce{{\mathbb C}}
\def\He{{\mathbb H}}

\def\Ze{{\mathbb Z}}

\def\hp#1#2{\langle #1, #2 \rangle}
\def\norm#1{\|#1\|}
\def\myspan#1{\langle #1 \rangle}

\def\overto#1{\stackrel{#1}{\to}}
\def\myarray#1#2{\left( \!\begin{array}{c} #1 \\ #2 \end{array}
\!\right)}

\def\k{\kappa}
\def\icd{\eta}

\def\iCe{\mathbb C_\icd}
\def\iRe{{{}_\icd\mathbb R}}

\def\Invol{\Pi}
\def\dual{{\cal D}}

\def\CKPointSpace{ S^2_{[\k_1],\k_2}}
\def\CeCKPointSpace{{\Ce}\CKPointSpace}
\def\iCeCKPointSpace{{\iCe}\CKPointSpace}
\def\CKLineSpace{{ S^2_{\k_1,[\k_2]}}}
\def\CeCKLineSpace{{\Ce}\CKLineSpace}
\def\iCeCKLineSpace{{\iCe}\CKLineSpace}

\def\c{C_{\k_1}}
\def\cc{C_{\k_2}}

\def\s{S_{\k_1}}
\def\ss{S_{\k_2}}

\def\T{T_{\k_1}}
\def\TT{T_{\k_2}}

\def\v{V_{\k_1}}
\def\vv{V_{\k_2}}

\def\ci{C_{\icd}}
\def\si{S_{\icd}}
\def\csa{C_{\icd\k_1^2}}
\def\ssa{S_{\icd\k_1^2}}
\def\csca{C_{\icd\k_2^2}}
\def\ssca{S_{\icd\k_2^2}}
\def\vsa{V_{\icd\k_1^2}}
\def\vsca{V_{\icd\k_2^2}}

\def\rph{\varepsilon}
    \def\rphi{\rph_i}
    \def\rphI{\rph_I}
\def\vecz{{\bf z}} 
\def\vecZ{{\bf Z}}

\def\dd{I}
\def\jj{J}
\def\mm{M}
\def\ll{B}
\def\tt{T_1}
\def\ttt{T_2}
\def\pp{P_1}
\def\ppp{P_2}
\def\qq{Q_1}
\def\qqq{Q_2}
\def\hh{H_1}
\def\hhh{H_2}

\def\om{\Omega}           
\def\Om{\omega}           

\def\symarea{{\cal{S}}}
\def\symcoarea{s}

\def\gramm{\triangle_g}
\def\Gramm{\triangle_G}
\def\grammR{\gamma}
\def\GrammR{\Gamma}

\let\oldxi=\xi

      \def\xi{x_i}
      \def\xj{x_j}
      \def\xk{x_k}
      \def\xI{X_I}
      \def\xJ{X_J}
      \def\xK{X_K}
\def\lp{\phi}            
      \def\lpi{\lp_i}
      \def\lpj{\lp_j}
      \def\lpk{\lp_k}
      \def\lpx{\lp_x}
      \def\lpa{\lp_a}
      \def\lpb{\lp_b}
      \def\lpc{\lp_c}
\def\ap{\Phi}            
      \def\apX{\ap_X}
      \def\apA{\ap_A}
      \def\apB{\ap_B}
      \def\apC{\ap_C}
      \def\apI{\ap_I}
      \def\apJ{\ap_J}
      \def\apK{\ap_K}
\def\hsi{\Psi}
      \def\hsiA{\hsi_A}
      \def\hsiB{\hsi_B}
      \def\hsiC{\hsi_C}
\def\holinc{\Upsilon}            
      \def\holincA{\holinc_A}
      \def\holincB{\holinc_B}
      \def\holincC{\holinc_C}
\def\fsa{\Lambda}                
      \def\fsaA{\fsa_A}
      \def\fsaB{\fsa_B}
      \def\fsaC{\fsa_C}
\def\fscia{\Theta}

\def\myexp#1{e^{#1}}

\def\apa{a P_a}
\def\bpb{b P_b}
\def\cpc{c P_c}
\def\AJA{A J_A}
\def\BJB{B J_B}
\def\CJC{C J_C}
\def\ata{\lpa T_a}
\def\btb{\lpb T_b}
\def\ctc{\lpc T_c}
\def\AIA{\apA I_A}
\def\BIB{\apB I_B}
\def\CIC{\apC I_C}

\def\ctesd{\oldxi}
\def\cteSD{\Xi}
\def\ctesS{\tau}

\begin{document}

\noindent\ 
\bigskip\bigskip

\begin{center} {\LARGE{\bf{Trigonometry of `complex Hermitian'
type homogeneous symmetric spaces.\\[0.2cm] }}}
\end{center}
\bigskip

\begin{center} Ram\'on Ortega and Mariano Santander
\end{center}

\begin{center} {\it Departamento de F\'{\i}sica Te\'orica,
Facultad de Ciencias\\ Universidad de Valladolid, E--47011
Valladolid, Spain}
\end{center}

\bigskip

\begin{abstract}
\noindent

This paper contains a thorough study of the trigonometry of the
homogeneous symmetric spaces in the Cayley-Klein-Dickson family of
spaces of  `complex Hermitian' type and rank-one.  The  complex
Hermitian elliptic $\Ce P^N\equiv SU(N+1)/(U(1)\otimes SU(N))$ and
hyperbolic $\Ce H^N\equiv SU(N,1)/(U(1)\otimes SU(N))$ spaces,
their analogues with indefinite Hermitian metric $SU(p+1,
q)/(U(1)\otimes SU(p,q))$ and the non-compact symmetric space
$SL(N+1, \Re)/(SO(1,1)\otimes SL(N, \Re))$ are the generic members
in this family; the remaining spaces are some contractions of the
former.  

The method encapsulates trigonometry for this whole family of
spaces into a single {\em basic trigonometric group equation}, and
has `universality' and `(self)-duality' as its distinctive traits.
All previously known results on the trigonometry of $\Ce P^N$ and
$\Ce H^N$ follow as particular cases of our general equations.
 
The following topics are covered rather explicitly: 0) Description
of the complete Cayley-Klein-Dickson family of rank-one spaces of
`complex type',  1) Derivation of the single basic group
trigonometric  equation,  2) Translation to the basic `complex
Hermitian' cosine, sine and dual cosine laws),  3) Comprehensive
exploration of the bestiarium of `complex Hermitian' trigonometric
equations,  4) Uncovering of a `Cartan' sector of Hermitian
trigonometry, related with triangle symplectic area and coarea,  
5) Existence conditions for a triangle in these spaces as
inequalities and 6) Restriction  to the two special cases of 
`complex' collinear and purely real triangles.

The physical Quantum Space of States of any quantum system 
belongs, as the complex Hermitian space member,  to this
parametrised family; hence its trigonometry appears as a rather
particular case of the equations we obtain. 

\end{abstract}

\newpage

\section{Introduction}

In a previous paper \cite{SpaceTimeTrig} the trigonometry of the
complete family of symmetric spaces of {\em rank one and real
type} was studied. These spaces are also called Cayley-Klein
(hereafter CK) real spaces and were first discussed by Klein
extending the Cayley idea of `projective metrics'. In two
dimensions there are {\em nine} real spaces with a real quadratic
symmetric metric of any (positive, zero, negative) constant
curvature, and any (positive definite, degenerate, indefinite)
signature \cite{Yaglom}. Further to this, the paper
\cite{SpaceTimeTrig} had a long run aim towards opening an avenue
for exploring the trigonometry of general symmetric homogeneous
spaces.

Next to the spaces of real type, there are spaces of `complex'
type. In `complex' dimension $N$ there are $3^{N+1}$ such
geometries
\cite{MSBurgos}; `complex' type means here that these spaces are
coordinatised by elements of a one-step extension of $\Re$ 
through a labelled Cayley-Dickson procedure $\Re \to {}_{\icd}\Re$
which adjoins a imaginary unit
$i$ with $i^2=-\icd$ to $\Re$, producing either the complex, dual
or split complex numbers according as $\icd>0, =0, <0$. We will
term  Cayley-Klein-Dickson (CKD) these spaces. They are
`Hermitian', since they are related to a scalar product with
`complex' values and  hermitian-like symmetry

Within this family, only the spaces coordinatised by ordinary
complex numbers (where $\icd>0$, which can be rescaled to $\icd=1$
and $i^2=-1$) are actually {\it complex} spaces;  after this
restriction, there are only
$3^{N}$ CK complex type geometries in complex dimension $N$
\cite{ExtSU}.  For $N=2$ there are nine 2D complex Hermitian CK
spaces, with constant holomorphic curvature (either $K_{hol}>0,
=0, <0$) and a Hermitian metric of  either definite, degenerate or
indefinite  signature. All these are Hermitian symmetric  spaces
with a complex structure, hence K\"ahlerian, but only the three
spaces with a definite positive Hermitian metric belong to the
restricted family of the so-called {\em two-point homogeneous
spaces} \cite{Wang}. These are the {\em elliptic Hermitian space},
i.e.\  the complex projective space $\Ce P^2$ with the
Fubini-Study metric, the {\em hermitian hyperbolic space} $\Ce
H^2$ which can be realized in the interior  of a Hermitian quadric
in $\Ce P^2$ like its real analogue, and the Hermitian euclidean
space $\Ce R^2$,  a two-dimensional Hilbert space, as the common
`limiting' space. 

If degenerate and indefinite Hermitian products are allowed, a CK
family of nine complex 2D spaces is obtained: the complex elliptic,
euclidean and hyperbolic Hermitian planes with a definite positive 
metric, the complex  co-euclidean, galilean and co-minkowskian 
(or Anti Newton-Hooke, Galilean and Newton-Hooke) Hermitian
planes  with a degenerate metric and finally the indefinite metric
complex co-hyperbolic, minkowskian and doubly hyperbolic (or Anti
De Sitter, Minkowskian and De Sitter) Hermitian planes.  The last
six spaces are the complex Hermitian analogues of the three
non-relativistic and three relativistic space-times.  In group
theoretical terms,  all these nine Hermitian spaces appear as {\em
four} generic cases and {\em five} non-generic ones within the
complex CK family. The generic cases are   
$SU(3)/(U(1)\otimes SU(2))$,
$SU(2,1)/(U(1)\otimes SU(2))$,
$SU(2,1)/(U(1)\otimes SU(1,1))$, the last hosting two different
spaces with `time like' and `space-like' complex lines
interchanged. Non-generic cases are contractions of these four
generic ones, with either curvature vanishing, metric degenerating
or both.  Within the full complex CK family of spaces of complex
type, results on trigonometry are only available, as far as we
know, for the Hermitian spaces with definite positive  metric and
constant (either positive or negative) holomorphic curvature
\cite{Coo, BlasTer, Ter, ShiPetRoz, Hsiang, Bre, RosHermTrig95,
RosGeoLieGroups}; a review of these results is included in Section
2.

The spaces in the subfamily $\icd<0$ are much less known;  they
have not a complex structure, but its `split complex'  analogue.
Its generic members correspond to the symmetric homogeneous space
$SL(3,
\Re)/(SO(1,1)\otimes SL(2,\Re))$, and the non-generic ones to some
of its contractions.  The trigonometry for such spaces has
apparently not been studied. The spaces with  $\icd=0$ appear as
common contractions from the spaces with $\icd>0$ and $\icd<0$.

In this paper we set out the task of studying in full detail the
{\em trigonometry of the complete family of spaces of `complex'
type}.  It clearly suffices to consider the two-dimensional case,
as a triangle in any CKD  such `complex'-type space in `complex'
dimension $N$ is fully contained in a totally geodesic subspace
with `complex' dimension $2$.

The approach distinctive traits are:  {\em First}, it  covers at
once the trigonometry in the whole family of
$3^3=27$  such geometries, parametrized by three real labels
$\icd; \k_1,
\k_2$. The study of the trigonometry in the family as a whole is
in fact easier than its study for just one space at a time. {\em
Second}, it gives a clear view of several {\em duality}
relationships between the trigonometry of these different spaces.
In particular it explicity displays the self-duality of the
Hermitian elliptic space (analogous  to the self-duality of the
real sphere $S^2$) that is  completely hidden in the
trigonometric  equations derived for this space in
\cite{ShiPetRoz, Hsiang, Bre,  RosGeoLieGroups}. {\em Third}, it
gives more than previously known: {\it all} previously known
equations appear as a rather small subset of the equations to be
derived here.  {\em Fourth}, it deals uniformly with the
(contracted) non-generic cases which  correspond to  curvature
vanishing $(\k_1=0)$ and/or (`Fubini-Study') metric degenerating
$(\k_2=0\hbox{ or }\icd=0)$. `Hermitian' analogues of the real
angular and lateral excesses vanish in these limits, and are
related to triangle symplectic area and its dual quantity. And
{\em fifth}, the presence of the additional Cayley-Dickson type
label
$\icd$ makes it possible the consideration of a new type of
contractions, encompassed by the limit $\icd\to 0$ whose physical
meaning is worth exploring. 

The paper is intended to be self-contained, but reference to
\cite{SpaceTimeTrig} may be helpful, specially for motivations and
general background. A condensed review of already known results on
the trigonometry of both the Hermitian elliptic and hyperbolic
spaces \cite{BlasTer, Ter, ShiPetRoz, Hsiang, Bre, RosHermTrig95,
RosGeoLieGroups} is given in Section 2. Here we display the  basic
equations which in some cases were originally given in terms of
angular invariants not considered in this paper (e.g.\ the
Fubini-Study angles and the holomorphy inclinations at each
vertex), and will be rewriten here so they can be easily
identified with the ones we obtain. The choice of basic invariants
adopted here affords the equations in a form which we believe
simpler than using any other choice. 

Information on CKD  spaces of `complex' type is given in Sections
3 and 4. Section 3 deals with the ordinary complex case, and
therefore refers to a much more familiar situation. Section 4
comments the main new traits appearing when complex numbers are
replaced by their `complex' parabolic and split (hyperbolic)
versions.

In Section 5 the approach to the trigonometry proposed in
\cite{SpaceTimeTrig}  is developed in depth for the {\em complete}
CKD family.   The whole of trigonometry for all these spaces is
encapsulated in a single {\em basic trigonometric group equation},
involving  sides, angles, and {\em lateral} and {\em angular}
`phases' and exhibiting explicitly (self)-duality in the whole
family.  This is mainly achieved due to a choice of triangle
invariants as  the  canonical parameters of two pairs of
commuting  isometries, a choice which should ring a bell to  any
physicist educated in Quantum Mechanics.  Dealing with many spaces
at once, this equation gives a  perspective on some  relationships
going far beyond any treatment devoted to the study of a single
space. The behaviour of trigonometry when either the curvature
vanishes or the metric degenerates is  explicitly described 
through the CKD constants
$\icd; \k_1, \k_2$. Duality is the main structural backbone in our
approach, and the requirement to explicitly maintaining duality in
all expressions and at all stages acts as a kind of method
`fingerprint'.  Cartan duality for  symmetric spaces
\cite{Helgason} appears here as the change of sign in either
$\icd, \k_1$ or $\k_2$.

The {\em basic trigonometric group equation} is an equation  for
the parameter-dependent group of motions. By writing it in the
fundamental `complex' representation, a set of nine `complex'
equations follows. With these equations as starting point, we will
explore in section 6 the rather unknown territory of `complex
hermitian' trigonometric equations. The background provided by
real trigono\-metry makes this exploration easier by deliberately
pursuing  the analogies, while at the same time the relevant
differences stand  out clearly. The most interesting difference 
with real trigonometry is the natural splitting of the equations
into two `sectors'. The first involves quantities linked to Cartan
generators of the motion group, where two new triangle invariants
appear in a rather natural way; they play a specially important
role since they are proportional to the symplectic area and coarea
of the triangle. For the {\em hermitian elliptic space $\Ce
P^2$},  these quantities were first found by  Blasckhe and
Terheggen \cite{BlasTer, Ter}.  The other `sector' is the
`complex' analogue of the whole real trigonometric set of
equations. Most results in the real case have (sometimes several
and/or partly) analogous in this `complex' trigonometry. The  {\em
family form} of {\em all} previously known equations is obtained
here,  together with a large number of new ones. In the $\Ce P^2$
case  ($\icd=1;\k_1=1,\k_2=1$) all trigonometric functions of
sides, angles,  lateral and angular phases are the ordinary
circular ones, and at a first look  the whole paper can be read by
restricting to this case; this may help  to grasp the key ideas,
while not losing view of the increased scope  afforded by the
possibility of zero or negative
$\icd; \k_1$ or $\k_2$.  

The {\it basic trigonometric identity} for the family of `complex 
Hermitian' spaces is also directly linked to other product
formulas which we believe new. They can be considered as a kind of
`complex Hermitian' Gauss-Bonnet formulas, and contain the
totality of `complex Hermitian' trigonometry in a nutshell, as
they are equivalent to the  {\em basic trigonometric identity}.
The subject of such `exponential product formulas' appears as an
step in our derivation (Section 5.1) but it can be further
developped by itself, and affords a number of new identities; this
will be discussed elsewhere. 

There is actually a strong link between this study, which
superficially seems like a work in geometry, and physics: the
mathematical structure underlying the Quantum State Space belongs
to the `complex hermitian' CKD family. So as a   byproduct of this
work we obtain the basic equations of the `{\em Trigonometry of
the Quantum State Space}' \cite{HermTrigBurgos, QSSTrigGoslar}.
Any  Hilbert space appears  in the CKD family as a Hermitian {\em
euclidean\/} space (thus with labels $\icd=1;\, \k_1=0,\,
\k_2=\k_3=\dots=1$) and its {\em projective} Hilbert space which
plays in any Quantum Theory the role of space of states, appears
as the Hermitian elliptic space ($\icd=1;\, \k_1=\k>0,\,
\k_2=\k_3=\dots =1$).   Geometric phases are related to
trigonometric quantities:  for the simplest `triangle type' loop
in the Quantum state space, the Anandan-Aharonov phase appears
intriguingly as one of the triangle invariants introduced by
Blaschke and Terheggen sixty years ago. The paper by Sudarshan,
Ananadan and Govindarajan
\cite{suda-anan-govi} gives a group theoretical derivation of the
Anandan-Aharonov phase (equal to triangle symplectic area) for an
infinitesimal triangle loop in $\Ce P^N$; this result appears as a 
particular case of our {\em exact} expressions linking triangle
elements for any {\em finite} triangle. The role of symplectic
area for geodesic triangles in connection with coherent states and
geometric phases has also been  recently discussed by Berceanu
\cite{Berceanu} and Boya, Perelomov and Santander \cite{BPS}.  A
separate, more physically oriented paper \cite{HT2} will be 
devoted to the trigonometry of the Quantum space of states, in
relation with geometric phases and in general, with the view
towards a more geometrical formulation of Quantum Mechanics
\cite{Anandan}. 


\section{A review  on Hermitian trigonometry of $\Ce P^2$ and $\Ce
H^2$}

The hermitian elliptic space, i.e.,
$\Ce P^N$  endowed with the natural Fubini-Study (FS) metric
induced  by the {\it real part}  of the hermitian canonical flat
product in
$\Ce^{N+1}$, is an homogeneous {\em hermitian symmetric space}. It
has a natural complex structure, and the FS metric  is 
k\"ahlerian and has constant holomorphic curvature; the K\"ahler
form is  induced by the {\it imaginary part} of the hermitian
canonical metric in
$\Ce^{N+1}$ \cite{Klinbengerg}. The standard choice of scale in
the metric makes the maximum distance in $\Ce P^N$ equal to
$\pi/2$, and the total length of any (closed) geodesic equal to
$\pi$. With this choice the constant {\em holomorphic} curvature
of $\Ce P^N$ is $K_{{hol}}=4$; the ordinary sectional curvature
$K$ of the FS metric in $\Ce P^N$  seen as a riemannian space of
real dimension $2N$ is {\em not} constant, and lies in the
interval $1 \leq K
\leq 4$. Complex projective geometry was studied  by Cartan
\cite{CarLGDC}, building over the works by Study \cite{Study} and
Fubini
\cite{Fubini}.

For the real projective space $\Re P^N$, trigonometry essentially
reduces to  spherical trigo\-nometry \cite{CoxeterNEG}). The
homogeneous symmetric character of $\Ce P^N$  makes also possible
an explicit study of its trigonometry, which is however much more
complicated than for $\Re P^N$.  A common trait in most of the
previous works on hermitian trigonometry is to introduce {\it a
single} real invariant for each side (which seems natural as $\Ce
P^N$  is a {\it rank one} space), but {\it two} real invariants
for each vertex, which also seems natural due to presence of {\em
two} commuting factors in the group theoretical description of the
hermitian elliptic space as the homogeneous space $SU(N+1)/(U(1)
\otimes SU(N))$.

For {\em  side invariants} the canonical choice are the distances
in the FS metric
$a$ (resp.\ $b, c$) between vertices $BC$ (resp.\ $CA,\  AB$). To
avoid non-generic special cases all papers quoted before enforce
the restrictions $a, b, c < \pi/2$; this means that each pair of
sides does not meet the cut locus of the common vertex. In both
the elliptic hermitian space $\Ce P^N$, and the hermitian
hyperbolic space $\Ce H^N$, identified with a suitable bounded
domain of $\Ce P^N$ with the hyperbolic FS metric, each point
$[\vecz]$ is a {\it ray} in the linear ambient space $\Ce^{N+1}$ 
\cite{Bre}. Even if we assume normalized the ambient position
vectors, 
$\hp{\vecz}{\vecz} = 1$, every ray in $\Ce^{N+1}$ still contains
infinitely many normalized vectors differing only by a phase
factor, $[\vecz] \equiv
\{ e^{i \varepsilon} \vecz \}$. Let $\vecz^A, \vecz^B, \vecz^C$
denote arbitrarily chosen normalized position vectors in
$\Ce^{N+1}$ (defined up to a phase factor) for the three vertices
$A \equiv [\vecz^A],\  B\equiv [\vecz^B],\ C\equiv[\vecz^C]$. Then
the length $a$ of the side $a\equiv BC$ can be obtained from the
hermitian product of the two (normalized) vectors
$\hp{\vecz^B}{\vecz^C}$ in the ambient linear space through $\cos
a \exp(i
\epsilon_a): = \hp{\vecz^B}{\vecz^C}$. The phase $\epsilon_a$ is
{\em not} a triangle invariant, as the vectors representing the
vertices can be still modified by arbitrary phase factors.

{\em Vertex invariants} are defined in terms of the tangent space
to the hermitian space which will be considered here  as a {\em
real} vector space with a {\em complex structure}. At each point a
vector tangent to a geodesic
$g$ is defined only up to a nonzero {\em real} factor. The tangent
space to a complex (projective) line $l$ at a point $O$ is a real
2D subspace of the tangent space invariant under the complex
structure and can be thus identified to a complex 1D subspace; this
subspace contains a one-parameter family of real 1D subspaces,
corresponding to a one-parameter family of FS geodesics through 
$O$ and contained in $l$. For {\em vertex invariants} several real
quantities can be used.  In terms of the tangent vectors $u, v$ to
two (real one-dimensional) FS geodesic sides at the vertex $C$,
these are:  1) the {\it hermitian} angle between the sides seen as
complex projective lines, denoted $C$; 2) the ordinary or FS angle
$\fsa$  between $u,v$ computed as usual in the natural riemannian
FS metric in $\Ce P^N$ or $\Ce H^N$ \cite{Nomizu}, 
$g(\,\,,\,):=Re\hp{ \,\ }{\  }$; 3) the FS angle between $iu, v$,
denoted here
$\fscia$. These are defined as:
\be C: = \arccos \left( \frac{\mid \hp{ u}{v} \mid}{\norm{u} \cdot
\norm{ v}}\right) \qquad
\fsa: = \arccos \left( \frac{Re \hp{ u}{v}}{\norm{ u} \cdot \norm{
v}} \right)
\qquad
\fscia: = \arccos \left( \frac{Im \hp{ u}{v}}{\norm{ u} \cdot
\norm{ v}}\right).
\mylabel{ht:DefAngles}
\ee Note $\Psi=\pi/2 -\fscia$ is the minimum value of the
riemannian  FS angle between the tangent vector $u$ and a totally
geodesic
$\Re P^2$  containing the geodesic with tangent vector $v$. 

In addition to, yet not independent from these, one can consider
also 4) the holomorphy `inclination' $\holinc$  of the real 2-flat
tangent to the triangle at the vertex $C$, also called  K\"ahler
angle, inclination angle, holomorphy angle, slant angle, etc.\
between $u$ and $v$ \cite{Schar}. This quantity depends only on
the real 2-flat determined by $u$ and $v$, and not on $u,v$
separately, thus the  names {\em holomorphy inclination} or {\em
K\"ahler inclination} seem more appropriate. In terms of two
vectors
$u, t$  which span the given 2-plane and are  furthermore FS
orthogonal, the {\em holomorphy inclination} $\holinc$ is given by:
\be
\holinc = \arccos \left( \frac{Re \hp{ iu}{t}}{\norm{ u} \cdot
\norm{ t}}\right)\qquad \hbox{where } \,\, t = \alpha u + \beta
v,\quad \alpha, \beta \in \Re, \quad Re
\hp{ u}{t}=0.
\mylabel{ht:DefHolInc}
\ee The holomorphy inclination measures how this real 2-flat
separates from the unique real 2-flat $\Ce_u$ containing $u$ and
{\em invariant} under the complex structure; the FS angle between
$iu$ and $t$ is a natural measure of the separation between these
two 2-planes, since
$\Ce_u=\myspan{u, iu}$ and the given 2-plane is spanned by $u,t$
(both pairs are FS orthogonal). Finally, 5) another angular
invariant $\ap$ of the pair of tangent vectors $u, v$, its
pseudoangle or Kasner angle \cite{Schar} is :
\be
\hp{u}{v} = \, \mid \hp{u}{v}\mid  e^{i \ap} 
\ee This angle has not been explicitly used in previous works on
trigonometry on
$\Ce P^N$ or
$\Ce H^N$; it is generically well defined for any two vectors in
the tangent space at each point of $\Ce P^N$ or
$\Ce H^N$ (i.e.\ between two intersecting FS geodesics with
tangent vectors $u, v$ at the intersection point) but becomes
indeterminate when $u, v$ are FS orthogonal. The angular invariant
$\ap$ is obviously meaningless between {\em complex} lines in
these spaces. 

To sum up, there are several different choices available for two
independent vertex invariants; see the review by Scharnhorst
\cite{Schar}. Authors studying trigonometry have made different
choices and the following relations will be useful for comparing
the proposed trigonometric equations (there are of course similar
relations for the corresponding `angular  invariants' at vertices
$A, B$): 
\bea
\cos\fsaC = \cos C \cos\apC &\qquad
\cos \holincC\, {\sin \fsaC} = \cos C \sin \apC
\mylabel{ht:AngRels1} \\
\sin C = \sin\fsaC \sin\holincC &\qquad
\sin \hsiC = \sin\fsaC \cos\holincC
\mylabel{ht:AngRels2} \\
\cos^2 C = \cos^2 \fsaC + \sin^2\fsaC \cos^2\holincC &\qquad
\sin^2 \fsaC = \sin^2 C + \sin^2 \hsiC
\mylabel{ht:AngRels3}
\eea

In choosing symbols for these angular invariants, we have tried to
conform to the majority usage, but nevertheless we have 
systematically changed to capital letters, which allows a clear
and systematic typographic rendering of the self-duality of the
equations we will propose, by means of the change upper/lower case
letters. 


\subsection{Trigonometry in the Hermitian spaces $\Ce P^2$ and
$\Ce H^2$}

The oldest general result is the Coolidge's (1921) {\em sine
theorem}
\cite{Coo}: the sides $a, b, c$ and angles $A, B, C$ between the
sides seen as complex lines are related by:
\be
\frac{\sin a}{\sin A}=\frac{\sin b}{\sin B}=\frac{\sin c}{\sin C}
\mylabel{ht:ElipCSin}
\ee

The papers by Blaschke and Terheggen (1939) \cite{BlasTer, Ter}
(hereafter BT) contained the first {\em complete} approach to
trigonometry in the elliptic hermitian space, identified with $\Ce
P^2$. Unlike the phase 
$\epsilon_a$ of $\hp{ \vecz^B}{\vecz^C}$ which is meaningless as a
quantity in
$\Ce P^2$, the combination 
$\om: = \epsilon_a +\epsilon_b +\epsilon_c$ is a triangle
invariant, as can be clearly seen in the relation $ \hp{
\vecz^A}{\vecz^B}\hp{ \vecz^B}{\vecz^C}\hp{
\vecz^C}{\vecz^A} =  \cos a\cos b\cos c \exp(i \om) $. BT named
$\omega$ this quantity, but for the reasons explained below we
will change this notation to $\om$.  Let us now consider the
(normalized) position vectors
$\vecZ^a,\ \vecZ^b,\ \vecZ^c$ of the poles $[\vecZ^a],\ [\vecZ^b],\
[\vecZ^c]$ of the three sides  $a, b, c$ defined by BT  in the
ambient space $\Ce^3$ through a `vector product'  as
$\overline{\vecZ^a} = \frac{\vecz^B \times \vecz^C}{\sin a}$, and
cyclically, where the vector product is defined exactly as in the
real case, without complex conjugation in any factor. Then the
dual procedure ($\cos A \exp(i \epsilon_A): =\hp{
\vecZ^b}{\vecZ^c}$ and $\Om: =
\epsilon_A+\epsilon_B +\epsilon_C$), applied to the poles of the
three sides, produces four invariants; three angles $A, B, C$
between sides seen as complex lines  and another quantity  $\Om$
which was called $\tau$  by BT; these four quantities are   dual
to $a, b, c, \om$. BT gave a complete set of equations for the
hermitian elliptic space trigonometry. One is the Coolidge's sine
law (\ref{ht:ElipCSin}), and there are two new equations, which we
will call Blaschke-Terheggen cosine theorem for sides and angles;
the need of {\em four} quantities (e.g.\ $a, b, c,
\om$) to determine a triangle up to isometry in the elliptic 
hermitian space follows from these equations:
\be
\cos^2 a = \frac{\cos^2A +\cos^2B\cos^2C -2\cos A\cos B\cos C \cos 
\Om} {\sin^2B \sin^2C}
\mylabel{ht:ElipBTcos}
\ee
\be
\cos^2 A = \frac{\cos^2a +\cos^2b\cos^2c -2\cos a\cos b\cos c \cos 
\om} {\sin^2b \sin^2c}
\mylabel{ht:ElipBTdualcos}
\ee

Another approach was put forward by Shirokov, in a paper published
posthumously by Rosenfeld. Shirokov took two angular invariants at
each vertex: the riemannian FS angle $\fsaC$ between the two sides
$a, b$ considered  as real-1D FS geodesics and the holomorphy
inclination
$\holincC$ of the real 2-flat spanned at the vertex $C$ by the
real tangent vectors to the two sides $a, b$. The equations which 
we shall call Shirokov-Rosenfeld (SR)  form of sine theorem,
double sine theorem for sides, cosine theorem for sides and double
cosine theorem for sides respectively are:
\vskip-10pt
\be
\frac{\sin a}{\sin \fsaA \sin \holincA}=
\frac{\sin b}{\sin \fsaB \sin \holincB}=
\frac{\sin c}{\sin \fsaC \sin \holincC}
\mylabel{ht:ElipSRsin}
\ee
\vskip-10pt
\be
\frac{\sin 2a}{\sin \fsaA \cos \holincA}=
\frac{\sin 2b}{\sin \fsaB \cos \holincB}=
\frac{\sin 2c}{\sin \fsaC \cos \holincC}
\mylabel{ht:ElipSRsin2}
\ee
\vskip-15pt
\be
\cos^2 a = \big( \cos b\, \cos c + \sin b \,\sin c \, \cos \fsaA
\big)^2 +
\,\sin^2 b\, \sin^2 c \, \cos^2 \holincA \,\sin^2 \fsaA
\mylabel{ht:ElipSRcos}
\ee
\be
\cos 2a = \cos 2b\, \cos 2c + \sin 2b\, \sin 2c \, \cos \fsaA  -
\, 2 \sin^2 b\, \sin^2 c \, \sin^2 \holincA\, \sin^2 \fsaA
\mylabel{ht:ElipSRcos2}
\ee as well as similar cosine and double cosine equations for the
sides
$b, c$. Among all these equations, only {\it five} are functionally
independent (for instance (\ref{ht:ElipSRcos}) and
(\ref{ht:ElipSRcos2}) are equivalent). Note SR sine theorem
(\ref{ht:ElipSRsin}) is equivalent to the Coolidge sine law as
consequence of (\ref{ht:AngRels2}).

In 1989 Wu-Yi Hsiang \cite{Hsiang} gave a new derivation valid
simultaneously for the trigonometry of the two-point homogeneous
rank-one spaces of real, complex, quaternionic and Cayley
octonionic type, both elliptic and hyperbolic. At each vertex, say
$C$, Hsiang uses the three invariants $C, \fsaC,
\fscia_C$  linked by a  relation (\ref{ht:AngRels3}) (recall
$\fscia_C=\pi/2-\hsiC$). In the elliptic/hiperbolic case he
obtained some equations which when translated to $C,
\fsaC, \hsiC$ are given below in (\ref{ht:ElipBrehm}),
(\ref{ht:HypBrehm}) as well as a rather complicated form of
`cosine theorem', not reproduced here and that should not be
considered as a `basic' equation.

In 1990 Brehm \cite{Bre} gave a fresh approach to trigonometry of
both elliptic $\Ce P^N$ and hyperbolic $\Ce H^N$ hermitian spaces.
In terms of three angular invariants $\fsaC, C, \hsiC$, only two
of which are independent, Brehm derived the following equations:
\be
\begin{array}{l}
\displaystyle
\frac{\sin a}{\sin A}=\frac{\sin b}{\sin B}=\frac{\sin c}{\sin
C}\qquad\qquad
\frac{\sin 2a}{\sin \hsiA}=\frac{\sin 2b}{\sin
\hsiB}=\frac{\sin2c}{\sin \hsiC} \\[9pt]
\cos 2a = \cos 2b \cos 2c + \sin 2b \sin 2c \cos \fsaA - 2 \sin^2
b \sin^2 c \sin^2 A
\mylabel{ht:ElipBrehm}
\end{array}
\ee
\be
\begin{array}{l}
\displaystyle
\frac{\sinh a}{\sin A}=\frac{\sinh b}{\sin B}=\frac{\sinh c} {\sin
C}
\qquad\qquad
\frac{\sinh 2a}{\sin \hsiA}=\frac{\sinh 2b}{\sin
\hsiB}=\frac{\sinh 2c}{\sin
\hsiC}\\[9pt]
\cosh 2a = \cosh 2b \cosh 2c + \sinh 2b \sinh 2c \cos \fsaA - 2
\sinh^2 b \sinh^2 c \sin^2 A.
\end{array}
\mylabel{ht:HypBrehm}
\ee The first two equations are Coolidge's sine law and the Hsiang
form of double sine law. The third ones turns out to be  simply the
Shirokov-Rosenfeld cosine double theorem for sides expressed in
terms of Brehm's angular variables. The need of {\em four}
quantities to determine a triangle up to isometry in
$\Ce P^N$ or $\Ce H^N$ was stressed by Brehm who introduced the
{\em shape invariant} $\sigma$ of the triangle, defined in the
elliptic case as 
$\sigma  := \hbox{\rm Re}
\hp{ \vecz^A}{\vecz^B}
\hp{  \vecz^B}{\vecz^C}
\hp{  \vecz^C}{\vecz^A} = \cos a\cos b\cos c \cos \om$, and as 
$\sigma = - \cosh a\cosh b\cosh c \cos\om$ in the hyperbolic case;
Brehm showed a triangle is completely determined up to isometry by
$a, b, c,
\sigma$ (recall in the elliptic case Brehm assumes $a, b, c<
\pi/2$), and gave  inequalities that must be fulfilled in order
the triangle  to exist, as well as a careful discussion on
congruence theorems. 

In 1994 Hangan and Masala \cite{HanMas} gave an interpretation of
$\om$ in the complex projective space $\Ce P^2$ as equal to twice
the symplectic area enclosed by the triangle. Symplectic area
comes from the K\"ahler structure of $\Ce P^2$, and is well
defined by the triangle `skelethon' itself, due to the closed
nature of the K\"ahler form which makes the symplectic area of any
surface with given boundary to depend only on the boundary.

The  existence of two distinguished, non generic types of
triangles is clear.  In $\Ce P^2$, when
$\holincA=0$, then $\holincB=\holincC=0$ follows and the
trigonometry equations reduce to those of a spherical triangle in
a sphere of curvature
$K=4$; this is seen in (\ref{ht:ElipSRsin2}) and
(\ref{ht:ElipSRcos2}) and corresponds to a triangle  completely
contained in a complex line $\Ce P^1$. When $\holincA=\pi/2$, then
$\holincB=\holincC=\pi/2$, the equations reduce (locally) to those
of  a  spherical triangle in curvature $K=1$; this can be seen in
(\ref{ht:ElipSRsin}) and (\ref{ht:ElipSRcos}) and corresponds to a
triangle  completely contained in a real projective subplane $\Re
P^2$, whose trigonometry comes from the spherical one after
antipodal identification. This case corresponds to {\em real}
values for
$\exp(i \om)$ and $\exp(i \Om)$, as implied by the
Blaschke-Terhegen equations (\ref{ht:ElipBTcos}) and
(\ref{ht:ElipBTdualcos}).

In these two special cases,
contained in a totally geodesic submanifold, whose sectional
curvature attains the extremal values $1$ and $4$. In all other
situations, the  triangle is {\em not} contained in a totally
geodesic submanifold. The sectional curvature of either $\Ce P^2$
or
$\Ce H^2$ along any real 2-direction depends only on its
holomorphy inclination
$\holinc$ and is:
\be K = \pm \Big(4 \cos^2\holinc + \sin^2\holinc \Big).
\ee


\section{The family of nine complex Hermitian Cayley--Klein 2D
geometries and their spaces}

\subsection{The nine complex Hermitian Cayley--Klein 2D geometries}

Let us consider a complex hermitian form $(z, w) \to \hp{ z}{ w} =
\sum_{i,j}^N \overline{z_i}\,\, \Lambda_{ij} \,w_j$ in an ambient
complex linear space $\Ce^{N+1}=(z^0,z^1\dots z^N)$, where the
symmetric real matrix $\Lambda$ is a diagonal matrix with entries
$\{ 1, \k_1, \k_1\k_2,
\dots, \k_1\k_2\cdots\k_N\}$, depending on $N$ real numbers
$\k_i$.  Linear isometries in $\Ce^{N+1}$ for such a hermitian
product close the special unitary CK families of groups
$SU_{\k_1,\k_2, \dots, \k_N}(N+1)$ with Lie algebras
$\goth{su}_{\k_1,\k_2, \dots, \k_N}(N+1)$.  The structure of
algebras in this family (any dimension) as well as the associated
unitary
$U_{\k_1,\k_2, \dots, \k_N}(N+1)$ and $\goth{u}_{\k_1,\k_2, \dots,
\k_N}(N+1)$ CK families is described in \cite{ExtSU}. 

When particularised for $N=2$ a two-parametric family
$SU_{\k_1,\k_2}(3)$ of groups is obtained; these are the
eight-dimensional linear isometry groups of a complex hermitian
form $(z, w) \to \hp{ z}{ w} = \sum_{i,j}^3
\overline{z_i} \Lambda_{ij} w_j$ in an ambient linear space
$\Ce^3=(z^0,z^1,z^2)$, with symmetric real matrix $\Lambda={\rm
diag}\{1, \k_1, \k_1\k_2\}$. In the natural CK basis 
$\{P_1,P_2,Q_1, Q_2; J, M, B, I\}$ the CK algebra
$\goth{su}_{\k_1,\k_2}(3)$   has the following (fundamental or
vectorial) 3D complex matrix representation, where $i$ stands for
the pure imaginary complex unit:
\bea P_1=\left( \begin{array}{ccc}
 0   & -\k_1  & 0   \\
 1  &  0   & 0  \\ 0  &  0  & 0 \\
\end{array} \right) \quad P_2=\left( \begin{array}{ccc}
   0 & 0   &  -\k_1\k_2 \\ 0  &  0   & 0  \\ 1  &   0 & 0 \\
\end{array} \right) \quad J=\left( \begin{array}{ccc} 0&  0   & 0 
\nonumber \\
          0 &0     &  -\k_2 \\
         0  &  1  & 0 \\
\end{array} \right) \\ Q_1=\left( \begin{array}{ccc}
    0& i\k_1  & 0   \\
 i  &   0  & 0  \\ 0  & 0   & 0 \\
\end{array} \right) \quad Q_2=\left( \begin{array}{ccc}
  0  &  0  &  i\k_1\k_2  \\ 0  &  0   & 0  \\ i  &  0  & 0 \\
\end{array} \right) \quad M=\left( \begin{array}{ccc} 0&  0   &
0   \\
        0   & 0 &  i\k_2 \\
       0    &  i  & 0 \\
\end{array} \right) \mylabel{ht:AlgFRep}\\ B=\left(
\begin{array}{ccc} 0&  0  &  0  \\
  0  &  -i & 0  \\ 0   & 0   &  i \\
\end{array} \right) \qquad I=\left( \begin{array}{ccc}
\frac{-2i}{3}&  0  &  0  \\
  0  & \frac{i}{3}& 0  \\
   0 &  0  &  \frac{i}{3} \\
\end{array} \right). \qquad\qquad\qquad
 \nonumber
\eea  When $\k_1, \k_2$ are different from zero,
$\goth{su}_{\k_1,\k_2}(3)$ is simple. This algebra is isomorphic
to $\goth{su}(3)$ when both $\k_1, \k_2$ are positive or to
$\goth{su}(2,1)$ when at least one is negative.

The generators $B, I$ span a two-dimensional Cartan subalgebra
(regardless of the values of $\k_i$ ); further references to {\em
the Cartan subalgebra} will mean to this fiducial subalgebra. Let
us introduce four new Cartan generators:
\vskip-13pt
\be T_1= \frac{1}{2}(I+B)\quad    T_2=\frac{1}{2}(I-B) \quad
H_1=\frac{1}{2}(3I-B) \quad H_2=\frac{1}{2}(3I+B) 
\mylabel{ht:CartanEDefs}
\ee
\vskip-5pt
\noindent whose representing matrices in the vector fundamental
representation are:
\bea T_1=\left( \begin{array}{ccc} -\frac{i}{3}&  0  &  0  \\
  0  &  -\frac{i}{3} & 0  \\ 0   & 0   &  \frac{2i}{3} \\
\end{array} \right) \qquad T_2=\left( \begin{array}{ccc}
-\frac{i}{3}&  0  &  0  \\
  0  &  \frac{2i}{3} & 0  \\ 0   & 0   &  -\frac{i}{3} \\
\end{array} \right) \nonumber \\[-6pt]
\mylabel{ht:CartanFRes} \\ H_1=\left( \begin{array}{ccc} -i&  0 
&  0  \\
  0  &  i & 0  \\ 0   & 0   &  0 \\
\end{array} \right) \qquad H_2=\left( \begin{array}{ccc} -i&  0 
&  0  \\
  0  & 0& 0  \\
   0 &  0  &  i\\
\end{array} \right). \quad\nonumber 
\eea   The Lie commutators of all these generators are given in 
(\ref{eq:HermCKDBiDConmRels}) for $\icd=1$. The CK algebras
$\goth{su}_{\k_1,\k_2}(3)$ can  be endowed with a ${\Ze}_2\otimes
{\Ze}_2$ group of commuting involutive automorphisms generated by:
\be
\begin{array}{ll}
\Invol_{(1)} \!\!\!\!&: (P_1,P_2,Q_1, Q_2; J, M, B, I)\to
(-P_1,-P_2,-Q_1, -Q_2; J, M, B, I)\cr
\Invol_{(2)} \!\!\!\!&: (P_1,P_2,Q_1, Q_2; J, M, B, I)\to
(P_1,-P_2,Q_1, -Q_2; -J, -M, B, I).
\mylabel{ht:Invols}
\end{array}
\ee The two remaining involutions are  the composition
$\Invol_{(02)}=\Invol_{(1)}\cdot \Invol_{(2)}$ and the identity.
Each involution $\Invol$ determines  a subalgebra of
$\goth{su}_{\k_1,\k_2}(3)$, denoted $\goth{h}$, whose elements are
invariant under $\Invol$; the subgroups generated $\goth{h}$ will
be denoted $H$, all with suitable subindices. By direct checking
one can assure that the  three Lie subalgebras $\goth{h}_{(1)}$,
$\goth{h}_{(2)}$ and $\goth{h}_{(02)}$ are  of unitary CK type,
$\goth{u}_{\k}(2) \equiv  \goth{u}(1) \oplus \goth{su}_{\k}(2)$
with
$\k=\k_2, \k_1, \k_1\k_2$ respectively. Namely:

\noindent$\bullet$ The subalgebra $\goth{h}_{(1)}$ is spanned by 
 $I; J, M, B$ which  close an
$\goth{u}_{\k_2}(2)$ (with
$I$ commuting with $J, M, B$). The group $H_{(1)}$ they generate is
isomorphic to $U(1) \otimes SU_{\k_2}(2)$.

\noindent$\bullet$ The subalgebra $\goth{h}_{(2)}$ is spanned by 
 $T_1; P_1, Q_1, H_1$ which close an $\goth{u}_{\k_1}(2)$ (with
$T_1$ commuting with $P_1, Q_1, H_1$). The group
$H_{(2)}$ they generate is isomorphic to $U(1) \otimes
SU_{\k_1}(2)$.

\noindent$\bullet$ The subalgebra $\goth{h}_{(02)}$ is spanned by
 $T_2; P_2, Q_2, H_2$ closing an
$\goth{u}_{\k_1\k_2}(2)$ (with $T_2$ commuting with $P_2, Q_2,
H_2$). The group $H_{(02)}$ they generate is isomorphic to $U(1)
\otimes SU_{\k_1\k_2}(2)$.

All these generators can be represented in a pictorial way in a
block triangular diagram,
\be
\begin{tabular}{ll}
$\pp \qq$&$\ppp \qqq$ \\
$ \tt \hh$&$ \ttt \hhh$  \\
\\[-5pt] &$\jj \mm$ \\ &$\dd\, \ll$
\end{tabular}
\mylabel{ht:GenDiagram}
\ee where each `block' involves the four generators of the
$\goth{u}_{\k}(2)$ subalgebras listed above; the generator in the
$\goth{u}(1)$ subalgebra inside the center of each unitary
subalgebra
$\goth{u}_{\k}(2)$ appears at the left-lower corner in each block.
The global block pattern reproduces the pattern 
\begin{tabular}{l}
$\pp \ppp$\\
$ \phantom{\pp}\jj$
\end{tabular} made by the three generators $P_1, P_2, J$ of a real
type CK algebra $\goth{so}_{\k_1,\k_2}(3)$. This diagram will be
extremely helpful for visualization of most properties discussed
below.

There is a single quadratic Lie algebra Casimir in
$\goth{su}_{\k_1,\k_2}(3)$. A good way of writing it, with each
group of terms corresponding to one of the three $su_{\k}(2)$-like
subalgebras is:
\be {\cal C}=\k_2 ((P_1^2 +Q_1^2)+ \k_1 H_1^2) +
         ((P_2^2+Q_2^2)+ \k_1\k_2 H_2^2)+
         \k_1 ((J^2 + M^2) + \k_2 B^2).
\mylabel{ht:Casimir}
\ee

The elements defining a  2D CK complex hermitian geometry are
analogous  to the ones in the real case \cite{SpaceTimeTrig,
SanHerrDubna96}. By a {\it two-dimensional complex CK geometry} we
will understand the set of three symmetric homogeneous spaces of
points and lines of first- and second--kind. 

\noindent
$\bullet$ The {\it plane} as the set of points corresponds to the
symmetric homogeneous space
\be
\CeCKPointSpace\equiv SU_{\k_1,\k_2}(3)/H_{(1)}\equiv
SU_{\k_1,\k_2}(3)/(U(1)\otimes SU_{\k_2}(2))
\qquad  H_{(1)}=\myspan{I; J, M, B}
\mylabel{ht:CKPointSpace}
\ee whose dimension (over $\Ce$) is $2$. The generators $I$ and 
$J, M, B$ leave a point $O$ (the origin) invariant, and so
generate a direct product $U(1) \otimes SU_{\k_2}(2)$ of
`rotations' about $O$. The involution $\Invol_{(1)}$ is the 
reflection around $O$ and $P_1, Q_1$ (resp. $P_2, Q_2$) move $O$
and generate translations along the (complex) {\it basic}
direction $l_1$ (resp. $l_2$.)

\noindent
$\bullet$ The set of first-kind {\it complex lines} is identified
to the symmetric homogeneous space
\be
\CeCKLineSpace\equiv SU_{\k_1,\k_2}(3)/H_{(2)}\equiv
SU_{\k_1,\k_2}(3)/(U(1)\otimes SU_{\k_1}(2))
\qquad  H_{(2)}=\myspan{T_1;  P_1, Q_1, H_1}
\mylabel{ht:CKLineSpace}
\ee with dimension 2 over $\Ce$. The generators $T_1$ and $P_1,
Q_1, H_1$ should be  interpreted in $\CeCKLineSpace$ as the
generators of `rotations' about the `origin'  line $l_1$, which is
left invariant by them. The point $O$ is  moved along two
(complex) basic  directions by  $J, M$ and $P_2, Q_2$. The
reflexion in $l_1$ is $\Invol_{(2)}$.  Complex lines obtained by
group motions from the basic fiducial line $l_1$ will be  called
first-kind lines.  

\noindent
$\bullet$ There is another set of complex lines, the complex-2D
symmetric homogeneous space
\be SU_{\k_1,\k_2}(3)/H_{(02)}\equiv
SU_{\k_1,\k_2}(3)/(U(1)\otimes SU_{\k_1\k_2}(2))
\qquad  H_{(02)}=\myspan{T_2;  P_2, Q_2, H_2}.
\mylabel{ht:CKLine2Space}
\ee In  this space $T_2$ and $P_2, Q_2, H_2$ leave invariant an
`origin' line $l_2$  while $J, M$ and $P_1, Q_1$ move it. The
reflexion in $l_2$ is $\Invol_{(02)}$.  These lines will be called
second-kind. They are  actually different from first-kind ones
only when $\k_2\leq 0$, since when $\k_2>0$ $P_1, Q_1$ and
$P_2, Q_2$ are conjugate within $\goth{su}_{\k_1,\k_2}(3)$.

Consideration of the spaces of first and second kind lines can be
bypassed, since  lines can be seen not as `points' in the spaces
$\CeCKLineSpace$ or
$SU_{\k_1,\k_2}(3)/H_{(02)}$, but alternatively as 1D complex
submanifolds of
$\CeCKPointSpace$, and all properties of the two spaces of lines
can be transcribed in terms of this space, in which  $l_1$ and
$l_2$ should be considered as two hermitian orthogonal complex
lines  intersecting at $O$ (see figure \ref{CKD:labelgenerators}
for $\icd=1$).  This space $\CeCKPointSpace$  has a {\em complex
hermitian metric}  with an associated {\em real `Fubini-Study' 
metric} (`FS') given by the real part of the hermitian product.
This `FS' metric can also be derived directly from the Casimir
(\ref{ht:Casimir}): at the origin $O$ the hermitian product is
given by the matrix 
$\mbox{diag}(1,\k_2)$, and the `FS' metric by
$\mbox{diag}(1,1,\k_2, \k_2)$ (basis ordering $P_1, Q_1, P_2,
Q_2$); at other points they are uniquely determined by
invariance.  This `FS' metric is definite positive when
$\k_2>0$, degenerate for
$\k_2=0$ and indefinite of real type $(2,2)$ for $\k_2<0$; when
$\k_2=1$ it is the ordinary FS metric (elliptic or hyperbolic) with
holomorphic curvature $4\k_1$. The line $l_1$ (resp.\ $l_2$)
contains two `FS' orthogonal geodesics through $O$, the orbits of
$O$ by the one-parameter subgroups generated by $P_1$ and $Q_1$
(resp.\
$P_2$ and $Q_2$).

Thus $\k_1$ is (one fourth of) the constant holomorphic curvature, 
and $\k_2$ determines the signature of both the hermitian metric
and the `FS'  metric, hereafter denoted as FS for any space in the
CKD family. The  canonical conexion of $\CeCKPointSpace$ as
homogeneous symmetric  space \cite{Nomizu} is compatible  with the
FS metric.   A suitable rescaling of generators $P_1, J_{12}$
allows to reduce 
$\k_1$ (resp.\ $\k_2$) to $\pm 1$. Thus nine 2D-complex hermitian
CK geometries are obtained; their groups of motion and  isotopy
subgroups are displayed in table \ref{table:NineCKGeometries}.

\begin{table}[h] {\footnotesize
 \noindent
\caption{{The nine two-dimensional complex hermitian CK
geometries. At each entry the group $G$ and the three subgroups
$H_{(1)}, H_{(2)}, H_{(02)} $ are
displayed}}\label{table:NineCKGeometries}
\smallskip
\noindent\hfill
\begin{tabular}{llll}
\hline
 &\multicolumn{3}{c}{Measure of distance}\\
\cline{2-4} Measure&Elliptic&Parabolic&Hyperbolic\\ of
angle&$\k_1=1$&$\k_1=0$&$\k_1=-1$\\
\hline &Hermitian Elliptic&Hermitian Euclidean&Hermitian
Hyperbolic\\ &$SU(3)$&$IU(2)$&$SU(2,1)$\\
Elliptic&$H_{(1)}=U(1)\otimes SU(2)$&$H_{(1)}=U(1)\otimes
SU(2)$&$H_{(1)}=U(1)\otimes SU(2)$\\
$\k_2=1$&$H_{(2)}=U(1)\otimes SU(2)$&$H_{(2)}=U(1)\otimes
IU(1)$&$H_{(2)}=U(1)\otimes SU(1,1)$\\ &$H_{(02)}=U(1)\otimes
SU(2)$&$H_{(02)}=U(1)\otimes IU(1)$&$H_{(02)}=U(1)\otimes
SU(1,1)$\\
\hline &Hermitian Co-Euclidean&Hermitian Galilean&Hermitian
Co-Minkowskian\\ &Hermitian Oscillating NH & &Hermitian Expanding
NH\\ &$IU(2)$&$IIU(1)$&$ISU(1,1)$\\ Parabolic&$H_{(1)}=U(1)\otimes
IU(1)$&$H_{(1)}=U(1)\otimes IU(1)$&$H_{(1)}=U(1)\otimes IU(1)$\\
$\k_2=0$&$H_{(2)}=U(1)\otimes SU(2)$&$H_{(2)}=U(1)\otimes
IU(1)$&$H_{(2)}=U(1)\otimes SU(1,1)$\\ &$H_{(02)}=U(1)\otimes
IU(1)$&$H_{(02)}=U(1)\otimes IU(1)$&$H_{(02)}=U(1)\otimes IU(1)$\\
\hline &Hermitian Co-Hyperbolic&Hermitian Minkowskian&Hermitian
Doubly Hyperbolic\\ &Hermitian Anti-de Sitter& &Hermitian De
Sitter\\ &$SU(2,1)$&$IU(1,1)$&$SU(2,1)$\\
Hyperbolic&$H_{(1)}=U(1)\otimes SU(1,1)$&$H_{(1)}=U(1)\otimes
SU(1,1)$&$H_{(1)}=U(1)\otimes SU(1,1)$\\
$\k_2=-1$&$H_{(2)}=U(1)\otimes SU(2)$&$H_{(2)}=U(1)\otimes
IU(1)$&$H_{(2)}=U(1)\otimes SU(1,1)$\\ &$H_{(02)}=U(1)\otimes
SU(1,1)$&$H_{(02)}=U(1)\otimes IU(1)$&$H_{(02)}=U(1)\otimes
SU(2)$\\
\hline
\end{tabular}\hfill}
\end{table}

A fundamental property of the whole scheme of CK geometries is the
existence of an  `automorphism' of each family, called {\em
ordinary duality} $\dual$. It is well defined for any dimension,
and for the 2D case it is given by the following family
automorphism:
\be
\begin{array}{l}
\dual :\cases{     
        (P_1,Q_1, P_2,Q_2; J, M, H_2, T_2) \to
        (-J,- M, -P_2,- Q_2; -P_1, - Q_1, H_2, - T_2)\\
        (\k_1, \k_2 ) \to (\k_2, \k_1) }.
\mylabel{ht:CKBiDDuality}
\end{array}
\ee Duality $\dual$ leaves the general commutation rules
(\ref{eq:HermCKDBiDConmRels}) invariant while it interchanges the
corresponding constants $\k_1\leftrightarrow\k_2$ and the space of
points with the space of first-kind lines,
$\CeCKPointSpace\leftrightarrow \CeCKLineSpace$, preserving the
space of second-kind lines. It relates in general {\em two}
different  geometries  placed in symmetrical positions relative to
the main diagonal in table \ref{table:NineCKGeometries}, just like
in the real case.  Duality also underlies the introduction of the
Cartan generators (\ref{ht:CartanEDefs}): $B, I$ form a natural
basis for the fiducial Cartan subalgebra ($B$ is the unique Cartan
generator in the $SU_{\k_2}(2)$ part, and $I$ in the $U(1)$ part,
of  the isotopy subalgebra of a point in $\CeCKPointSpace$), $H_1,
T_1$ appear as their duals, and $H_2, T_2$ have the simplest
behaviour under duality.  In terms of the block-triangular
arrangement (\ref{ht:GenDiagram}), duality  corresponds to a
`block reflection' along the secondary diagonal and an eventual
sign change. For Cartan generators duality can be depicted by 
(figure \ref{cartandualidad}), related to the  $\goth{su}(3)$ root
diagram. More details on the geometric interpretation of the
Cartan subalgebra generators $B, I, T_1, T_2, H_1, H_2$ will be
given later. 


\setlength{\unitlength}{0.1mm}
\begin{figure}[th]
\begin{center}
\begin{picture}(100,320)(65,-10)
\put(0,200){\circle*{9}}
\put(0,300){\circle*{9}}
\put(200,0){\circle*{9}}
\put(300,0){\circle*{9}}
\put(100,200){\circle*{9}}
\put(200,200){\circle*{9}}
\put(200,100){\circle*{9}}
\put(0,0){\line(0,1){300}}
\put(0,0){\line(1,0){300}}
\put(0,0){\line(1,1){300}}
\put(100,0){\line(0,1){300}}
\put(200,0){\line(0,1){300}}
\put(300,0){\line(0,1){300}}
\put(0,100){\line(1,0){300}}
\put(0,200){\line(1,0){300}}
\put(0,300){\line(1,0){300}}
\put(100,100){\vector(1,-1){90}}
\put(100,100){\vector(-1,1){90}}
\put(150,150){\vector(1,-1){45}}
\put(150,150){\vector(-1,1){45}}
\qbezier(290,10)(190,190)(10,290)
\put(290,10){\vector(1,-1){2}}
\put(10,290){\vector(-1,1){2}}
\put(-20,300){\makebox(0,0){$B$}}
\put(-35,218){\makebox(0,0){$- T_2$}}
\put(80,218){\makebox(0,0){$T_1$}}
\put(175,218){\makebox(0,0){$H_2$}}
\put(185,80){\makebox(0,0){$I$}}
\put(80,108){\makebox(0,0){$\bf 0$}}
\put(200,-23){\makebox(0,0){$T_2$}}
\put(300,-23){\makebox(0,0){$H_1$}}
\end{picture}
\caption{Fiducial Cartan subalgebra generators and their behaviour
under duality.}
\label{cartandualidad}
\end{center}
\end{figure}
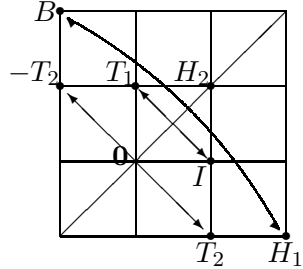
\vspace{-.5cm}


\subsection{Realization of the spaces of points in the complex
Hermitian Cayley--Klein spaces}

Exponentiation of the matrix representation (\ref{ht:AlgFRep}) and
(\ref{ht:CartanFRes}) of
$\goth{su}_{\k_1,\k_2}(3)$  produces a representation of
$SU_{\k_1,\k_2}(3)$ as a linear transformations group  in an
ambient linear space $\Ce^3=(z^0,z^1,z^2)$. The one-parametric
subgroups generated by $P_1, P_2, Q_1, Q_2, J$ and $M$ are:
\be
\begin{tabular}{ll}
$e^{\pp x} = \left(\matrix{C_{\k_1}(x)&-\k_1 S_{\k_1}(x)&0\cr
S_{\k_1}(x)&C_{\k_1}(x)&0\cr 0&0&1\cr}\right)$ &
$e^{\ppp x} = \left(\matrix{C_{\k_1\k_2}(x)&0&-\k_1\k_2 
S_{\k_1\k_2}(x)\cr 0&1&0\cr
S_{\k_1\k_2}(x)&0&C_{\k_1\k_2}(x)\cr}\right)$
\\
$e^{\qq x} = \left(\matrix{C_{\k_1}(x)&i\k_1 S_{\k_1}(x)&0\cr i
S_{\k_1}(x)&C_{\k_1}(x)&0\cr 0&0&1\cr}\right)$ &
$e^{\qqq x} = \left(\matrix{C_{\k_1\k_2}(x)&0&i\k_1\k_2 
S_{\k_1\k_2}(x)\cr 0&1&0\cr
iS_{\k_1\k_2}(x)&0&C_{\k_1\k_2}(x)\cr}\right)$
\\ \mylabel{ht:CKBiDClasSubgroups} \\
$e^{\jj x}=\left( \matrix{1&0&0\cr0&C_{\k_2}(x)&-\k_2
S_{\k_2}(x)\cr 0&S_{\k_2}(x)&C_{\k_2}(x)\cr}\right)$ &
$e^{\mm x}=\left( \matrix{1&0&0\cr0&C_{\k_2}(x)&i\k_2
S_{\k_2}(x)\cr 0&iS_{\k_2}(x)&C_{\k_2}(x)\cr}\right)$
\end{tabular}
\ee where the cosine $C_\k(x)$ and sine $S_\k(x)$ functions with
label $\k$ are defined by:
\be C_{\k}(x) : =\left\{
\begin{array}{ll}
  \cos {\sqrt{\k}\, x} & \cr
  1  & \cr
\cosh {\sqrt{-\k}\, x} &
\end{array}\right.
\quad  S_{\k}(x) : =\left\{
\begin{array}{ll}
    \frac{1}{\sqrt{\k}} \sin {\sqrt{\k}\, x} &\quad  \k >0 \cr
  x &\quad \k  =0 \cr
\frac{1}{\sqrt{-\k}} \sinh {\sqrt{-\k}\, x} &\quad \k <0
\end{array}\right. .
\mylabel{ht:CKSineCosine}
\ee These functions coincide with the circular and hyperbolic
trigonometric ones for $\k=1$ and $\k=-1$; the case $\k=0$ 
provides the so called `parabolic' or galilean functions:
$C_{0}(x)=1$,
$S_{0}(x)=x$. General properties of these functions are given in
the Appendix in \cite{SpaceTimeTrig}.

The exponentials of the Cartan subalgebra generators $\ll, \dd,
\tt,
\ttt, \hh, \hhh$ are:
\be
\begin{tabular}{lll}
$e^{\ll x}=\left(
\matrix{1&0&0\cr0&e^{-ix}&0\cr0&0&e^{ix}\cr}\right)$
 &
$e^{\dd x}=\left(
\matrix{e^{\frac{-
2ix}{3}}&0&0\cr0&e^{\frac{ix}{3}}&0\cr0&0&e^{\frac{ix}{3}
}\cr}\right)$
\\[20pt]
$e^{\tt x}=\left(
\matrix{e^{\frac{-ix}{3}}&0&0\cr0&e^{\frac{-
ix}{3}}&0\cr0&0&e^{\frac{2ix}{3 }}\cr}\right)$
 &
$e^{\ttt x}=\left(
\matrix{e^{\frac{-ix}{3}}&0&0\cr0&e^{\frac{2ix}{3}}&0\cr0&0&e^{\frac{-
ix}{3 }}\cr}\right)$
\mylabel{ht:CKBiDCartSubgroups}
\\[20pt]
$e^{\hh x}=\left(
\matrix{e^{-ix}&0&0\cr0&e^{ix}&0\cr0&0&1\cr}\right)$
 &
$e^{\hhh x}=\left(
\matrix{e^{-ix}&0&0\cr0&1&0\cr0&0&e^{ix}\cr}\right).$
\end{tabular}
\ee Any element $U\in SU_{\k_1,\k_2}(3)$, satisfying
$U^{\dagger}\, \Lambda\, U =\Lambda, \ 
\det U =1$  where
$U^{\dagger}=\overline{U^{{T}}}$, can be written as a product of
matrices (\ref{ht:CKBiDClasSubgroups}) and two commuting `Cartan'
transformations in (\ref{ht:CKBiDCartSubgroups}). The action of
$SU_{\k_1,\k_2}(3)$ on
$\Ce^3$ is linear but not transitive, since it conserves the
hermitian form
$|z^0|^2+\k_1 |z^1|^2+ \k_1\k_2 |z^2|^2$. The isotopy subgroup of 
$O=(1,0,0)$ is the three parameter subgroup $SU_{\k_2}(2)$
generated by $\myspan{J, M, B}$, while the
$U(1)$ subgroup generated by $I$ multiplies $O$ by a phase factor.
Hence the homogeneous symmetric space $\CeCKPointSpace\equiv
SU_{\k_1,\k_2}(3)/(U(1) \otimes SU_{\k_2}(2))$ can be identified
to the orbit of the {\em ray} $[O]$ of the vector $O$ under the
action of  $SU_{\k_1,\k_2}(3)$.  This orbit is the domain of  $\Ce
P^2$ determined by  
$|z^0|^2+\k_1 |z^1|^2+ \k_1\k_2 |z^2|^2 > 0$, and when
$\k_1>0, \k_2>0$ it is the full complex projective space $\Ce
P^2$.  The coordinates $(z^0, z^1, z^2)$ can be called {\em
Weierstrass coordinates};  they are linked by
$|z^0|^2+\k_1 |z^1|^2+
\k_1\k_2 |z^2|^2=1$ and still are defined up to a common
unimodular complex factor which can be used to make $z^0$ real and
non negative; these are the natural coordinates in the {\em vector
models} of the Hermitian CK spaces, since  the motion groups act
linearly on them. Also in analogy with the real case,
$(z^1/z^0,z^2/z^0)$ are called {\em Beltrami coordinates}; the
groups act on these coordinates by complex fractional linear
transformations.

The non-generic situation where $\k_1, \k_2$ vanishes corresponds
to an In\"on\"u--Wigner  contraction \cite{IW}.  The limit
$\k_1\to 0$ is a local contraction (around a point); it carries
the first and third columns of table
\ref{table:NineCKGeometries} to the flat middle one. The limit
$\k_2\to 0$ is an axial contraction (around a line), carrying
geometries of first and third rows to the middle one. 


\section{The complete family of `complex Hermitian'
Cayley--Klein--Dickson 2D geometries and their spaces}

The previous section has been written so it can be re-read with
minimal {\em mutatis mutandis} changes to suit the description of
the full family of `complex type' spaces. The new fact is the
explicit  Cayley-Dickson (CD) label
$\icd$ in the `labelled' CD doubling $\Re \to \iRe$: $x \to x+i
y$, where $i^2=-\icd$. There are three different cases: $\icd>1$
can be rescaled to $\icd=1$ and gives the division algebra of
ordinary complex numbers ${}_{+1}\Re \equiv \Ce$;
$\icd=0$ gives the {\em dual} or {\em Study} numbers
${}_{0}\Re\equiv \Ce_{0}$; and $\icd<0$, which can be rescaled to
$\icd=-1$ the {\em split complex} numbers
${}_{-1}\Re \equiv \Ce_{-1}$, also called {\em double} numbers,
hyperbolic complex numbers, Lorentz numbers or perplex numbers.
These are three instances of a one-parameter system 
${}_{\icd}\Re\equiv\Ce_{\icd}$, the notation duplicity stressing
either the `complex' nature of the numerical system here obtained
($\Ce_{\icd}$) or its character as a CD  extension of $\Re$
$({}_{\icd}\Re)$. 

\vspace{-3mm}
\subsection{The `complex Hermitian' Cayley--Klein--Dickson 2D
geometries}

The groups behind these geometries are the linear isometry groups
of a `complex hermitian' form $(z, w) \to \hp{ z}{ w} = \sum_{i,j}
\overline{z_i}\,\, \Lambda_{ij} w_j$ in the $N+1$ dimensional
`complex' ambient linear space $\iCe^{N+1}\equiv
\iRe^{N+1}=(z^0,z^1,\dots z^N)$ with the same $\Lambda$ as in the
complex case. For
$N=2$ the CKD  algebra, denoted
${}_{\icd}\goth{su}_{\k_1,\k_2}(3)$ is eight dimensional, and its
fundamental or vectorial 3D `complex'  representation is given by
$3\times 3$ matrices (\ref{ht:AlgFRep}, \ref{ht:CartanFRes}), 
where now entries are in $\iCe$ and $i$ stands for the pure
imaginary `complex' unit in $\iCe$. This form has `hermitian'
symmetry $\hp{ w}{z} = \overline{\hp{ z}{ w}}$, with `complex'
conjugation in $\iCe$: $z=a+ib \to
\overline{z} = a-ib$, for real $a, b$ and the form $(z, w) \to Im
\hp{z}{w}$ is still real and antisymmetric in $z, w$, and
therefore is a symplectic form in the real space $\Re^{2(N+1)}$
underlying to $\iRe^{N+1}$. Rosenfeld
\cite{RosGeoLieGroups} uses the word hermitian without qualifying,
but to prevent misunderstandings, we will keep the term {\em
hermitian} for the truly complex case, and we will put quotes in
`complex' and `hermitian' when refering to the general `complex'
numbers
$\iCe$ with Cayley-Dickson label $\icd$. A simple scale change may
reduce simultaneously
$\icd$ and $\k_1,
\k_2$  to either
$1, 0, -1$. When
$\icd=1$, the CKD algebra ${}_{+}\goth{su}_{\k_1,\k_2}(3)$ is
isomorphic to the CK Lie algebra $\goth{su}_{\k_1,\k_2}(3)$; it is
simple when $\k_1,
\k_2$ are different from zero. When $\icd=-1$ and for any non-zero
values of $\k_1, \k_2$,  the CKD algebra
${}_{-}\goth{su}_{\k_1,\k_2}(3)$ is also simple and isomorphic to
the Lie algebra
$\goth{sl}(3, \Re)$.  Generators $B, I, T_1, T_2, H_1, H_2$ still
belong to  the fiducial two-dimensional Cartan subalgebra of
${}_{\icd}\goth{su}_{\k_1,\k_2}(3)$.

For any value of $\icd$ the CKD algebras in the family
${}_{\icd}\goth{su}_{\k_1,\k_2}(3)$ can be endowed with a
${\Ze}_2\otimes {\Ze}_2$ group of commuting involutive
automorphisms  generated by
$\Invol_{(1)}, \Invol_{(2)}$ (\ref{ht:Invols}); denoting
everything  as in the former section, the  three Lie subalgebras
$\goth{h}_{(1)}$,
$\goth{h}_{(2)}$ and
$\goth{h}_{(02)}$ spanned by the generators with the same name as
in the complex case (the Lie algebra elements invariant under the
involutions with the same indices), turn out to be of CKD type,
${}_\icd\goth{u}(1)
\oplus
\goth{su}_{\k}(2)$ with $\k=\k_2, \k_1,
\k_1\k_2$ respectively. The groups they generate are isomorphic to 
${}_{\icd}U(1) \otimes {}_{\icd}SU_{\k_2}(2)$, \  
${}_{\icd}U(1) \otimes {}_{\icd}SU_{\k_1}(2)$ and  
${}_{\icd}U(1) \otimes {}_{\icd}SU_{\k_1\k_2}(2)$.   In all these
expressions, the Lie algebras of CKD `unitary type' are 
${}_{\icd}\goth{u}_{\k}(2)
\equiv {}_\icd\goth{u}(1) \oplus {}_{\icd}\goth{su}_{\k}(2)$, and
for the groups of `unimodular complex numbers' ${}_\icd U(1)$ we
have two generic cases  ${}_{+}U(1)
\equiv U(1)\equiv SO(2)$, ${}_{-}U(1) \equiv SO(1,1)$ and one
limiting case ${}_{0}U(1)
\equiv ISO(1)\equiv \Re$. 

The Lie algebra ${}_{\icd}\goth{su}_{\k_1,\k_2}(3)$  is given by 
the following Lie conmutators:
\vspace{-5mm}
\be
\begin{tabular}{llll}
 $[\pp,\ppp] = \k_1 \jj$ &$[\ppp,\qq] =\k_1 \mm$  &$[\qq,\qqq]
=\icd \k_1
\jj$ \\
 $[\pp,\qq] = 2 \k_1 \hh$&$[\ppp,\qqq] =2 \k_1\k_2\hhh$& \\
 $[\pp,\qqq] =\k_1 \mm$  &                          &
\\\vspace{-2mm}
\\
 $[\pp,\jj]=-\ppp$&$[\ppp,\jj]=\k_2\pp$&$[\qq,\jj]=-
\qqq$&$[\qqq,\jj]=\k_2\qq$\\
$[\pp,\mm]=-\qqq$&$[\ppp,\mm]=-\k_2\qq$
&$[\qq,\mm]=\icd\ppp$&$[\qqq,\mm]=\icd\k_2\pp$\\
$[\pp,\ll]=\qq$&$[\ppp,\ll]=-\qqq$
&$[\qq,\ll]=-\icd\pp$&$[\qqq,\ll]=\icd\ppp$\\
 $[\pp,\dd]=-\qq$&$[\ppp,\dd]=-\qqq$&$[\qq,\dd]=\icd\pp$&$
[\qqq,\dd]=\icd\ppp$\\
 $ [\pp,\tt]=0$&$[\ppp,\tt]=-\qqq$&$[\qq,\tt]=0$&$
[\qqq,\tt]=\icd\ppp$\\
 $ [\pp,\ttt]=-\qq$&$[\ppp,\ttt]=0 $&$[\qq,\ttt]=\icd\pp$&$
[\qqq,\ttt]=0$\\
$[\pp,\hh]=-2\qq$&$[\ppp,\hh]=-\qqq$
&$[\qq,\hh]=2\icd\pp$&$[\qqq,\hh]=\icd\ppp$\\
 $[\pp,\hhh]=-\qq$&$[\ppp,\hhh]=-2\qqq$
&$[\qq,\hhh]=\icd\pp$&$[\qqq,\hhh]=2\icd\ppp$\\
\vspace{-3mm}
\\
 $[\jj,\mm] =2\k_2 \ll$&  &
\\\vspace{-3mm}
\\ 
$[\jj,\ll] =-2\mm$& $[\jj,\dd] =0$ & $   [\mm,\ll] =2\icd\jj$ &
$[\mm,\dd] =0 $ \\
$[\jj,\tt] =-\mm$ & $  [\jj,\ttt] =\mm$ & $[\mm,\tt] =\icd\jj$ &
$  [\mm,\ttt] =-\icd\jj $ \\
$[\jj,\hh] =\mm$  & $  [\jj,\hhh] =-\mm$& $[\mm,\hh] =-\icd\jj$ &
$  [\mm,\hhh] =\icd\jj $.
\end{tabular} \label{eq:HermCKDBiDConmRels}
\ee

The elements defining a  2D CK `complex hermitian' geometry can be
now described: 

\noindent
$\bullet$ The {\it plane} as the set of points corresponds to the
symmetric homogeneous space
\vspace{-3mm}
\be
\iCeCKPointSpace\equiv {}_{\icd}SU_{\k_1,\k_2}(3)/H_{(1)}\equiv
{}_{\icd}SU_{\k_1,\k_2}(3)/({}_{\icd}U(1) \otimes
{}_{\icd}SU_{\k_2}(2))
\quad  H_{(1)}=\myspan{I; J, M, B}.
\mylabel{ht:CKDPointSpace}
\ee
\vspace{-3mm}
\noindent
$\bullet$ The set of (first-kind) {\it `complex' lines} is
identified to the symmetric homogeneous space
\be
\iCeCKLineSpace\equiv {}_{\icd}SU_{\k_1,\k_2}(3)/H_{(2)}\equiv
{}_{\icd}SU_{\k_1,\k_2}(3)/({}_{\icd}U(1) \otimes
{}_{\icd}SU_{\k_1}(2))
\quad  H_{(2)}=\myspan{T_1;  P_1, Q_1, H_1}.
\mylabel{ht:CKDLineSpace}
\ee Both spaces have again dimension 2 over $\iCe$, and all
comments made in the complex case can be easily rephrased. The
definition of the space of second kind `complex' lines can be also
suitably adapted. 

\setlength{\unitlength}{0.25mm}
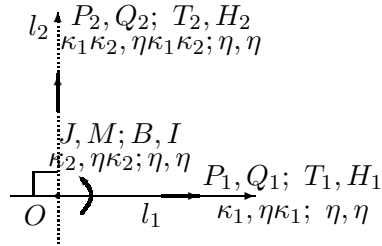
\begin{figure}[ht]
\begin{center}
\begin{picture}(100,100)(0,25)
\put(25,25){\circle*{3}}
\put(13,15){\makebox(0,0){$O$}}
\put(12,25){\line(0,1){13}}
\put(12,38){\line(1,0){13}}
\put(0,25){\vector(1,0){130}}
\put(75,15){\makebox(0,0){$l_1$}}
\put(150,35){\makebox(0,0){$P_1, Q_1; \ T_1, H_1$}}
\put(150,15){\makebox(0,0){$\k_1, \icd\k_1;\  \icd,\icd$}}
\qbezier[50](25,0)(25,60)(25,120)
\put(25,120){\vector(0,1){3}}
\put(15,115){\makebox(0,0){$l_2$}}
\put(80,120){\makebox(0,0){$P_2, Q_2; \ T_2, H_2$}}
\put(80,105){\makebox(0,0){$\k_1\k_2, \icd\k_1\k_2; \icd,\icd$}}
\put(58,57){\makebox(0,0){$J, M; B, I$}}
\put(58,43){\makebox(0,0){$\k_2,\icd\k_2; \icd,\icd$}}
\linethickness{1pt}
\qbezier(37,16)(47,26)(37,34)
\put(37,35){\vector(-1,1){1}}
\put(25,70){\vector(0,1){20}}
\put(80,25){\vector(1,0){20}}
\end{picture}
\end{center}
\label{CKD:labelgenerators}
\caption{Generators and their associated labels  in a  `complex
hermitian'-2D CKD geo\-metry. Lines $l_1$, and $l_2$ are
`complex', thus two-dimensional from a real point of view}
\end{figure}

By a {\it two-dimensional `complex hermitian' CKD geometry} we will
mean the set of three symmetric homogeneous spaces of points, lines
of first-kind and lines of second-kind. The group
${}_{\icd}SU_{\k_1,\k_2}(3)$ acts transitively on each of these
spaces. The fundamental {\em ordinary duality} $\dual$
(\ref{ht:CKBiDDuality}) extends, by simply assuming $\dual:\icd
\to
\icd$, to an `automorphism' of the complete CKD family and  leaves
the general commutation rules (\ref{eq:HermCKDBiDConmRels})
invariant.  In general $\dual$ relates  {\em two} different
`complex hermitian' CKD geometries  with the same label $\icd$ but
$\k_1, \k_2$ interchanged. Figure 2 displays the generators (with
their labels) as related to the three fiducial elements $O, l_1,
l_2$. 

The quadratic Lie algebra Casimir in
${}_{\icd}\goth{su}_{\k_1,\k_2}(3)$ can be written grouping the
terms  which correspond to the three
${}_{\icd}\goth{su}_{\k}(2)$-like subalgebras:
\be {\cal C}=((\icd P_2^2+Q_2^2)+ \k_1\k_2 H_2^2)+
         \k_2 ((\icd P_1^2 +Q_1^2)+ \k_1 H_1^2) +
         \k_1 ((\icd J^2 + M^2) + \k_2 B^2).
\mylabel{ht:CKDCasimir}
\ee From  this Casimir we can easily derive the invariant FS
metric  in the space $\iCeCKPointSpace$ which is given, at the
origin and in the basis $P_1, Q_1, P_2, Q_2$ by the matrix
$\mbox{diag}(1,\icd,\k_2,\icd\k_2)$, coming also as the real part
of the hermitian product whose matrix at $O$ is
$\mbox{diag}(1,\k_2)$.

We should mention that the CKD algebras in the family
${}_{\icd}\goth{su}_{\k_1,\k_2}(3)$ can actually be endowed with a
${\Ze}_2\otimes({\Ze}_2\otimes {\Ze}_2)$ group of commuting
involutive automorphisms; in addition to $\Invol_{(1)},
\Invol_{(2)}$ given by (\ref{ht:Invols}), the extra involution 
${}_{(1)}\Invol$  will not play any explicit role in what follows,
but it is mentioned for completeness; the homogeneous spaces
$SU(3)/SO(3)$,
$SU(2,1)/SO(2,1)$ and $SL(3,\Re)/SO(3)$, $SL(3,\Re)/SO(2,1)$
appear by mimicking the former construction using this  involution:
\be
\begin{array}{ll} {}_{(1)}\Invol \!\!\!\!&: (P_1,P_2,Q_1, Q_2; J,
M, B, I)\to (P_1,P_2,-Q_1, -Q_2; J, -M, -B, -I).
\mylabel{ht:ExtraInvols}
\end{array}
\ee

\vspace{-3mm}
\subsection{Realization of spaces of points in the `complex
hermitian' Cayley--Klein--Dickson spaces}

When $\icd$ is present, the fundamental 3D `complex' matrix
representation (\ref{ht:AlgFRep}, \ref{ht:CartanFRes}) 
exponentiates to a representation of ${}_{\icd}SU_{\k_1,\k_2}(3)$
as a  linear transformations group in the ambient linear space
$\iCe^3$. One-parametric subgroups corresponding to the generators
so far considered  are given again by
(\ref{ht:CKBiDClasSubgroups}, \ref{ht:CKBiDCartSubgroups}). where
now the exponential $e^{ix}$ is related to the sine and  cosine
with label $\icd$ by a Euler-like formula:
\be e^{ix}=\ci(x) + i \si(x).
\label{ht:CDEuler}
\ee Again the action of ${}_{\icd}SU_{\k_1,\k_2}(3)$ on $\iCe^3$
is linear but not transitive, since it conserves the `hermitian'
form $|z^0|^2+\k_1 |z^1|^2+ \k_1\k_2 |z^2|^2$. The isotopy
subgroup of the point $O$ whose position vector is $O\equiv
(1,0,0)$ is easily seen to be the three parameter subgroup
${}_{\icd}SU_{\k_2}(2)$ generated by $J, M, B$, while the
${}_{\icd}U(1)$ subgroup generated by $I$ multiplies this vector
by a unimodular  `complex' phase factor. Hence the homogeneous
symmetric space
$\iCeCKPointSpace\equiv {}_{\icd}SU_{\k_1,\k_2}(3)/({}_{\icd}U(1)
\otimes {}_{\icd}SU_{\k_2}(2))$ can be identified to the orbit of
the {\em ray}
$[O]$ under the action of the group 
${}_{\icd}SU_{\k_1,\k_2}(3)$.

The geometry behind the case $\icd<0$ differs greatly from the one
in the ordinary complex case: for $\icd<0, \k_2\neq0$ the FS
metric is always indefinite and of $(2,2)$ real type, no matter of
the sign of $\k_2$.  The four spaces of points with $\icd<0$,
$\k_1\neq0, \k_2\neq0$ are essentially the same, though the
choices of lines and FS geodesics of first- and second-kind are
interchanged; this is tantamount to what happens in the real case
for the 1+1 Anti-DeSitter and DeSitter spaces. These four spaces
can be realized as spaces of
$0$-pairs in $\Re P^2$ (pairs made from a point and an hyperplane
(here line) in the {\em real} projective plane
$\Re P^2$), and the distance between two $0$-pairs $(X;\alpha)$,
$(Y;\beta)$ is related to the cross ratio of the four points $X,
Y; Z, T$, where $Z, T$ are the intersections of the line
determined by $X, Y$ with the hyperplanes $\alpha, \beta$ (see
Rosenfeld book \cite{RosGeoLieGroups}, theorems 2.39 and 4.21). 

The non-generic situation where a coefficient $\icd; \k_1, \k_2$
vanishes corresponds to an In\"on\"u--Wigner  contraction
\cite{IW}.  The limit
$\k_1\to 0$ is a local-contraction (around a point). The limit
$\k_2\to 0$ is a  line-contraction (around a whole `complex' line).
Finally, the limit $\icd\to 0$ corresponds to a new kind of
contraction around a purely real submanifold, the  projectivized
real $\k_1, \k_2$ CK space. Contractions are  built-in in the
expressions associated to the `complex hermitian' CK geometries
and groups, simply  by making zero any of the constants  $\icd;
\k_1,
\k_2$ (determining the curvature and signature of the space). 


\section{The compatibility conditions for a triangular loop}

In this section we discuss the approach to the trigonometry of the
twenty seven `complex hermitian type'  CKD spaces, and we
introduce the `complex hermitian' compatibility equations, loop
equations and the basic trigonometric identity. Some general
comments on this approach are given in
\cite{SpaceTimeTrig} and will not be repeated here; specially we
refer to the choice of `external angles' at the vertex $A$ and the
fact that the standard angular excess appears without the explicit
presence of the measure of twice a quadrant of angle (which equals
$\pi$ when
$\k_2=1$).

A triangle in a `complex hermitian' CKD space can be seen either
as  a triangle point loop or dually as a triangle line loop (see
figure  (\ref{fig:TriangleLoops})). In the first case a point
$C$ is considered to move to a different point $B$ `translating' 
either along the geodesic segment $CB$, or along the two geodesic
segments $CA$ and $AB$. Dually, the geodesic $c\equiv AB$ is
considered to move to a different geodesic $b\equiv CA$
`rotating'  either about the vertex $A\equiv bc$, or about the
vertices $B\equiv ca$ and then $C\equiv ab$. There exists though a
very important difference with the real 2D case: the
`translations' along a geodesic are not  uniquely defined by the
geodesic only. Thus to make out sense of the idea of triangle loop
a closer analysis of the geometry is required.  Any geodesic $g$
through $C$ determine a well defined `complex' line  $\iCe g$
containing $g$. Thus for the two geodesics 
$a, b$ intersecting at the vertex $C$ there are two uniquely 
determined `complex' lines
$\iCe a$, $\iCe b$ through $C$. These two `complex' lines  will
lie on a (generically) well defined line-geodesic $G_C$ (also
called a line-chain). This is dual to the determination of a
(generically) well defined geodesic $g_a$ through two different
points $C, B$. This subtlety is not neccesary in the real case, as
then the set of all lines through a point $C$ is one-dimensional,
while in the complex case, the set of `complex' lines through
$C$ is two-dimensional.

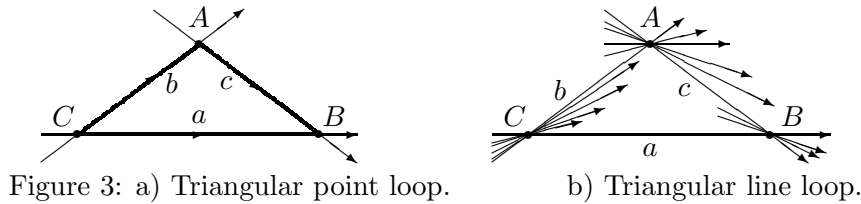
\setlength{\unitlength}{0.3mm}
\begin{figure}[h]
\begin{center}
\begin{picture}(350,40)(0,40)
\put(0,25){\vector(1,0){140}}
\put(0,13){\vector(4,3){90}}
\put(50,80){\vector(4,-3){90}}
\put(16,25){\circle*{3}}
\put(123,25){\circle*{3}}
\put(70,65){\circle*{3}}
\put(10,33){\makebox(0,0){$C$}}
\put(130,33){\makebox(0,0){$B$}}
\put(70,77){\makebox(0,0){$A$}}
\put(70,33){\makebox(0,0){$a$}}
\put(82,48){\makebox(0,0){$c$}}
\put(58,48){\makebox(0,0){$b$}}
\put(200,25){\vector(1,0){150}}
\put(200,13){\vector(4,3){85}}
\put(250,80){\vector(4,-3){90}}
\put(200,21){\vector(4,1){40}}
\put(200,20){\vector(3,1){50}}
\put(200,17){\vector(2,1){60}}
\put(200,14){\vector(3,2){65}}
\put(250,60){\vector(4,1){45}}
\put(250,65){\vector(1,0){55}}
\put(250,72){\vector(3,-1){65}}
\put(250,75){\vector(2,-1){75}}
\put(300,33){\vector(3,-1){48}}
\put(300,37){\vector(2,-1){45}}
\put(216,25){\circle*{3}}
\put(323,25){\circle*{3}}
\put(270,65){\circle*{3}}
\put(210,33){\makebox(0,0){$C$}}
\put(333,33){\makebox(0,0){$B$}}
\put(270,77){\makebox(0,0){$A$}}
\put(270,18){\makebox(0,0){$a$}}
\put(285,45){\makebox(0,0){$c$}}
\put(230,45){\makebox(0,0){$b$}}
\put(69,25){\vector(1,0){3}}
\put(48,49){\vector(4,3){3}}
\put(94,47){\vector(4,-3){3}}
\linethickness{1pt}
\qbezier(16,25)(42,25)(68,25)
\qbezier(69,25)(96,25)(123,25)
\qbezier(16,25)(32,37)(48,49)
\qbezier(48,49)(59,57)(70,65)
\qbezier(70,65)(82,56)(94,47)
\qbezier(94,47)(109,36)(123,25)
\end{picture}
\end{center}
\caption{a) Triangular point loop.\qquad\qquad b) Triangular line
loop.}
\label{fig:TriangleLoops}
\end{figure}
 
To start with the study of trigonometry we will take as `sides'
and `angles', the canonical parameters of certain one-parameter
subgroup elements associated  to some algebra generators. To
explain this choice, we first select a  (real) flag $O \subset g_1
\subset l_1 \subset G_1$ as follows: $O$ is the origin point
$O=[(1,0,0)]$, $g_1$ is the orbit of $O$ under the one-dimensional
subgroup generated by $P_1$ (thus the `complex' line  $l_1$ is the
orbit of $O$ under the subgroup generated by $P_1, Q_1$) and the
line-geodesic $G_1$, is the orbit of $l_1$ under the subgroup
generated by
$J$; this flag is determined by singling out the generators $P_1,
Q_1, J$. Now move the triangle to the canonical position where $C$
coincides with $O$, the side $a$ is on $g_1$ and the side $b$ lies
on a geodesic in $G_1$. This `canonical' position guarantees that
the  side
$b$ is obtained from $a$ by means of two {\em commuting} rotations
generated by
$J$ and
$I$, where the {\it phase rotation generator} $I$ is the unique
generator in the fiducial Cartan subalgebra {\it commuting} with
$J$. As angular invariants take the canonical parameters $C$
(`Hermitian' or `pure' angle  between `complex' lines) and $\apC$
({\em angular phase} between real-1D geodesics within a  `complex'
line) of the two `rotations' whose product
\be
\myexp{C \jj}\myexp{\apC I}
\mylabel{ht:ComplRot}
\ee carries the side $a$ to coincide with $b$; this products will
be called {\it complete rotations} about the vertex $C$. 

Since `hermitian' CKD spaces are rank-one, and therefore each pair
of points have  a single invariant, it would seem enough to
consider the FS distance $a$ between the points
$C, B$ as the unique moduli of sides. This is what was done in the
previous works on hermitian trigonometry 
\cite{ShiPetRoz, Hsiang, Bre, RosGeoLieGroups}. But the formal
duality requirement  prompts the consideration of translation
partners to both
$J; I$ and since duality maps $J;I$ into $-P_1; -T_1$, this
suggests the use of the following {\em complete  translations}:
\be
\myexp{a P_1}\myexp{\lpa T_1}.
\mylabel{ht:ComplTras}
\ee As well as the duality requirement, there are geometrical
reasons  for the use of the extra `translation' $\myexp{\lpa
T_1}$:  the `pure' translation 
$\myexp{a P_1}$ carries the vertex $C$ to
$B$, but this alone {\em does not} carry the unique `complex
line'-geodesic at the vertex $C$ determined by 
$\iCe a, \iCe b$, to the `complex' line-geodesic at $B$ determined
by $\iCe b, \iCe c$; an additional $\myexp{\lpa T_1}$  is required
to bring them into each other.

If we now consider complete rotations at each vertex, and complete
translations along each side, duality is manifestly restored: at
each vertex a complete rotation is required to bring into
coincidence simultaneously the sides seen both as `complex' lines
and as point-geodesic sides. And along each side, a complete
translation is required to bring into coincidence simultaneously
both the vertices as points and the `complex' line-geodesics
determined at each vertex by the two sides. 

This choice of two {\it conmmuting} generators is very natural
from a  Quantum Mechanics wiewpoint and afford  six `vertex'
quantities (three Hermitian or pure angles $A, B, C$ and three
{\it angular phases}, $\apA, \apB, \apC$), and six `side'
quantities  (three lengths, $a, b, c$ and three {\it lateral
phases} $\lpa,
\lpb, \lpc$). All these invariants appear as canonical parameters
of pairs of  commuting isometries, respectively generated by
$J_A, J_B, J_C$; $I_A, I_B, I_C$ and $P_a, P_b, P_c$; $T_a, T_b,
T_c$. At each side $P_a, P_b, P_c$ are pure translation generators
that perform the canonical parallel transport along their FS
geodesic axes, and
$T_a, T_b, T_c$ are the only  Cartan generators in the isotopy
subalgebras of the sides $a, b, c$  commuting with  $P_a, P_b,
P_c$ respectively. 

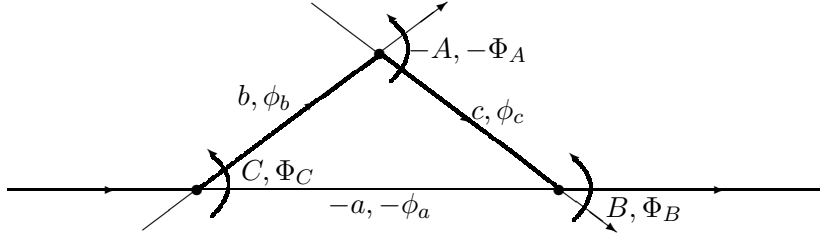
\begin{figure}[ht]
\setlength{\unitlength}{0.45mm}
\begin{center}
\begin{picture}(235,60)(0,20)
\put(-10,25){\line(1,0){224}}
\put(30,13){\vector(4,3){90}}
\put(80,80){\vector(4,-3){90}}
\put(46,25){\circle*{3}}
\put(153,25){\circle*{3}}
\put(100,65){\circle*{3}}
\put(70,30){\makebox(0,0){$C, \apC$}}
\put(178,19){\makebox(0,0){$B, \apB$}}
\put(126,66){\makebox(0,0){$-A, -\apA$}}
\put(135,48){\makebox(0,0){$c, \lpc$}}
\put(66,52){\makebox(0,0){$b, \lpb$}}
\put(100,20){\makebox(0,0){$-a, -\lpa$}}
\put(78,49){\vector(4,3){3}}
\put(124,47){\vector(4,-3){3}}
\linethickness{1pt}
\qbezier(46,25)(62,37)(78,49)
\qbezier(78,49)(89,57)(100,65)
\qbezier(100,65)(112,56)(124,47)
\qbezier(124,47)(139,36)(153,25)
\qbezier(153,25)(168,25)(198,25)
\qbezier(199,25)(214,25)(229,25)
\put(199,25){\vector(1,0){3}}
\qbezier(-10,25)(4,25)(18,25)
\qbezier(19,25)(33,25)(46,25)
\put(19,25){\vector(1,0){3}}
\qbezier(50,17)(60,27)(50,35)
\put(50,36){\vector(-1,1){1}}
\qbezier(157,16)(167,26)(157,34)
\put(157,35){\vector(-1,1){1}}
\qbezier(103,57)(113,67)(103,75)
\put(103,76){\vector(-1,1){1}}
\end{picture}
\end{center}
\caption{A `Complex Hermitian' triangular loop as a single curve.}
\label{fig:singlecurve}
\end{figure}

Cartan generators exponentiate to  somewhat `hybrid' 
transformations. The Cartan subalgebra is contained in the  isotopy
subalgebra of $O$, so its elements generate `rotations' about $O$
and  conjugates of them rotations about other points. The phase
`rotation' part $e^{\lpx T_1}$ of the complete fiducial
translation $e^{xP_1} e^{\lpx T_1}$ apparently breaks the scheme
symmetry between rotations and translations. Nevertheless, since
any Cartan transformation as $e^{\lp T_1}$ leaves pointwise
invariant the `complex' line $l_1$  it should also be considered a
`translation' along $l_1$. Thus Cartan transformations are both
rotations about a point and translations along a `complex' line.
This could have been read from the diagram (\ref{ht:GenDiagram})
as the {\em whole} Cartan subalgebra is contained in each of the
three blocks, the isotopy subalgebras of $l_1, l_2, O$
respectively. 

From now on everything follows the real pattern
\cite{SpaceTimeTrig}, and the commutativity between both
components of a complete transfomation allows the extension of the
basic real identities to  `complex' ones: {\it compatibility}
identities, {\it point loop and side loop} equations, and {\it
basic trigonometric identity}.

The generators $P_a, T_a, P_b, T_b, P_c, T_c; J_A, I_A, J_B, I_B,
J_C, I_C$ are not independent. They are related by several {\em
compatibility conditions}:
\be
\begin{array}{ll}
\myarray{P_b}{T_b}=\myexp{\CJC}\myexp{\CIC}\myarray{P_a}{T_a}\myexp{-
\CIC}
\myexp{-\CJC} \ &\
\myarray{J_B}{I_B}
=\myexp{\cpc}\myexp{\ctc}\myarray{J_A}{I_A}\myexp{-\ctc}\myexp{-\cpc}
\cr
\myarray{P_c}{T_c}=\myexp{-\AJA}\myexp{-
\AIA}\myarray{P_b}{T_b}\myexp{\AIA}
\myexp{\AJA} \ &\
\myarray{J_C}{I_C}=\myexp{-\apa}\myexp{-
\ata}\myarray{J_B}{I_B}\myexp{\ata}
\myexp{\apa}
\cr
\myarray{P_a}{T_a}=\myexp{\BJB}\myexp{\BIB}\myarray{P_c}{T_c}\myexp{-
\BIB}\myexp{-\BJB} \ &\
\myarray{J_A}{I_A}=\myexp{\bpb}\myexp{\btb}\myarray{J_C}{I_C}\myexp{-
\btb}\myexp{-\bpb}
\end{array}
\mylabel{ht:CompatConds}
\ee which can be considered as an implicit group theoretical
definition for the three sides, the three angles, the three
lateral phases and the three angular phases.

All the trigonometry of the `complex' CKD space is {\em
completely} contained in these equations, which have as  a
remarkable property  their explicit invariance under the {\em 
duality} interchange $a, b, c \leftrightarrow  A,   B,  C$ and
$\lpa, \lpb, \lpc \leftrightarrow  \apA,   \apB,  \apC$ for
triangle group theoretical invariants  (sides $\leftrightarrow$
angles, lateral phases
$\leftrightarrow$ angular phases) and $P
\leftrightarrow  J$, $T  \leftrightarrow  I$ for generators; this
duality is a consequence of the fact that $\dual$
(\ref{ht:CKBiDDuality}) is an automorphism of the  family of CKD
algebras which interchanges $P_1
\leftrightarrow -J$, and $T_1 \leftrightarrow -I$. These equations
ressemble their real analogues: the real rotation
$\myexp{\AJA}$ or translation $\myexp{\apa}$ are replaced by the
`complete' products
$\myexp{\AJA}\myexp{\AIA}$ or $\myexp{\apa}\myexp{\ata}$.

Each equation in (\ref{ht:CompatConds}) is actually a pair
relating both components of each `complete' translation or
rotation. As in
\cite{SpaceTimeTrig} we will refer to them as $P_b(P_a),T_b(T_a)$,
etc.\ (or $P_a(P_b), T_a(T_b)$ when the equation is read
inversely). By cyclic substitution in the three pairs of equations
$P_a(P_c), T_a(T_c)$;
$P_c(P_b), T_c(T_b)$ and $P_b(P_a), T_b(T_a)$ we find the identity
\be
\myexp{\BJB}\myexp{\BIB} \myexp{-\AJA}\myexp{-\AIA}
\myexp{\CJC}\myexp{\CIC} \myarray{P_a}{T_a}
\myexp{-\CIC} \myexp{-\CJC}
\myexp{\AIA}\myexp{\AJA}\myexp{-\BIB}\myexp{-\BJB} = 
\myarray{P_a}{T_a}
\mylabel{ht:cyclicPa}
\ee
$P_c, T_c$,
are
completely parallel process gives:
\be
\myexp{-\apa}\myexp{-
\ata}\myexp{\cpc}\myexp{\ctc}\myexp{\bpb}\myexp{\btb}
\myarray{J_C}{I_C}
\myexp{-\btb}
\myexp{-\bpb}\myexp{-\ctc}\myexp{-\cpc}\myexp{\ata}\myexp{\apa} =
\myarray{J_C}{I_C}.
\mylabel{ht:cyclicJC}
\ee Equations (\ref{ht:cyclicPa}) and  (\ref{ht:cyclicJC}) can be
written alternatively as:
\be
\begin{array}{l}
\myexp{\BJB}\myexp{\BIB} \myexp{-\AJA}\myexp{-\AIA}
\myexp{\CJC}\myexp{\CIC}
               \hbox{ must commute with $ P_a$ and $T_a$} ,\\
\myexp{-\apa}\myexp{-
\ata}\myexp{\cpc}\myexp{\ctc}\myexp{\bpb}\myexp{\btb}
               \hbox{ must commute with $J_C$ and $I_C$}.
\mylabel{ht:MustCommute}
\end{array}
\ee


\subsection{Loop excesses and  loop equations}

Had we not considered lateral phases, the `loop' product
$\myexp{-\apa}
\myexp{\cpc} \myexp{\bpb}$ would have been a natural object
associated to the three `pure' translations along the triangle
sides $C\overto{b} A \overto{c} B
\overto{-a} C$. This transformation is the ordinary holonomy
associated to the triangle, as each factor $\myexp{\apa}$ is the
ordinary parallel transport operator in the canonical connection
of the hermitian space
$\iCeCKPointSpace$. It  moves the base point $C$ along the
triangle and returns it back to its original position, so it must
be a {\em rotation} about $C$. This rotation is a product of an
${}_{\icd}U(1)$ phase part and a ${}_{\icd}SU_{\k_2}(2)$ part, but
the explicit expression for the ${}_{\icd}SU_{\k_2}(2)$ part is
rather involved. 

The use of angular and lateral phases in this self-dual approach 
affords some simple and apparently new results for certain
similar  `loop' operators. The guideline is the pattern
established in the real case, replacing every traslation or
rotation generators $T$ / $J$  by its `complete' versions $P, T$ /
$J, I$. We start with the equation which gives $P_c(P_b),
T_c(T_b)$ in the set (\ref{ht:CompatConds}), replace
$\myarray{P_c}{T_c}$ by
$\myexp{-\cpc}\myexp{-\ctc}\myarray{P_c}{T_c}\myexp{\ctc}\myexp{\cpc}$
and then substitute  $P_c(P_a), T_c(T_a)$ from the compatibility
equations to obtain:
\be
\myexp{-\AJA} \myexp{-\AIA}\myarray{P_b}{T_b}
\myexp{\AIA}\myexp{\AJA} =
\myexp{-\cpc} \myexp{-\ctc}\myexp{-\BJB}\myexp{-\BIB}
\myarray{P_a}{T_a}\myexp{\BIB}
\myexp{\BJB}\myexp{\ctc}\myexp{\cpc}.
\ee We introduce $J_B(J_C), I_B(I_C)$ and trivially simplify and
rearrange.  Now we use $J_A(J_C), I_A(I_C)$, simplify, and finally
substitute
$P_b(P_a), T_b(T_a)$. This gives:
\be
\begin{array}{l}
\myexp{B J_C}\myexp{\apB I_C}
\myexp{-\apa}\myexp{-
\ata}\myexp{\cpc}\myexp{\ctc}\myexp{\bpb}\myexp{\btb}
\myexp{-A J_C}\myexp{-\apA I_C}
\myexp{C J_C}\myexp{\apC I_C}    \myarray{P_a}{T_a} \times \\
\quad\myexp{-\apC I_C} \myexp{-C J_C} \myexp{\apA I_C}  \myexp{A
J_C}\myexp{-\btb}\myexp{-\bpb} 
\myexp{-\ctc}\myexp{-\cpc}\myexp{\ata}\myexp{\apa}
\myexp{-\apB I_C} \myexp{-B J_C}= \myarray{P_a}{T_a} .
\mylabel{ht:PaPa}
\end{array}
\ee The three {\em complete} translations along the triangle appear
in the former relation in a single piece
$\myexp{-\apa}\myexp{-\ata}\myexp{\cpc}
\myexp{\ctc}\myexp{\bpb}\myexp{\btb}$,
while the  three {\em complete} rotations are all  about the base
point
$C$. Now we can go a bit further: (\ref{ht:MustCommute}) implies
that
$\myexp{-\apa}\myexp{-\ata}\myexp{\cpc}
\myexp{\ctc}\myexp{\bpb}\myexp{\btb}$
must commute with $J_C$ and with $I_C$, so it will commute with
{\em any} complete rotation $\myexp{-\apX I_C}\myexp{-X J_C}$
about $C$, for any values of the complete angle $X,\apX$. Then we
can commute the whole complete translation piece
$\myexp{-\apa}\myexp{-\ata}\myexp{\cpc}
\myexp{\ctc}\myexp{\bpb}\myexp{\btb}$
in (\ref{ht:PaPa}) with the rotations about $C$ and collect these 
altogether; as both components of the complete rotation do
commute, we get:
\be
\begin{array}{l}
\myexp{-\apa}\myexp{-
\ata}\myexp{\cpc}\myexp{\ctc}\myexp{\bpb}\myexp{\btb}
 \myexp{(- A + B + C) J_C} \myexp{(- \apA + \apB + \apC) I_C} 
\hbox{ must commute with $P_a$ and $T_a$} .
\mylabel{ht:aComm}
\end{array}
\ee

We had already derived that 
$\myexp{-\apa}\myexp{-\ata}\myexp{\cpc}
\myexp{\ctc}\myexp{\bpb}\myexp{\btb}
\myexp{X J_C}\myexp{\apX I_C}$ must commute with $J_C$ and $I_C$
for {\em any} `complete angle' $(X, \apX)$ (see
(\ref{ht:MustCommute})). Since  this expression also commutes with
$P_a, T_a$ for the special values
$X=- A + B + C, \ \apX=- \apA + \apB + \apC$, we can conclude:
\be
\myexp{-\apa}\myexp{-
\ata}\myexp{\cpc}\myexp{\ctc}\myexp{\bpb}\myexp{\btb}
\myexp{(- A + B + C) J_C} \myexp{(- \apA + \apB + \apC) I_C}  = 1
\ee because the identity is the only element of
${}_{\icd}SU_{\k_1,\k_2}(3)$ commuting with two such pairs of 
generators as $P_a, T_a$ and $J_C, I_C$. This equation can be also
written as:
\be
\myexp{-\apa}\myexp{-
\ata}\myexp{\cpc}\myexp{\ctc}\myexp{\bpb}\myexp{\btb} =
 \myexp{-(- A + B + C) J_C}\myexp{-(- \apA + \apB + \apC) I_C}.
\mylabel{ht:GB}
\ee A similar procedure (or direct use of (\ref{ht:CompatConds}) in
(\ref{ht:GB})) allows us to derive two analogous equations:
\be
\begin{array}{l}
\myexp{\bpb}\myexp{\btb}  \myexp{-\apa} \myexp{-\ata}\myexp{\cpc}
\myexp{\ctc}
   = \myexp{-(- A + B + C) J_A} \myexp{-(- \apA + \apB + \apC)
I_A}\cr
\myexp{\cpc} \myexp{\ctc}\myexp{\bpb}\myexp{\btb}
\myexp{-\apa}\myexp{-\ata}
   = \myexp{-(- A + B + C) J_B}\myexp{-(- \apA + \apB + \apC) I_B}.
\end{array}
\mylabel{ht:GBab}\ee The quantities defined as
\be
\Delta := - A + B + C \qquad \Delta_\ap := - \apA + \apB + \apC 
\mylabel{ht:AngExcess}
\ee will be called the {\em  (Hermitian) angular excess} and the
{\em angular phase excess} of the triangle loop. The {\em
complete} angular excess $(\Delta, \Delta_\ap)$ fits into the view
of the geodesic line loop as a geodesic starting on $a$, and
successively rotating by complete angles $(C, \apC)$, $(-A,
-\apA)$ and
$(B, \apB)$ about the three vertices of the triangle; thus 
$(\Delta, \Delta_\ap)$ should be looked as the (oriented) total
complete angle turned by the geodesic line loop. Equations
(\ref{ht:GB}) or (\ref{ht:GBab}),  to be called the {\em `complex
hermitian' point loop equations}, express the product of the three
{\em complete} translations along the oriented sides of the
triangle loop as a complete rotation about the loop base point.
These equations are the closest `complex Hermitian' analogues of
the  Gauss-Bonnet triangle theorem (see
\cite{SpaceTimeTrig}) but we have found no reference to such a
simple  result in the literature.

The explicit duality of the starting equations
(\ref{ht:CompatConds}) under the interchanges $a, b, c
\leftrightarrow  A,   B,  C$ and $\lpa, \lpb, \lpc
\leftrightarrow  \apA,   \apB,  \apC$ for sides, angles and
phases, and $P 
\leftrightarrow  J$, $T \leftrightarrow  I$ for generators
immediately implies that the dual process leads to the dual
partners of (\ref{ht:GB}) and (\ref{ht:GBab}):
\be
\begin{array}{l}
\myexp{-\AJA} \myexp{-\AIA}\myexp{\CJC}\myexp{\CIC} \myexp{\BJB}
\myexp{\BIB}
    =  \myexp{-(-a + b + c) P_c} \myexp{-(-\lpa + \lpb + \lpc)
T_c}\cr
\myexp{\BJB}\myexp{\BIB}  \myexp{-\AJA}\myexp{-\AIA} \myexp{\CJC}
\myexp{\CIC}
    =  \myexp{-(-a + b + c) P_a} \myexp{-(-\lpa + \lpb + \lpc)
T_a}\cr
\myexp{\CJC}\myexp{\CIC} \myexp{\BJB}\myexp{\BIB}  \myexp{-\AJA}
\myexp{-\AIA}
    =  \myexp{-(-a + b + c) P_b}\myexp{-(-\lpa + \lpb + \lpc) T_b}
\mylabel{ht:GBpolarABC}
\end{array}
\ee so that the two quantities
\be
\delta=-a+b+c \qquad \delta_\lp = -\lpa + \lpb + \lpc
\mylabel{ht:LatExcess}
\ee play the role of {\em lateral excess} and {\em lateral phase
excess} of the triangle loop.  
lateral
orientations
`phased' along
(\ref{ht:GBpolarABC}) give the  product of the three oriented 
complete rotations about the three vertices of a triangle as a
complete translation along the base line of the loop.


\subsection{The basic trigonometric identity}

Each one of  the equations (\ref{ht:GB}), (\ref{ht:GBab}) or
(\ref{ht:GBpolarABC}) contains all  the relationships between
triangle sides, lateral phases, angles and angular phases in any
CKD `complex hermitian' space. However, all twelve elements appear
in these equations not only explicitly as canonical parameters, but
also implicitly inside the complete translation and rotation
generators. This prompts the search for another relation,
equivalent to the previous ones but more suitable to display the
trigonometric equations; this new equation is indeed the bridge
between the former equations and the trigonometry of the space. 

The idea is to  express {\em all} the generators as suitable
conjugates of {\em one} pair of a translation and a phase
translation generator  and {\em one} pair of a rotation generator
and a phase rotation generator, which we will take as primitive
{\em independent}  generators. A natural choice is to take the two
pairs $P_a, T_a$ and $J_C, I_C$ as `basic' independent generators.
Next by using (\ref{ht:CompatConds}) we {\em define} the remaining
triangle pairs of generators $P_b, T_b; J_A, I_A; P_c, T_c;  J_B,
I_B$ in terms of the previous ones and sides and angles, lateral
phases and angular phases  as:
\be
\begin{array}{l}
\myarray{P_b}{T_b}:=\myexp{\CJC}\myexp{\CIC}
\myarray{P_a}{T_a}\myexp{-\CIC} \myexp{-\CJC}\cr
\myarray{J_A}{I_A}:=\myexp{\bpb}\myexp{\btb}\myarray{J_C}{I_C}
\myexp{-\btb} \myexp{-\bpb}\cr
\myarray{P_c}{T_c}:=\myexp{-
\AJA}\myexp{-\AIA}\myarray{P_b}{T_b}\myexp{\AIA} \myexp{\AJA}\cr
\myarray{J_B}{I_B}:=\myexp{\cpc}\myexp{\ctc}
\myarray{J_A}{I_A}\myexp{-\ctc}\myexp{-\cpc}
\end{array}
\ee which after full expansion and simplification gives:
\be
\begin{array}{ll}
\myarray{P_b}{T_b}:=&\myexp{\CJC}\myexp{\CIC}
                    \myarray{P_a}{T_a}
                    \myexp{-\CIC} \myexp{-\CJC}  \cr
\myarray{J_A}{I_A}:=&\myexp{\CJC}\myexp{\CIC}\myexp{b
P_a}\myexp{\lpb T_a}
                    \myarray{J_C}{I_C}
                    \myexp{-\lpb T_a}\myexp{-b P_a}
\myexp{-\CIC}\myexp{-\CJC} \cr
\myarray{P_c}{T_c}:=&\myexp{\CJC}\myexp{\CIC}\myexp{b
P_a}\myexp{\lpb T_a}\myexp{-A J_C}\myexp{-\apA I_C}
                    \myarray{P_a}{T_a}
                    \myexp{\apA I_C}\myexp{A J_C}\myexp{-\lpb
T_a}\myexp{-b P_a}\myexp{-\CIC}\myexp{-\CJC}\cr
\myarray{J_B}{I_B}:=&\myexp{\CJC}\myexp{\CIC}\myexp{b
P_a}\myexp{\lpb T_a} \myexp{-A J_C}\myexp{-\apA I_C}
\myexp{\lpc T_a}\myexp{c P_a}
                    \myarray{J_C}{I_C} \times \\
                    & \qquad \myexp{-\lpc T_a}\myexp{-c
P_a}\myexp{\apA I_C}\myexp{A J_C}\myexp{\lpb T_a}\myexp{-b P_a}
\myexp{-\CIC}\myexp{-\CJC}.
\end{array}
\mylabel{ht:GenCanDefs}
\ee (Note the highly ordered pattern in these expressions). By
direct substitution in the equation (\ref{ht:GB}) and after
obvious cancellations which  due to the commutativity of each
member in the pairs of the complete transformations fully mimic
the pattern found in the real case, we find:
\be
\myexp{-a P_a}\myexp{-\lpa T_a}  \myexp{C J_C}  \myexp{\apC
I_C}\myexp{b P_a}\myexp{\lpb T_a}
 \myexp{-A J_C}\myexp{-\apA I_C} \myexp{c P_a} \myexp{\lpc
T_a}\myexp{B J_C}\myexp{\apB I_C} = 1.
\mylabel{ht:basicEq}
\ee The same process starting from any equation in (\ref{ht:GBab})
or (\ref{ht:GBpolarABC}) leads again to the same  equation. This
justifies to call (\ref{ht:basicEq}) the {\em basic trigonometric
equation}. We sum up in:

\medskip
\noindent {\bf Theorem 1.} \em Sides $a, b, c$, lateral phases
$\lpa, \lpb,
\lpc$, angles $A, B, C$  and angular phases $\apA, \apB, \apC$ of
any triangle loop in the complex CKD space $\iCeCKPointSpace$ are
linked by a single group identity  called the {\em basic `complex
hermitian' trigonometric identity}
\be
\myexp{-a P}\myexp{-\lpa T}  \myexp{C J}  \myexp{\apC I}\myexp{b
P}\myexp{\lpb T}
 \myexp{-A J}\myexp{-\apA I} \myexp{c P} \myexp{\lpc T}\myexp{B
J}\myexp{\apB I} = 1
\mylabel{ht:BasicIdent}
\ee where $P,T$ are the generators of translations and phase
translations along {\em any} fixed fiducial geodesic  $g$, and $J,
I$ are the generators of rotations and phase rotations about any 
fixed fiducial `complex' line-geodesic $G$ containing the
`complex' line $l_g$ and about $O$.\rm

\medskip
\noindent {\em Proof}: A group motion can be used to move the
triangle to a canonical  position described before
(\ref{ht:ComplRot}) for the flag $O \subset g \subset l \subset
l_g$.  Then the theorem statement is simply (\ref{ht:basicEq}).
\medskip

\noindent {\bf Theorem 2.} \em Let us consider a triangle loop in
the `complex hermitian' CKD space $\iCeCKPointSpace$, and let
$P_a, P_b, P_c; T_a, T_b, T_c$ be the generators of {\em
translations} and {\em phase translations} along the three
triangle geodesic sides, whose lengths and lateral phases  are $a,
b, c$ and $\lpa, \lpb, \lpc$. Let $J_A, J_B, J_C;$ $I_A, I_B, I_C$
be the generators  of {\em rotations} and {\em phase rotations}
about the three geodesic line vertices of the triangle, whose
angles and angular phases are $A, B, C$, and $\apA, \apB, \apC$.
These quantities are related by two sets of identities,
(\ref{ht:GB}, \ref{ht:GBab}) and (\ref{ht:GBpolarABC})  called the
{\em `complex hermitian' point loop} and the {\em `complex
hermitian' line loop} triangle equations, each equation being
equivalent to the identity in Theorem 1. \rm 

Several points are worth highlighting. First, each term in the
basic identity is either a complete translation along a fixed
geodesic $g$ (through the fixed point $O$)  or a complete rotation
along a fixed line-geodesic $G$ (about  $O$). The canonical
parameters of these translations or rotations are exactly the
triangle  sides, angles, lateral and  angular phases. In the point
loop or line loop equations the transformations involved are the
translations  along the sides or the rotations about the vertices.

Second, these equations are analogue to the ones found in the real
case, with the consistent replacement of every translation or
rotation by its `complete' version, made up of two commuting
factors. The point loop equations follow from a point travelling
along the triangle according to the obvious shorthand
$\overline{a}\,b\,c$ (see figure
\ref{fig:TriangleLoops}a). The line loop equations follow from a
line looping about the triangle like $B\, \overline{A}\,C$; it 
starts in a base line (say $a$) and successively rotates by an
angle $C$, then by an angle $-A$ and finally by an angle $B$ about
the corresponding vertices thus ending up back on the starting
position $a$ (see figure \ref{fig:TriangleLoops}b).  The basic
equation follows from the pattern
$\overline{a}\,C\,b\,\overline{A}\,c\,B$, which keeps track of
both  sides and vertices found when looping around the triangle.

Third, the (three) point loop equations and the (three) line loop
equations are mutually dual sets; the single basic equation is
clearly self-dual. And fourth, these equations hold in the same
explicit form for {\em all} the twenty seven  2D `complex
hermitian' CKD geometries, as neither an {\em explicit\/}
Cayley-Dickson label $\icd$ nor a Cayley-Klein one $\k_1, \k_2$
ever appears in them.


\section{The basic equations of trigonometry for any  `complex
hermitian' two-dimensional Cayley--Klein--Dickson space}

To obtain the trigonometric equations for the `complex hermitian'
CKD space we start with the basic trigonometric identity
(\ref{ht:BasicIdent}), for the triangle in its canonical position,
so $P_a, T_a$ and $J_C, I_C$ can be taken  exactly as $P_1, T_1$
and $J,I$. For notational clarity we  will omit even  the subindex
and will denote
$P_1 , T_1$ here simply as $P, T$. We first write
(\ref{ht:BasicIdent})  in the equivalent form $\overline{A}\,c\,B =
\overline{b}\,\overline{C}\,a$:
\be e^{-A J}e^{-\apA I}e^{c P}e^{\lpc T}e^{B J}e^{\apB I}= e^{-b
P}e^{-\lpb T}e^{-C J}e^{-\apC I}e^{a P}e^{\lpa T}
\mylabel{ht:BasicId2}
\ee By considering this identity in the fundamental 3D vector
representation of the motion group (\ref{ht:CKBiDClasSubgroups})
and (\ref{ht:CKBiDCartSubgroups}) we obtain an equality between $
3 \times 3$ complex matrices, giving rise to nine `complex'
identities:
\be
\begin{array}{ll}
\c(c) e^{i \frac{-2\apA+2\apB+\lpc}{3}}=
\c(a)\c(b)e^{i \frac{\lpa-\lpb-2\apC}{3}}+\k_1 \s(a)\s(b)\cc(C)e^{i
\frac{\lpa-\lpb+\apC}{3}}
\\[6pt]
\cc(C) e^{i \frac{-2\lpa+2\lpb+\apC}{3}}=
\cc(A)\cc(B)e^{i \frac{\apA-\apB-2\lpc}{3}} +\k_2
\ss(A)\ss(B)\c(c)e^{i \frac{\apA-\apB+\lpc}{3}}
\\[6pt]
\s(c)\ss(A)e^{i \frac{\apA+2\apB+\lpc}{3}}=
               \s(a)\ss(C)e^{i \frac{\lpa+2\lpb+\apC}{3}}
\\[6pt]
\s(c)\ss(B)e^{i \frac{-2\apA-\apB+\lpc}{3}}=
               \s(b)\ss(C)e^{i \frac{-2\lpa-\lpb+\apC}{3}}
\\[6pt]
\s(c) \cc(A)e^{i \frac{\apA+2\apB+\lpc}{3}}= -\c(a)\s(b)e^{i
\frac{\lpa-\lpb-2\apC}{3}} + \s(a)\c(b)\cc(C)e^{i
\frac{\lpa-\lpb+\apC}{3}}
\\[6pt]
\s(c) \cc(B)e^{i \frac{-2\apA-\apB+\lpc}{3}} =\c(b)\s(a)e^{i
\frac{\lpa-\lpb-2\apC}{3}} - \s(b)\c(a)\cc(C)e^{i
\frac{\lpa-\lpb+\apC}{3}}
\\[6pt]
\ss(C) \c(a)e^{i \frac{\lpa+2\lpb+\apC}{3}}= -\cc(A)\ss(B)e^{i
\frac{\apA-\apB-2\lpc}{3}}+
\ss(A)\cc(B)\c(c)e^{i \frac{\apA-\apB+\lpc}{3}}
\\[6pt]
\ss(C) \c(b)e^{i \frac{-2\lpa-\lpb+\apC}{3}}=
\cc(B)\ss(A)e^{i \frac{\apA-\apB-2\lpc}{3}}-
\ss(B)\cc(A)\c(c)e^{i \frac{\apA-\apB+\lpc}{3}}
\\[6pt]
\k_2\ss(A) \ss(B)e^{i
\frac{\apA-\apB-2\lpc}{3}}+\cc(A)\cc(B)\c(c)e^{i
\frac{\apA-\apB+\lpc}{3}}=
\\ \qquad\qquad
\k_1\s(a)\s(b)e^{i \frac{\lpa-\lpb-2\apC}{3}}+ \c(a)\c(b)\cc(C)e^{i
\frac{\lpa-\lpb+\apC}{3}}
\end{array}
\mylabel{ht:HermTrigEqns}
\ee which contain the trigonometry of the space
$\iCeCKPointSpace={}_{\icd}SU_{\k_1,\k_2}(3)/ ({}_{\icd}U(1)
\otimes {}_{\icd}SU_{\k_2}(2))$.

Each equation in this set either is self-dual or appears in a
mutually dual pair;  this  could have been expected due to the
self-duality of the starting equation. We should recall again that
all along this section
$i$ denotes the imaginary unit of the Cayley-Dickson `complex'
numbers
$\iCe$, so that $i^2=-\icd$; the labelled sine and cosine with
label $\icd$ are related to the exponential $e^{i x}$ by
(\ref{ht:CDEuler}).

The association between sides (resp.\ angles) and the labels
$\k_1$ (resp.\
$\k_2$) found in the equations of the real space
$\CKPointSpace=SO_{\k_1,\k_2}(3)/ SO_{\k_2}(2)$ extends to the
Hermitian complex analogues so lengths $a, b, c$  are associated
to $\k_1$, and angles $A, B, C$ to $\k_2$.  The lateral $\lpa,
\lpb, \lpc$ and angular phases $\apA,\apB, \apC$ have $\icd$ as
its label.  The elements ($-a, -\lpa, -A, -\apA$) always appear in
the equations with a  minus sign as compared with $(b,\lpb, B,
\apB), (c, \lpc, C, \apC)$;  this follows from the structure of
the basic equation (\ref{ht:BasicId2}),   in which the side $a$
and vertex $A$ are traversed or rotated backwards.

The set (\ref{ht:HermTrigEqns}) is equivalent to two other similar
sets, obtained by starting with the basic identity splitted in the
two equivalent forms symbolically denoted as $C\,b\,\overline{A} =
c\,B\,\overline{a}$ and  $b\,\overline{A}\,c =
B\,\overline{a}\,C$.  In order to present all these equations in a
concise way, it proves adequate to introduce a compact notation, 
following the pattern explained in
\cite{SpaceTimeTrig}. The sides and angles, lateral phases and
angular phases  will be denoted  as $\xi, \xI, \lpi, \apI$, $i,
I=1, 2, 3$  according to
\be
\begin{array}{llllll} x_1=-a  & x_2=b   & x_3=c \quad  &  X_1=-A  
&  X_2=B  & X_3=C  \\
\lp_1=-\lpa & \lp_2=\lpb  & \lp_3=\lpc \quad & 
\ap_1=-\apA  & \ap_2=\apB  &\ap_3=\apC .  \\
\end{array}
\mylabel{ht:CompNot}
\ee The built-in minus sign in $\xi, \lpi, \xI, \apI$ when $i=I=1$
is natural when the triangle is considered as a point loop with
the side $a,
\lpa$ traversed backwards, or as a side loop with the angle $A,
\apA$ rotated backwards; this choice absorbs the signs related to
$a, A$ and confer an uniform appearance to the equations
(\ref{ht:HermTrigEqns}). In particular, the angular and lateral
excesses appear in this notation as $\Delta=\xI+\xJ+\xK$ and
$\delta=\xi+\xj+\xk$. The basic equation (\ref{ht:BasicIdent}) now
reads:
\be e^{\xi P}e^{\lpi T}e^{\xK J}e^{\apK I}e^{\xj P}e^{\lpj T}
e^{\xI J}e^{\apI I}e^{\xk P}e^{\lpk T}e^{\xJ J}e^{\apJ I}=1
\mylabel{ht:BasicEqCN}
\ee where $i=I, j=J, k=K$ are {\em any} cyclic permutation of
$123$. This basic equation can be very simply recalled: replace in
the shorthand 
$iKjIkJ$ each letter by the associated complete translation or
rotation.  From now on we will adopt this convention which makes
all equations of trigonometry explicitly invariant under cyclic
permutations of the `oriented' complete sides $\xi, \lpi$ and
angles $\xI, \apI$. Capital indices will also help in
distinguishing between mutually dual pairs $\xi, \xI$ and $\lpi,
\apI$. 


\subsection{The trigonometric equations in the `Cartan sector'}

Each equation in (\ref{ht:HermTrigEqns}) is `complex', and phases,
both lateral and angular, appear through unimodular `complex'
factors  $e^{i\lp}, e^{i\ap}$, while `pure' sides and angles $\xi,
\xI$ appear through their labelled sines or cosines, which are
always real. The equations in the third line split into an
equation for the modulus and another for the argument; this last
part is:
\be -\lpj+2\lpk+\apI=-\apJ+2\apK+\lpi
\ee Writing the same equation for the choice of indices $i,j,k \to
j, k, i$ and comparing we get:
\be
\lpi-\apI=\lpj-\apJ.
\mylabel{ht:PhasesLaw}
\ee These three equations, only two of which are independent, are
self-dual  and hold for all the `complex' CKD spaces. A
consequence of these very simple {\em linear} relations is:
\be -\lpa+\apB+\apC=-\apA+\lpb+\apC=-\apA+\apB+\lpc; 
\ee the common value in this formula turns out to be a quantity 
first introduced for the complex hermitian elliptic space by
Blaschke-Terheggen. We will call this $\om$:
\be
\om:=\apI+\apJ+\lpk.
\mylabel{ht:omegaDef}
\ee In a similar dual way and starting also from 
(\ref{ht:PhasesLaw}) we  find another {\em linear} relation
between phases:
\be -\apA+\lpb+\lpc=-\lpa+\apB+\lpc=-\lpa+\lpb+\apC
\ee whose value turns out to be the invariant dual to $\om$ and
which for the complex hermitian elliptic case was also introduced
by Blaschke-Terheggen:  
\be
\Om:=\lpi+\lpj+\apK.
\mylabel{ht:OmegaDef}
\ee

Another quantities linked to $\lpi, \apI$ are the {\em angular
phase excess} $\Delta_{\ap}$ (\ref{ht:AngExcess}) and {\em lateral
phase excess}
$\delta_{\lp}$ (\ref{ht:LatExcess}),  which appear in the hermitian
analogues of Gauss-Bonnet triangle theorems. In terms of the
compact notation (\ref{ht:CompNot}) they are:
\be
\Delta_{\ap}:=\apI+\apJ+\apK \qquad
\delta_{\lp}:=\lpi+\lpj+\lpk
\mylabel{ht:ExcessesCN}
\ee The invariants $\om$ (resp.\ $\Om$) are thus two kinds of {\it
mixed phase excesses}, with dominance of angular (resp.\ lateral)
phases.  The departure from the BT notation $\omega, \tau$ to ours
$\om, \Om$ conforms to the typographical convention upper/lower
case in order to stress duality and to convey that each mixed
excess $\om, \Om$ has dominance of either angular or lateral
phases.  These invariants and the two phase excesses
$\Delta_{\ap}, \delta_{\lp}$, which appear in the point and line
loop equations are related by:
\be
\Delta_{\ap}=2 \om-\Om \qquad  \delta_{\lp}=2\Om-\om
\mylabel{ht:ExcsOmeg}
\ee

Thus there is a sector of hermitian trigonometry involving {\it
only} phases and completely decoupled from `pure' sides $\xi$ and
angles $\xI$. This sector holds in {\em exactly the same form} in
the twenty seven `complex' CKD spaces, as no explicit labels
$\icd; \k_1, \k_2$ appear. Since the triangle invariants $\lpi,
\apI$ are related to the Cartan subalgebra, we will call these
equations  the `Cartan' sector of `complex hermitian' trigonometry.
This `Cartan sector' has no analogue in the trigonometry of real
spaces, and their equations are purely linear witnessing the
abelian character of Cartan subalgebra.


\subsection{The complete set of `complex hermitian' trigonometric
equations}

Now by exploiting the `Cartan' relations between phases
(\ref{ht:PhasesLaw}), and introducing explicitly the invariants
$\om, \Om$, it turns out possible to simplify the equations
(\ref{ht:HermTrigEqns}) by multiplying each one of them by some
suitably chosen unimodular `complex' factor.  This leads to the
{\em full} set of trigonometric equations coming from the basic
trigonometric group identity as:
\be
\begin{array}{ll} 0ij \equiv 0IJ &\apI-\lpi=\apJ-\lpj \qquad \big(
\Rightarrow
\om:=\apI+\apJ+\lpk,  \qquad \Om:=\lpi+\lpj+\apK
\big)
\\[6pt] 1i &\c(\xi)e^{i
\om}=\c(\xj)\c(\xk)-\k_1\s(\xj)\s(\xk)\cc(\xI)e^{i
\apI}
\\[4pt] 1I&\cc(\xI)e^{i
\Om}=\cc(\xJ)\cc(\xK)-\k_2\ss(\xJ)\ss(\xK)\c(\xi)e^{i
\lpi}
\\[4pt]
\displaystyle 2ij\equiv 2IJ&\displaystyle \frac{\s(\xi)}{\ss(\xI)}=
\frac{\s(\xj)}{\ss(\xJ)}
\\[10pt] 3iJ&\s(\xi)\cc(\xJ)e^{i \lpk}=-\c(\xj)\s(\xk)e^{-i
\apI}-\s(\xj)\c(\xk)\cc(\xI)
\\[4pt] 3Ij&\ss(\xI)\c(\xj)e^{i \apK}=-\cc(\xJ)\ss(\xK)e^{-i
\lpi}-\ss(\xJ)\cc(\xK)\c(\xi)
\\[4pt] 4ij\equiv 4IJ
&-\k_1\s(\xi)\s(\xj)+\c(\xi)\c(\xj)\cc(\xK)e^{i\apK}\\
        &\qquad
=-\k_2\ss(\xI)\ss(\xJ)+\cc(\xI)\cc(\xJ)\c(\xk)e^{i\lpk}.
\end{array}
\mylabel{ht:TrigFinalEqns}
\ee

These equations will be referred to by a tag, and are either
self-dual (for instance 
$2ij \equiv 2IJ$) or appear in mutually dual pairs (as $1i, 1I$).
Equations with tag $0$ allow the introduction of the `symmetric'
invariants $\om$ and $\Om$ and are in the `Cartan' sector. The
remaining tags are intentionally made to match the ones used in
\cite{SpaceTimeTrig}; the equations with tags $1, 2, 3, 4$ are in
most respects the closest `complex  hermitian' analogues to the
equations found in the real case, as far as their mutual relations,
dependence or independence, etc.\  are concerned. Therefore the
trigonometry of real spaces provide a rough first guide in the
exploration of the  whole forest of `complex hermitian'
trigonometric equations.


\subsection{The `complex hermitian' trigonometric bestiarium}

Taking the equations (\ref{ht:TrigFinalEqns})  as starting point,
we now perform a fully explicit study of `complex hermitian'
trigonometry, including some comments.  As the scheme enjoys
self-duality, those equations which are not self-dual will have a
dual partner, obtained by exchange in capitalization of names and
indices: $x\leftrightarrow X$, $\lp
\leftrightarrow \ap$, 
$i \leftrightarrow I$ and in CK constants $\k_1 \leftrightarrow
\k_2$; in these cases we will only sketch the derivation of one
member of the dual pair, but we will write each pair together, to
emphasize  self-duality as the main trait of this approach. The
label $\icd$ does not change under duality. 

These equations will  hold for {\em all twenty seven}  `complex'
CKD spaces with arbitrary $\icd; \k_1, \k_2$.  In the degenerate
cases $\k_1=0$ (flat `complex hermitian' spaces) and/or $\icd=0
\hbox{ or } \k_2=0$ (degenerate `complex' `Hermitian' metric) some
equations may collapse or even reduce to trivial identities; these
cases will be  discussed  later but for the moment we will stay in
the general situation where $\icd; \k_1, \k_2$ are assumed to have
{\em any} values. All equations found in the literature for the
elliptic (hyperbolic) complex hermitian spaces will follow from
this set after  we specialize $\icd=1; \k_1=1\  (\k_1=-1),
\k_2=1$; in those cases, the equations we found will be allocated
a suitable name. 

\noindent$\bullet$ The {\em Cartan sector} equations $0IJ\equiv
0ij$ will be called the {\em `complex hermitian' phases theorem}.
They are self-dual and involve only the triangle Cartan
invariants.  They allow the introduction of two symmetric triangle
invariants
$\om$ and $\Om$ after (\ref{ht:omegaDef}) and (\ref{ht:OmegaDef}):
\be 0IJ\equiv 0ij \qquad\qquad \lpi-\apI=\lpj-\apJ=\lpk-\apK
=\Om-\om.
\mylabel{ht:CKsineArg}
\ee There are two such  independent equations, thus {\em four}
independent quantities among the six lateral and angular phases. 
This number equals the number of essential independent triangle
invariants; this is not accidental (see the comment at the end of
Sect. 6.5).

\noindent$\bullet$ The  equations $2iJ\equiv 2jI$, taken together
will be called the {\em hermitian sine theorem}.
\be 2IJ\equiv 2ij \qquad\qquad
\frac{\s(\xi)}{\ss(\xI)}=\frac{\s(\xj)}{\ss(\xJ)}=\frac{\s(\xk)}{\ss(\xK
)}.
\mylabel{ht:CKsine}
\ee This self-dual relation has two independent equations linking
the six `pure' sides and angles. The hermitian phases theorem can
be written in terms of the phase factors $e^{i\lpi},e^{i\apI}$ and
has the same form as the sine theorem; this is so because phases
theorem (\ref{ht:CKsineArg}) and sine theorem (\ref{ht:CKsine}) are
the modulus and argument of the same `complex' equality.

\noindent$\bullet$ Each of the {\em `complex hermitian' cosine
theorems}
$1i$ and $1I$ is a `complex' equation. By  splitting the hermitian
cosine theorem $1i$ into real and imaginary parts, we get:
\be 1i \qquad \begin{array}{l}
\c(\xi)\ci(\om)=\c(\xj)\c(\xk)-\k_1
\s(\xj)\s(\xk)\cc(\xI)\ci(\apI) 
\\[3pt]
\c(\xi)\si(\om)=-\k_1 \s(\xj)\s(\xk)\cc(\xI)\si(\apI)
\end{array}
\mylabel{ht:HermCosRI}
\ee called real and imaginary Hermitian cosine laws (for sides).
Their duals are the  real and imaginary Hermitian dual cosine laws
(for angles):
\be 1I \qquad \begin{array}{l}
\cc(\xI)\ci(\Om) =  \cc(\xJ)\cc(\xK)-\k_2\ss(\xJ)\ss(\xK)\c(\xi)
\ci(\lpi) 
\\[3pt]
\cc(\xI)\si(\Om) = -\k_2 \ss(\xJ)\ss(\xK)\c(\xi)\si(\lpi).
\end{array}
\mylabel{ht:HermDualCosRI}
\ee

\noindent$\bullet$ By equating the modulus of both sides of the
hermitian cosine theorem $1i$ we get:
\be
\begin{array}{l}
\c^2(\xi)=
\Big(\c(\xj)\c(\xk)-\k_1\s(\xj)\s(\xk)\cc(\xI)\ci(\apI)\Big)^2\\
\qquad +\,\icd\, \k_1^2 \,\s^2(\xj)\s^2(\xk) \cc^2(\xI)
\si^2(\apI).
\mylabel{ht:SRcos}
\end{array}
\ee For the complex elliptic case this is the Shirokov-Rosenfeld
cosine theorem (\ref{ht:ElipSRcos}) \cite{ShiPetRoz} yet expressed
in terms of the angular variables $\xI$ and $\apI$, instead of the
ones used in \cite{ShiPetRoz}. In the general case we will also
call this equation Shirokov-Rosenfeld cosine theorem. This admits
another form, starting from 
$\c(2\xi) + 1 = 2 \c^2(\xi)$, substituting (\ref{ht:SRcos}) and
expanding the squared sines of sides:
\be
\begin{array}{l}
\c(2\xi)= \c(2\xj)\c(2\xk) -\k_1 \s(2\xj)\s(2\xk)\cc(\xI)\ci(\apI)
\\
\qquad  -2\, \k_1^2 \k_2\, \s^2(\xj)\s^2(\xk)\ss^2(\xI)
\mylabel{ht:SRcos2}
\end{array}
\ee to be called Shirokov-Rosenfeld cosine double theorem because
in the complex elliptic case it reduces to (\ref{ht:ElipSRcos2}).
The dual is the Shirokov-Rosenfeld cosine theorem (for angles):
\be
\begin{array}{l}
\cc^2(\xI)=
\Big(\cc(\xJ)\cc(\xK)-\k_2\ss(\xJ)\ss(\xK)\c(\xi)\ci(\lpi)\Big)^2\\
\qquad +\icd\, \k_2^2 \,\ss^2(\xJ)\ss^2(\xK) \c^2(\xi) \si^2(\lpi)
\mylabel{ht:SRdualcos}
\end{array}
\ee and Shirokov-Rosenfeld dual cosine double theorem (for angles):
\be
\begin{array}{l}
\cc(2\xI)=
\cc(2\xJ)\cc(2\xK) -\k_2 \ss(2\xJ)\ss(2\xK)\c(\xi)\ci(\lpj)
\\ \qquad -2\,\k_1\k_2^2\, \ss^2(\xJ)\ss^2(\xK)\s^2(\xi). \nonumber
\mylabel{ht:SRdualcos2}
\end{array}
\ee

\noindent$\bullet$ By building up the term $\k_1 
\s(2\xj)\s(2\xk)\cc(\xI)\ci(\apI)$ in (\ref{ht:HermCosRI}) and
sustituting it into (\ref{ht:SRcos2}), expanding and simplifying
we obtain:
\be
\c^2(\xi)= -\c^2(\xj)\c^2(\xk)+\k_1^2\,\s^2(\xj)\s^2(\xk)\cc^2(\xI)
+2\c(\xi)\c(\xj)\c(\xk)\ci(\om)
\mylabel{ht:BTcos}
\ee which will be called the Blaschke-Terheggen cosine theorem for
sides \cite{BlasTer, Ter}. Its dual is:
\be
\c^2(\xI)=-\cc^2(\xJ)\cc^2(\xK)+\k_2^2\,\ss^2(\xJ)\ss^2(\xK)\c^2(\xi)
+2\cc(\xI)\cc(\xJ)\cc(\xK)\ci(\Om)
 \mylabel{ht:BTdualcos}
\ee In the complex hermitian elliptic case (\ref{ht:BTcos}) and
(\ref{ht:BTdualcos}) reduce directly to the Blaschke-Terheggen
cosine (\ref{ht:ElipBTcos}) and dual cosine theorems
(\ref{ht:ElipBTdualcos}).

\noindent$\bullet$ By multiplying both sides of (\ref{ht:CKsine})
by 
$1/\si(\om)$ and using the second equation in (\ref{ht:HermCosRI})
we obtain:
\be
\frac{\s(2\xi)}{\si(\apI) \cc(\xI)} = \frac{\s(2\xj)}{\si(\apJ)
\cc(\xJ)}
\mylabel{ht:SRsin2}
\ee called Shirokov-Rosenfeld double sine theorem, because in the
complex elliptic case reduces to the SR double sine law
(\ref{ht:ElipSRsin2}) after changing to the angular variables used
by SR. Its dual is:
\be
\frac{\ss(2\xI)}{\si(\lpi) \c(\xi)} = \frac{\ss(2\xJ)}{\si(\lpj)
\c(\xj)}.
\mylabel{ht:CKSRdualsin2}
\ee

\noindent$\bullet$ By multiplying (\ref{ht:SRsin2}) and
(\ref{ht:CKSRdualsin2}) we get the self-dual equation:
\be
\frac{\s(\xi)\s(\xI)}{\si(\lpi)\si(\apI)} =
\frac{\s(\xj)\s(\xJ)}{\si(\lpj)
\si(\apJ)}.
\mylabel{ht:SS}
\ee

\noindent$\bullet$ By taking quotient between the double sine
theorem (\ref{ht:SRsin2}) and the sine theorem (\ref{ht:CKsine})
we get:
\be
\frac{\c(\xi)\TT(\xI)}{\si(\apI)}=\frac{\c(\xj)\TT(\xJ)}{\si(\apJ)}
\mylabel{ht:cT}
\ee whose dual is:
\be
\frac{\cc(\xI)\T(\xi)}{\si(\lpi)} =
\frac{\cc(\xJ)\T(\xj)}{\si(\lpj)}.
\mylabel{ht:Ct}
\ee

\noindent$\bullet$ Another equations derive from the equations
with tags
$3iJ$ and $3Ij$. In particular, by splitting the equations $3iJ$
into their real and imaginary parts we obtain:
\be 3iJ \qquad \begin{array}{l}
\s(\xi)\cc(\xJ)\ci(\lpk) = -\c(\xj)\s(\xk)\ci(\apI)-
\s(\xj)\c(\xk)\cc(\xI)
\\[3pt]
\s(\xi)\cc(\xJ)\si(\lpk) = \c(\xj)\s(\xk)\si(\apI)
\mylabel{ht:Eq3RI}
\end{array}
\ee whose duals are:
\be 3Ij \qquad \begin{array}{l}
\ss(\xI)\c(\xj)\ci(\apK)= -\cc(\xJ)\ss(\xK)\ci(\lpi)-
\ss(\xJ)\cc(\xK)\c(\xi) 
\\[3pt]
\ss(\xI)\c(\xj)\si(\apK) = \cc(\xJ)\ss(\xK)\si(\lpi).
\mylabel{ht:Eq3RIDual}
\end{array}
\ee

\noindent$\bullet$ The same splitting for the equations $4ij \equiv
4IJ$ leads to the pair of self-dual equations:
\be 4ij\equiv 4 IJ \qquad \begin{array}{l} -\k_2\ss(\xI)
\ss(\xJ)+\cc(\xI)\cc(\xJ)\c(\xk)\ci(\lpk) \\[2pt]
\qquad\qquad =-\k_1\s(\xi)\s(\xj)+ \c(\xi)\c(\xj)\cc(\xK)\ci(\apK)
\\[4pt]
\cc(\xI)\cc(\xJ)\c(\xk)\si(\lpk)=\c(\xi)\c(\xj)\cc(\xK)\si(\apK)
\mylabel{ht:Eq4RI}
\end{array}
\ee

\noindent$\bullet$ Starting from the real and imaginary parts of
the `complex hermitian' cosine theorem (\ref{ht:HermCosRI}),
expanding the  trigonometric functions of $\om= \apI + \lpj +\apK$
by considering it as a sum of two phases  and eliminating  the
term containing $\ci (\apI+\lpj)$ we get:
\be -\frac{\c(\xi)}{\si(\apI)}=\frac{\c(\xj)\c(\xk)}{\si (\lpj
+\apK)} = \frac{\c(\xj)\c(\xk)}{\si(\om-\apI)}.
\mylabel{ht:Eqcc}
\ee Its dual is:
\be -\frac{\cc(\xI)}{\si(\lpi)}=\frac{\cc(\xJ)\cc(\xK)}{\si (\apJ
+\lpk)} = \frac{\cc(\xJ)\cc(\xK)}{\si(\Om-\lpi)}.
\mylabel{ht:EqCC}
\ee where we have used the relations $\apJ +\lpk = \om - \apI =
\Om - \lpi$ which follow from the equations in the `Cartan' 
sector and the definitions of
$\om$ and $\Om$.

\noindent$\bullet$ By dividing the equation (\ref{ht:cT}) by
(\ref{ht:Eqcc}) we get:
\be -\frac{\TT(\xI)}{\si(\apI)}= \frac{\TT(\xK)
\c(\xj)}{\si(\apI+\lpj)} = \frac{\TT(\xK) \c(\xj)}{\si(\om-\apK)} 
\mylabel{ht:EqTc}
\ee whose form  for another suitable choice of indices is:
\be -\frac{\TT(\xK)}{\si(\apK)}= \frac{\TT(\xJ)
\c(\xj)}{\si(\apK+\lpj)} = \frac{\TT(\xJ) \c(\xj)}{\si(\om-\apI)}.
  \mylabel{ht:EqTc2}
\ee The duals of these equations are:
\be -\frac{\T(\xi)}{\si(\lpi)} =  \frac{\T(\xk)
\cc(\xJ)}{\si(\lpi+\apJ)} = \frac{\T(\xk) \cc(\xJ)}{\si(\Om-\lpk)}
\mylabel{ht:EqtC}
\ee
\be -\frac{\T(\xk)}{\si(\lpk)}= \frac{\T(\xj)
\cc(\xJ)}{\si(\lpk+\apJ)} = \frac{\T(\xj) \cc(\xJ)}{\si(\Om-\lpi)}.
 \mylabel{ht:EqtC2}
\ee

\noindent$\bullet$ By eliminating the angles $\xI, \xK$ between
(\ref{ht:EqTc}) and (\ref{ht:EqTc2}):
\be
\c^2(\xk) = \frac{\si (\apI + \lpk)\si (\apJ +
\lpk)}{\si(\apI)\si(\apJ)} =
\frac{\si(\om-\apI)\si(\om-\apJ)}{\si(\apI)\si(\apJ)}
\mylabel{ht:cEuler}
\ee whose dual is:
\be
\cc^2(\xK) = \frac{\si (\lpi + \apK)\si (\lpj +
\apK)}{\si(\lpi)\si(\lpj)}
 =  \frac{\si (\Om-\lpi)\si (\Om-\lpj)}{\si(\lpi)\si(\lpj)}.
\mylabel{ht:CEuler}
\ee These equations  give the cosine of each  side (angle) in
terms of the angular (lateral) phases {\em only}. They somehow
ressemble real trigonometry Euler equations for the cosine of half
the sides (angles) in terms of angles (sides). In these hermitian
`Euler-like' equations, pure sides (angles) are however given in
terms of angular phases and $\om$ (lateral phases and $\Om$).

\noindent$\bullet$ By expansion of sines of sums or differences and
elementary manipulation, we finally get the expresion for the
squared sines of the sides:
\be
 \s^2(\xk) = -\frac{\si(\lpk)\frac{\si(\om)}{\k_1}}
{\si(\apI)\si(\apJ)}
\mylabel{ht:sEuler}
\ee whose dual equation is:
\be
\ss^2(\xK)  =  -\frac{\si(\apK)\frac{\si(\Om)}{\k_2}}
{\si(\lpi)\si(\lpj)}.
\mylabel{ht:SEuler}
\ee

As we shall see shortly, and in spite of the presence of $\k_1,
\k_2$  in denominators, these equations are still meaningful when
$\k_1 \to 0$ or
$\k_2 \to 0$.


\subsection{Symplectic area and coarea}

For real CK spaces the angular excess $\Delta$ shares three
properties: $\Delta$ goes to zero with
$\k_1$, it is {\it proportional} (coefficient $\k_1$) to  triangle
area, and satisfies Gauss-Bonnet type equations. These three
properties are splitted in the `complex' case:  In the hermitian
point loop equations (\ref{ht:GB}) and (\ref{ht:GBab}), the
`complete excess' $(\Delta, \Delta_\ap)$ plays a role partly
analogue of the real angular excess, yet it may not vanish with
$\k_1$. There are {\it two different} independent  hermitian
triangle quantities which vanish with
$\k_1$. One of them is the Blaschke-Terheggen invariant $\om$.
This follows directly from the equations already derived. The
situation for the other vanishing quantity is not so obvious (see
however the comments in the next Section). Dually, while  the real
excess $\delta$ is proportional to the coarea, vanishes with
$\k_2$, and satisfies dual Gauss-Bonnet type equations, the
complete lateral excess ($\delta,
\delta_\lp$) appears in (\ref{ht:GBpolarABC}) and may not vanish
with
$\k_2$, while $\Om$ vanishes with $\k_2$.

In the real case, the three cosine equations $1i$ $(1I)$ turned
into trivial identities when $\k_1=0$ ($\k_2=0$). In the `complex
hermitian' case, the {\it three `complex'} independent  equations
$1i$ $(1I)$, which are independent when $\k_1\neq0$ ($\k_2\neq0$) 
collapse when $\k_1=0$ ($\k_2=0$) into a {\em single real} one in
the `Cartan' sector, and as far as pure sides  and angles are
concerned become trivial:
\bea 1i \hbox{ when $\k_1=0$}\qquad e^{i\om}=\ci(\om) +
i\si(\om)=1 \hbox{\ implying 
$\ci(\om)=1$, $\si(\om)=0$}
\mylabel{ht:Degen1}
\\ 1I \hbox{ when $\k_2=0$}\qquad e^{i\Om}=\ci(\Om) +
i\si(\Om)=1\hbox{\ implying 
$\ci(\Om)=1$, $\si(\Om)=0$}.
\mylabel{ht:Degen2}
\eea

The behaviour of the quotient $\frac{\si(\om)}{\k_1}$ as $\k_1 \to
0$ can be derived both from the imaginary part of the Hermitian
cosine theorem (\ref{ht:HermCosRI}) and from equations
(\ref{ht:sEuler}):
\be
\frac{\si(\om)}{\k_1}=-\frac{
\s(\xj)\s(\xk)\cc(\xI)\si(\apI)}{\c(\xi)}  = -\frac
{\si(\apI)\si(\apJ)\s^2(\xk)}{\si(\lpk)}
\mylabel{ht:omegaFlat}
\ee and since this quotient remains {\em finite} as $\k_1\to 0$,
$\om$ behaves like the real case angular  excess $\Delta=-A+B+C$.
Dually, 
$\Om$ behaves as the real pure lateral excess
$\delta=-a+b+c$:
\be
\frac{\si(\Om)}{\k_2} = -
\frac{\ss(\xJ)\ss(\xK)\c(\xi)\si(\lpi)}{\cc(\xI)} = -\frac
{\si(\lpi)\si(\lpj)\ss^2(\xK) }{\si(\apK)}
\mylabel{ht:OmegaFlat}
\ee

The real excesses $\Delta, \delta$ are  {\em proportional}, with
coefficients
$\k_1$ and $\k_2$, to the triangle area and coarea respectively.
In the elliptic hermitian space $\Ce P^2$ Hangan and Masala
\cite{HanMas} found for the  {\em symplectic} triangle area
$\symarea$ the relation $\symarea=-\om/2$ (the inessential minus
sign comes form their definition of symplectic form). For any
member of the CKD family of the `complex Hermitian' spaces, the
{\em definitions}  for triangle symplectic area and coarea :
\be
\symarea:=\frac{\om}{2\k_1} \qquad  \symcoarea:=\frac{\Om}{2\k_2}
\mylabel{ht:DefSymArea}
\ee (note the factor $2$) are in full agreement with the standard
definition of symplectic area as the integral of the symplectic
form  over any surface dressing the triangle; this form is closed
so by  the Stokes theorem the integral depends only on the
boundary.

Therefore all appearances of $\om$ or $\Om$ in trigonometric the
equations  could be rewritten in terms of trigonometric functions
of the {\em symplectic area} $\symarea$ with label $\icd \k_1^2$
(the symplectic area  goes like the product of lengths along the
geodesics generated by $P_1$ and $Q_1$,  whose labels are $\k_1$
and $\icd\k_1$), and {\em symplectic coarea} $\symcoarea$, with
label $\icd \k_2^2$.
\be
\csa(2\symarea) = \ci(\om) \quad \ssa(2\symarea) =
\displaystyle\frac{\si(\om)}{\k_1} \qquad 
\csca(2\symcoarea) = \ci(\Om) \quad  \ssca(2\symcoarea) =
\displaystyle\frac{\si(\Om)}{\k_2}
\mylabel{ht:SymAreaOmega}
\ee When $\k_1=0$, $\om$  vanishes but $\symarea$ keeps some
finite value, a kind of `residue' of the generically non vanishing
mixed phase excess $\om$.  Dually, the same happens for $\Om$ and
$\symcoarea$ as $\k_2\to0$.   


\subsection{Dependence and basic equations}
  
In the complex hermitian and hyperbolic spaces a triangle is known
to be determined by {\em four} independent quantities. Since we
have found {\em eight} generic independent relations between the
twelve sides, angles and phases, this is still true in the generic
CKD space $\iCe\CKPointSpace$. Two such relations are the
hermitian phase theorems (\ref{ht:CKsineArg}). The other six
happen to be exactly  twice as many independent equations as in
the real case, due to their {\it `complex'} nature. This allows to
split the dependence disscusion into  the Cartan sector and the
equations with a real analogue.

The Cartan sector includes six phases, to which we will add
symplectic area  and coarea. For {\em any} value of the labels
$\k_1, \k_2, \icd$, there are {\em four} independent equations
between the eight quantities $\lpi, \apI, \symarea,
\symcoarea$: 
\be
\begin{array}{c}
\lpi-\apI=\lpj-\apJ=\lpk-\apK \ (=\Om-\om)\\[2pt] (\om=)\
\apI+\apJ+\lpk = \k_1 2\symarea, \qquad  (\Om=)\ \lpi+\lpj+\apK =
\k_2 2\symcoarea
\end{array}
\mylabel{ht:CKsineArg2}
\ee
 so in any case there are always four independent such Cartan
sector quantities. Generically the triangle is almost determined
by these quantities (see
\ref{ht:sEuler}, \ref{ht:SEuler}).

In order to discuss the dependence of the remaining equations, 
let us consider:
\be
\begin{array}{l}
\!\!\gramm:={1-\c^2(\xi)-\c^2(\xj)-
\c^2(\xk)+2\c(\xi)\c(\xj)\c(\xk)\ci(\om)}
 \\[3pt]
\!\!\Gramm:={1-\cc^2(\xI)-\cc^2(\xJ)-
\cc^2(\xK)+2\cc(\xI)\cc(\xJ)\cc(\xK)\ci(\Om)}.
\mylabel{ht:GrammDef}
\end{array}
\ee

The quantity $\gramm$ is the determinant of the Gramm matrix whose
elements are the `hermitian' products  of the vectors
corresponding to the vertices in the linear ambient space, and
$\Gramm$ its dual quantity. From (\ref{ht:Degen1}) and
(\ref{ht:CKSineCosine}) it follows that
$\gramm$ vanishes when $\k_1 \to 0$; dually the same happens for
$\Gramm$ when
$\k_2\to 0$. Further, the quotient $\gramm/\k_1^2$ (resp.
$\Gramm/\k_2^2$) tends to a well defined finite limit when
$\k_1\to 0$ (resp. $\k_2\to 0$), although still goes to zero when
$\k_2\to 0$ (resp.\ when $\k_1\to 0$). To see this, simplify
(\ref{ht:GrammDef}) by using (\ref{ht:BTcos}) or
(\ref{ht:BTdualcos}) to obtain:
\be
\gramm = \k_1^2 \k_2 \s^2(\xi)\ss^2(\xJ)\s^2(\xk) \qquad 
\Gramm = \k_1 \k_2^2 \ss^2(\xI)\s^2(\xj)\ss^2(\xK). 
\ee This suggests to introduce two new `renormalized' quantities
$\grammR, \GrammR$ in a way similar to (\ref{ht:DefSymArea}):
\be
\grammR:=\frac{\gramm}{\k_1^2}, \qquad 
\GrammR:=\frac{\Gramm}{\k_2^2},
\mylabel{ht:GrammRDef}
\ee Relations between
$\GrammR, \grammR$ and  symplectic area and  coarea $\symarea,
\symcoarea$ holding for any $\icd; \k_1,
\k_2$ follow by expressing sines of sides and angles in
(\ref{ht:GrammRDef}) by means of (\ref{ht:sEuler}) and
(\ref{ht:SEuler}):
\be
\frac{\grammR}{\k_2}  = \s^2(\xi)\ss^2(\xJ)\s^2(\xk) =
-\frac{\ssa^2(2\symarea) \ssca(2\symcoarea)}{\si(\apI) \si(\apJ)
\si(\apK)}
\label{ht:gammaLados}
\ee
\be
\frac{\GrammR}{\k_1}  = \s^2(\xI)\ss^2(\xj)\s^2(\xK) =
-\frac{\ssa(2\symarea) \ssca^2(2\symcoarea)}{\si(\lpi) \si(\lpj)
\si(\lpk)}
\label{ht:GammaAngulos}
\ee By direct substitution using (\ref{ht:gammaLados}) and
(\ref{ht:GammaAngulos}) we can also derive the following relations
between $\GrammR, \grammR$ and the pure sides and angles:
\be
\frac{\GrammR}{\k_1} = \frac{\left( {\grammR}/{\k_2}
\right)^2}{\s^2(\xi)\s^2(\xj)\s^2(\xk) }, \qquad
\frac{\grammR}{\k_2} = \frac{\left( {\GrammR}/{\k_1}
\right)^2}{\ss^2(\xI)\ss^2(\xJ)\ss^2(\xK) }
\ee

Let us now discuss the dependence issue in the generic case
$\k_1\neq0, \k_2\neq0$, where all the  equations with tags
$2$, $3$ and $4$ follow from $1$, {\em exactly alike} in the real
case. This means that a triangle is completely determined by the
three sides $a, b, c$ and  $\om$, or by the three angles $A, B, C$
and $\Om$.   The proofs are also a verbatim translation of the
real ones, with a single hermitian {\em caveat}: sometimes the
`complex' conjugate of an equation $1I$ or $1i$ should be used. For
instance, let us derive the hermitian sine theorem from the dual
cosine theorems $1I$ in the case $\k_2\neq0$. Start from the
identity $
\s^2(\xi)\ss^2(\xJ)=
\s^2(\xi){ \left(1- \cc^2(\xJ) \right)}/ {\k_2}$, replace one
factor $\cc(\xJ)$ by its expression taken from $1J$, and the other
$\cc(\xJ)$  by the complex conjugate; then expand:
\be
\s^2(\xi)\ss^2(\xJ)=\frac{1-\c^2(\xi)-\c^2(\xj)-
\c^2(\xk)+2\c(\xi)\c(\xj )\c(\xk)\ci(\om)} {\k_1^2\k_2\s^2(\xk)} .
\ee which by using (\ref{ht:GrammDef}) and (\ref{ht:GrammRDef})
can be rewritten as: 
\be
\s^2(\xi)\ss^2(\xJ)\s^2(\xk)=\frac{\grammR}{\k_2}
\mylabel{ht:SSSgramm}
\ee As $\k_2\neq 0$, the r.h.s of (\ref{ht:SSSgramm}) is well
defined and is clearly symmetric in the indices $ijk$ (this was
also clear from (\ref{ht:GrammRDef})). Therefore
$\s^2(\xi)\ss^2(\xJ)\s^2(\xk) =
\s^2(\xj)\ss^2(\xK)\s^2(\xi)$ leads to the sine theorem. Dually,
when
$\k_1\neq0$ the sine theorem follows also from the three cosine
theorems $1i$ .  By following the real pattern, the dual hermitian
cosine theorem and equations with tag $3$ and $4$ can also be
derived. 

Therefore, in the generic  $\k_1\neq0, \k_2\neq0$ case, the three
`complex hermitian' cosine theorems $1i$ seen as {\em six}
idependent real equations are a set of basic equations. By
duality, the same applies to the three dual cosine theorems $1I$.
By adding to either choice the four independent `Cartan sector'
phase equations relating the six phases, $\symarea, \symcoarea$, 
we get a complete set of ten equations relating fourteen
quantities. A triangle in the hermitian spaces $\k_1\neq0,
\k_2\neq0$ is characterized by {\em four} independent quantities,
for instance either
$a, b, c, \om$ or $A, B, C, \Om$, which are a dual pair; this was
already known for hermitian elliptic or hyperbolic spaces, but
holds for the complete family of `complex' CKD spaces as we shall
see in the next section. Another choice are  the six phases linked
by the two relations (\ref{ht:CKsineArg}). 


\subsubsection{Alternative forms of Hermitian Cosine Equations when
$\k_1=0$ or $\k_2=0$} 

The collapse of the Hermitian cosine equations (\ref{ht:HermCosRI})
to (\ref{ht:Degen1})  when $\k_1=0$ can be circumvented by writing
(\ref{ht:HermCosRI}) in an alternative form. The imaginary  part
can be rewritten in terms of the symplectic area $\symarea$, by
using (\ref{ht:SymAreaOmega}).  For the real part, the procedure
mimicks the real one \cite{SpaceTimeTrig}: write all cosines of
the sides {\it and} of
$\om$ in terms of versed sines $V_\k(x):= (1-C_\k(x))/\k$
\be
\c(\xi)=1-\k_1 \v(\xi) \qquad 
\ci(\om) = \csa(2\symarea) = 1 - \icd\, \k_1^2 \vsa(2\symarea)
\ee and then substitute in $1i$, expand, cancel a common factor
$\k_1$ and use the identities for the versed sines of a sum. Thus
the pair of equations (\ref{ht:HermCosRI}) can be rewritten as:
\be 1i'\quad \begin{array}{l}
\v(\xi) - \v(\xj+\xk) = \s(\xj)\s(\xk) \big(
\cc(\xI)\ci(\apI)-1\big) -
\icd\, \k_1 \vsa(2\symarea) \c(\xi)\\[3pt]
\c(\xi)\ssa(2\symarea)=-\s(\xj)\s(\xk)\cc(\xI)\si(\apI)
\end{array}
\mylabel{ht:HermCosRIreform}
\ee a form meaningful for any value of $\k_1$ (even if all other
labels are equal to zero). When $\k_1=0$ they reduce to 
\be
\begin{array}{l}
\xi^2 = \xj^2 + \xk^2 + 2 \xj \xk \cc(\xI)\ci(\apI)\\[3pt]
2\symarea=-\xj\xk \cc(\xI)\si(\apI).
\end{array}
\mylabel{ht:HermCosRIreformk1null}
\ee

The dual  cosine equations $1I$ allow a similar reformulation:
\be 1I'\quad \begin{array}{l}
\vv(\xI) - \vv(\xJ+\xK) = \ss(\xJ)\ss(\xK) \big(
\c(\xi)\ci(\lpi)-1\big) -
\icd\, \k_2 \vsca(2\symcoarea) \cc(\xI)\\[3pt]
\cc(\xI)\ssca(2\symcoarea)=-\ss(\xJ)\ss(\xK)\c(\xi)\si(\lpi).
\end{array}
\mylabel{ht:HermCosRIdualreform}
\ee which is meaningful for any value of $\k_2$

The quantities $\grammR$ ($\GrammR$) can be also given in terms of
sides and symplectic area, (angles and symplectic coarea) by 
expressions which are still meaningful when $\k_1=0$ ($\k_2=0$):
\be
\begin{array}{ll} & \grammR =  2\big( \v(\xi)\v(\xj) +
\v(\xj)\v(\xk) + \v(\xk)\v(\xi)\big)  
       -\v^2(\xi) -\v^2(\xj) -\v^2(\xk) \nonumber \\ 
      & \qquad -2 \,\icd\, \c(\xi)\c(\xj)\c(\xk)\vsa(2\symarea) 
        -2 \k_1 \v(\xi)\v(\xj)\v(\xk)  \\[3pt] &
\!\!\!\!\!\!\GrammR=  2 \big( \vv(\xI)\vv(\xJ) + \vv(\xJ)\vv(\xK) +
\vv(\xK)\vv(\xI)\big) 
       \!-\!\vv^2(\xI) \!-\!\vv^2(\xJ) \!-\!\vv^2(\xK) \nonumber
\\ 
      & \qquad -2 \,\icd\,
\cc(\xI)\cc(\xJ)\cc(\xK)\vsca(2\symcoarea) 
        -2 \k_2 \vv(\xI)\vv(\xJ)\vv(\xK). 
\end{array}
\ee

Thus when $\k_1=0$ but $\k_2\neq0$, the six real equations $1i'$
are independent and all the remaining equations follow from the
Cartan sector  equations (\ref{ht:CKsineArg2}) and from the three
pairs
$1i'$ just as everything followed from $1i$ in the case $\k_1\neq
0$. Dually, mutatis mutandis from $1I'$ when $\k_2=0$ but
$\k_1\neq0$. 

We finally discuss the situation when
$\k_1=\k_2=0$. The lateral and angular phases are equal:
$\apI=\lpi$ and this provides three independent equations. Two
further equations are the sine theorem which in this case cannot
be derived from $1i'$ or $1I'$. This makes {\it five} independent
equations:
\be
\apI=\lpi, \qquad
\frac{\xi}{\xI}=\frac{\xj}{\xJ}=\frac{\xk}{\xK},\quad 
\mylabel{ht:HermCosRIanddualreformallzero}
\ee The remaining details depend on whether $\icd$ is zero or not. 
If $\icd\neq0$, the equations $1i'/1I'$ read: 
\be
\begin{array}{ll}
\hbox{Re}1i'\quad \xi^2 = \xj^2 + \xk^2 + 2 \xj \xk \ci(\apI)
\quad  &\hbox{Re}1I'\quad\xI^2 =\xJ^2 + \xK^2 + 2 \xJ \xK
\ci(\lpi)\\[3pt]  
\hbox{Im}1i'\quad 2\symarea=-\xj\xk \si(\apI)  &\hbox{Im}1I'\quad
2\symcoarea=-\xJ\xK\si(\lpi)
\end{array}
\mylabel{ht:zeroicdno}
\ee Taking into account (\ref{ht:HermCosRIanddualreformallzero}),
the groups of three  equations $\hbox{Re}1i'$ and $\hbox{Re}1I'$
are equivalent; either of them can be taken as three further
equations. Any of these sets imply the relation (whose general
form is (\ref{ht:cT}))
\be
\frac{\xi}{\si(\apI)}=\frac{\xj}{\si(\apJ)}=\frac{\xk}{\si(\apK)}
\ee which shows that the three equations either $\hbox{Im}1i'$ or
$\hbox{Im}1I'$ collapse to a single equation. Taken altogether,
these provide another {\it five} independent equations in
(\ref{ht:zeroicdno}). When
$\icd=0$, i.e., in the most contracted case,  the three equations
$\hbox{Re}1i'$ collapse to a single equation better written as
$\xi+\xj+\xk=0$ and likewise
$\hbox{Re}1I'$ collapse to $\xI+\xJ+\xK=0$; these two equations are
however not independent in view of the sine theorem. In this case
the most contracted form of  (\ref{ht:cT}) cannot be derived from
previous equations and have to be added as two further independent
equations, in either of the two forms:
\be
\frac{\xi}{\apI}=\frac{\xj}{\apJ}=\frac{\xk}{\apK}
\qquad \hbox{ or } \qquad 
\frac{\xI}{\lpi}=\frac{\xJ}{\lpj}=\frac{\xK}{\lpk};
\ee using these equations, each group of three equations
$\hbox{Im}1i'$ or
$\hbox{Im}1I'$ collapses to a single equation. This makes again
{\it five}  independent additional equations altogether in: 
\be
\xi+\xj+\xk=0,\  \xI+\xJ+\xK=0,\  
\frac{\xi}{\apI}=\frac{\xj}{\apJ}=\frac{\xk}{\apK},\ 
2\symarea=-\xj\xk \apI, \  2\symcoarea=-\xJ\xK \lpi. 
\mylabel{ht:HermCosRIanddualreformallzeroicdzero}
\ee

\medskip
\noindent {\bf Theorem 3.} \em The full set of equations of
`complex Hermitian' trigonometry linking the fourteen qantities
$\xi, \xI, \lpi, \apI, \symarea,
\symcoarea$  contains for any value of $\icd, \k_1$, $\k_2$
exactly {\em ten} independent equations. Any other equation in the
set is a consequence of them. When $\k_1$ or $\k_2$ are different
from zero, four such equations are the two phases equations
$0ij\equiv 0IJ$ and the two relations $\om=\k_1 2\symarea, 
\ \Om=\k_2 2\symcoarea$. The remaining six independent equations
are:

$\bullet$ When $\k_1\neq 0$ and $\k_2\neq 0$, any $\icd$, either
the equations $1i$ or
$1I$.

$\bullet$ When $\k_1=0$ but $\k_2\neq 0$, any $\icd$, the
equations $1i'$.

$\bullet$ When $\k_1\neq 0$  but $\k_2=0$, any $\icd$, the
equations
$1I'$. 

\noindent When both $\k_1=\k_2=0$, the independent equations are:

$\bullet$ When  $\icd\neq 0$, the ten independent equations in 
(\ref{ht:HermCosRIanddualreformallzero}) and (\ref{ht:zeroicdno}).

$\bullet$ When $\icd=0$, the ten independent equations in 
(\ref{ht:HermCosRIanddualreformallzero}) and
(\ref{ht:HermCosRIanddualreformallzeroicdzero}). \rm


\begin{table}[h] {\footnotesize { \noindent
\caption{Complex Hermitian sine theorems and relations between
symplectic area $\symarea$, coarea $\symcoarea$ and mixed phase
excesses $\om,
\Om$ for the twenty seven `complex Hermitian' (`CH') CKD spaces. 
The Table is arranged with columns labeled by $\k_1=1, 0, -1$ and
rows by $\k_2=1, 0, -1$; $(\icd; \k_1, \k_2)$ are explicitly
displayed at each entry. All relations in this Table hold in the
same form no matter of the value of $\icd$. The group description
of the homogeneous spaces is given in the CKD type notation.}
\label{table:htSineTrigEqns}
\bigskip
\noindent\hskip-10pt
\begin{tabular}{ccc}
\hline &&\\[-8pt] `CH' Elliptic\quad $(\icd; +1,+1)$& `CH'
Euclidean\quad $(\icd; 0,+1)$& `CH' Hyperbolic\quad $(\icd;
-1,+1)$\\
${}_{\icd}SU(3)/{}_{\icd}U(1){\otimes}{}_{\icd}SU(2)$&
${}_{\icd}IU(2)/{}_{\icd}U(1){\otimes}SU(2)$&
${}_{\icd}SU(2,1)/{}_{\icd}U(1){\otimes}{}_{\icd}SU(2)$\\[4pt]
$\om=2\symarea$&$\om=0$&$\om=-2\symarea$\\[2pt]
$\displaystyle{\frac{\sin a}{\sin A}=\frac{\sin b}{\sin B}
  =\frac{\sin c}{\sin C}}$&
$\displaystyle{\frac{a}{\sin A}=\frac{ b}{\sin B}
  =\frac{ c}{\sin C}}$&
$\displaystyle{\frac{\sinh a}{\sin A}=\frac{\sinh b}{\sin
B}=\frac{\sinh c}{\sin C}}$\\[8pt]
$\apA-\lpa\!=\!\apB-\lpb\!=\!\apC-\lpc\!=\!\om-\Om$
&$\apA-\lpa\!=\!\apB-\lpb\!=\!\apC-\lpc\!=\!-\Om$
&$\apA-\lpa\!=\!\apB-\lpb\!=\!\apC-\lpc\!=\!\om-\Om$\\[4pt]
$\Om=2\symcoarea$&$\Om=2\symcoarea$&$\Om=2\symcoarea$\\[4pt]
\hline &&\\[-8pt] `CH' Co-Euclidean\quad $(\icd; +1,0)$& `CH'
Galilean\quad $(\icd; 0,0)$& `CH' Co-Minkowskian\quad $(\icd;
-1,0)$\\ `CH' Oscillating NH\quad & 
\ & `Complex Hermitian' Expanding NH \\
${}_{\icd}IU(2)/{}_{\icd}U(1){\otimes}{}_{\icd}IU(1)$&
${}_{\icd}IIU(1)/{}_{\icd}U(1){\otimes}{}_{\icd}IU(1)$&
${}_{\icd}IU(1,1)/{}_{\icd}U(1){\otimes}{}_{\icd}IU(1)$\\[4pt]
$\om=2\symarea$&$\om=0$&$\om=-2\symarea$\\[2pt]
$\displaystyle{\frac{\sin a}{ A}=\frac{\sin b}{ B}
    =\frac{\sin c}{ C}}$&
$\displaystyle{\frac{a}{ A}=\frac{ b}{ B}=\frac{ c}{ C}}$&
$\displaystyle{\frac{\sinh a}{ A}=\frac{\sinh b}{ B}
    =\frac{\sinh c}{ C}}$\\[8pt]
$\apA-\lpa\!=\!\apB-\lpb\!=\!\apC-\lpc\!=\!\om$
&$\apA-\lpa\!=\!\apB-\lpb\!=\!\apC-\lpc\!=\!0$
&$\apA-\lpa\!=\!\apB-\lpb\!=\!\apC-\lpc\!=\!\om$\\[4pt]
$\Om=0$&$\Om=0$&$\Om=0$\\[4pt]
\hline &&\\[-8pt] `CH' Co-Hyperbolic\quad $(\icd; +1,-1)$& `CH'
Minkowskian\quad $(\icd; 0,-1)$& `CH' Doubly Hyperbolic\quad
$(\icd; -1,-1)$\\ `CH' Anti-de Sitter&
\ &  `Complex Hermitian' De Sitter\\[4pt]
${}_{\icd}SU(2,1)/{}_{\icd}U(1){\otimes}{}_{\icd}SU(1,1)$&
${}_{\icd}IU(1,1)/{}_{\icd}U(1){\otimes}{}_{\icd}SU(1,1)$&
${}_{\icd}SU(2,1)/{}_{\icd}U(1){\otimes}{}_{\icd}SU(1,1)$\\[4pt]
$\om=2\symarea$&$\om=0$&$\om=-2\symarea$\\[2pt]
$\displaystyle{\frac{\sin a}{\sinh A}=\frac{\sin b}{\sinh B}
    =\frac{\sin c}{\sinh C}}$&
$\displaystyle{\frac{a}{\sinh A}=\frac{ b}{\sinh B}
    =\frac{ c}{\sinh C}}$&
$\displaystyle{\frac{\sinh a}{\sinh A}
   =\frac{\sinh b}{\sinh B}=\frac{\sinh c}{\sinh C}}$\\[8pt]
$\apA-\lpa\!=\!\apB-\lpb\!=\!\apC-\lpc\!=
\!\om-\Om$&$\apA-\lpa\!=\!\apB-\lpb\!=
\!\apC-\lpc\!=\!-\Om$&$\apA-\lpa\!=
\!\apB-\lpb\!=\!\apC-\lpc\!=\!\om-\Om$\\[4pt]
$\Om=-2\symcoarea$&$\Om=-2\symcoarea$&$\Om=-2\symcoarea$\\[4pt]
\hline
\end{tabular} }}
\end{table}

Tables 2 and 3,4,5 display the basic equations for the twenty seven
CKD spaces, written in conventional notation. As neither the Cartan
sector equations, nor the sine theorem involve the CD label
$\icd$, these equations are displayed in a single  Table 2,
according to the values of the CK constants $(\k_1,\k_2)$. The
remaining basic equations, i.e.\  Hermitian cosine and dual cosine
theorems, which involve the CD label $\icd$ are given in an
Appendix threefold display (Tables 3,4,5), one table for each
value of $\icd=1, 0, -1$.  


\subsection{Symmetric invariants and existence conditions}

Several Hermitian trigonometric equations are or can be rewritten
as a relation belonging to one of two  types. The first type has a
structure similar to the sine theorem: a `one-element' expression
involving  only  one index (vertex, opposite side) has the same
value for the two remaining ones:
\be
\apI-\lpi= \om-\Om \quad
\frac{\s (\xi)}{\ss (\xI)} =: \ctesS \quad
\frac{\s (2 \xi)}{\si(\apI)\cc (\xI)} =: \ctesd \quad
\frac{\ss (2 \xI)}{\si(\lpi) \c (\xi)} =: \cteSD.
\ee Under duality $\ctesS \leftrightarrow \frac{1}{\ctesS}$ and
$\ctesd \leftrightarrow \cteSD$.  Other such `one-element' type
equations have values which can be expressed in terms of the three
triangle invariants $\ctesS,\, \ctesd,\, \cteSD$: 
\be
\frac{\s (\xi) \ss (\xI)}{\si(\lpi) \si(\apI)} = \frac{1}{4}\,
\ctesd
\,\cteSD \quad
\frac{\c(\xi)\TT(\xI)}{\si(\apI)} = \frac{1}{2}
\frac{\ctesd}{\ctesS}
\quad
\frac{\cc(\xI)\T(\xi)}{\si(\lpi)} =\frac{1}{2}\, \cteSD\, \ctesS.
\ee

The second type has a structure like formulas allowing the
introduction of $\om, \Om$; a `cyclic' expression   invariant
under any cyclic permutation of the three indices it involves:
\be
\begin{array}{l}
\apI + \apJ  + \lpk =: \om = \k_1 2\symarea \qquad \lpi + \lpj  +
\apK =:
\Om = \k_2 2\symcoarea \\
\ss(\xI)\ss(\xJ)\s(\xk) = {\sqrt{{\GrammR}/{\k_1}}} \qquad
\s(\xi)\s(\xj)\ss(\xK) = {\sqrt{{\grammR}/{\k_2}}}.
\end{array}
\ee

There is {\em no} essential difference between the `one-element'
and `cyclic'  types of equations, and it turns out to be possible
to express the `one-element' invariants in an explicitly `cyclic'
form:
\be
\begin{array}{l}
\displaystyle
\frac{\s(\xi)}{\ss (\xI)} =: \ctesS =
      \frac{\s(\xi)\s(\xj)\s(\xk)}
           {\sqrt{{\grammR}/{\k_2}}} =
      \frac{ \sqrt{{\grammR}/{\k_2}} }{  \sqrt{{\GrammR}/{\k_1}} }
\\
\displaystyle
\frac{\ss(\xI)}{\s (\xi)} =:\frac{1}{ \ctesS} =
     \frac{\ss(\xI)\ss(\xJ)\ss(\xK)}{\sqrt{{\GrammR}/{\k_1}}} =
      \frac{ \sqrt{{\GrammR}/{\k_1}} }{  \sqrt{{\grammR}/{\k_2}}
}  \\
\displaystyle
   \frac{\s (2 \xi)}{\si(\apI)\cc (\xI)} =: \ctesd =
    -2 \frac{\s(\xi)\s(\xj)\s(\xk)}{\ssa(2\symarea)} =
    -2 \frac{{\grammR}/{\k_2}}{  \sqrt{{\GrammR}/{\k_1}}  }
\frac{1}{\ssa^2(2\symarea) }  \\
\displaystyle
\frac{\ss (2 \xI)}{\si(\lpi)\c (\xi)} =: \cteSD =-2
    \frac{\ss(\xI)\ss(\xJ)\ss(\xK)}{\ssca(2\symcoarea)} =
    -2 \frac{{\GrammR}/{\k_1}}{  \sqrt{{\grammR}/{\k_2}}  }
\frac{1}{\ssca^2(2\symcoarea) }.  \\
\end{array}
\label{xxm}
\ee

The two quantities ${\sqrt{{\grammR}/{\k_2}}}$ and
${\sqrt{{\GrammR}/{\k_1}}}$ must be real in order the triangle to
exist. Therefore, any triangle must satisfy the inequalities:
\be {\grammR}/{\k_2} \geq0 \qquad {\GrammR}/{\k_1} \geq0
\mylabel{ht:GrammRIneq}
\ee which  apply to {\em any} member of the CKD family of `complex
Hermitian' spaces, notwithstanding any restriction for the sides
and angles. Brehm's inequalities under which a triangle with
prescribed values for sides and shape invariant exists in the
elliptic and hyperbolic complex hermitian spaces are simply the
transcription of the condition $\frac{\grammR}{\k_2}
\geq0$ to the complex CK spaces with $\icd=1; \k_1\neq 0, \k_2=1$.
Thus 
$\grammR \geq 0$, or, equivalently
$\gramm=\k_1^2 \grammR \geq 0$; by using (\ref{ht:GrammDef}) this
gives:
\be
\gramm={1-\c^2(\xi)-\c^2(\xj)-
\c^2(\xk)+2\c(\xi)\c(\xj)\c(\xk)\cos\om}
 \geq 0 \\
\ee which covers simultaneously the inequalities given by Brehm 
for the elliptic ($\k_1>0$) and hyperbolic ($\k_1<0$) hermitian
spaces (remark Brehm calls $\omega$ our $\om$). 
inequalities
fourth element
$\Om$

It is also worth highlighting the translation of the  inequalities
(\ref{ht:GrammRIneq}) by using (\ref{ht:cEuler}); this brings them
in terms of angular and lateral phases, and symplectic area and
coarea:
\be
 -\frac{\ssca(2\symcoarea)}{\si(\apI)\si(\apJ)\si(\apK)} \geq 0
\qquad
 -\frac{\ssa(2\symarea)}{\si(\lpi)\si(\lpj)\si(\lpk)} \geq 0.
\label{ht:GrammRInequalitiesTrans}
\ee

The same translation can be done in the expressions (\ref{xxm}),
thereby expressing the values $\ctesS, \ctesd, \cteSD$ in terms of
lateral and angular phases, symplectic area and coarea:
$$
\frac{\s(\xi)}{\ss (\xI)} =: \ctesS =
      \sqrt{\frac{\si(\lpi)\si(\lpj)\si(\lpk)\ssa(2\symarea)}
                 {\si(\apI)\si(\apJ)\si(\apK)\ssca(2\symcoarea)}}
$$
\be
\begin{array}{l}
\displaystyle
   \frac{\s (2 \xi)}{\si(\apI) \cc (\xI)} =: \ctesd =
        \sqrt{\frac{-4\si(\lpi)\si(\lpj)\si(\lpk)\ssa(2\symarea)}
{\si^2(\apI)\si^2(\apJ)\si^2(\apK)}}=2 \ctesS\sqrt{\frac{-  
\ssca(2\symcoarea)}{\si(\apI)\si(\apJ)\si(\apK)}}\\
\displaystyle
\frac{\ss (2 \xI)}{\si(\lpi) \c (\xi)} =: \cteSD =
        \sqrt{\frac{-
4\si(\apI)\si(\apJ)\si(\apK)\ssca(2\symcoarea)}
{\si^2(\lpi)\si^2(\lpj)\si^2(\lpk)}}=
\frac{2}{\ctesS}\sqrt{\frac{-
\ssa(2\symarea)}{\si(\lpi)\si(\lpj)\si(\lpk)}}. \\
\end{array}
\ee in whose form the existence inequalities
(\ref{ht:GrammRInequalitiesTrans}) are evident.

The inequalities (\ref{ht:GrammRIneq}) are analogous to the
existence conditions $\frac{E}{\k_1}\leq 0$,  $\frac{e}{\k_2}\leq
0$ for the half-excesses $E=\Delta/2, e=\delta/2$ appearing in the
trigonometry of real spaces. A way to derive such real
inequalities, alternative to the one used in \cite{SpaceTimeTrig},
is to introduce in the real case the determinants $\gramm,
\Gramm$  of the Gramm matrices built up from the real symmetric
scalar products of vectors corresponding to vertices or  to poles
of sides; these are given by (\ref{ht:GrammDef}) with
$\om=0,\ \Om=0$, and also vanish when $\k_1, \k_2 \to 0$. If we
introduce again $\grammR, \GrammR$ by  (\ref{ht:GrammRDef}),  in
absence of the factors
$\ci(\om),
\ci(\Om)$, the identity A.30 in the appendix of
\cite{SpaceTimeTrig} allow a factorization of $\grammR, \GrammR$,
translating the conditions  $\frac{\grammR}{\k_2} \geq0,
\frac{\GrammR}{\k_1} \geq0$ into inequalities for angular and
lateral excesses $\frac{E}{\k_1}\leq 0$,  $\frac{e}{\k_2}\leq 0$.
This last step cannot be done in the Hermitian case and the
inequalities stay in the form (\ref{ht:GrammRIneq}).


\subsection{The three special cases: collinear triangles,
concurrent triangles and purely real triangles}

Browsing through the equations we have given, we find several
pairs of equations which can be stated in two similar variant
forms, one involving sides (angles) and the other involving twice
the sides (angles): examples of such pairs are (\ref{ht:SRcos},
\ref{ht:SRcos2}) or (\ref{ht:CKsine}, \ref{ht:SRsin2}) and their
duals. This fact insinuates the existence of two special
non-generic types of triangles, for which the appropriate generic
equation reduces to a (known) simpler form.  

\subsubsection{Complex Collinear triangles}

The first special case corresponds to a triangle determined by
three `complex' collinear vertices, hence collapsing  from the
`complex'-2D CK hermitian space ${}_{\icd}SU_{\k_1,
\k_2}(3)/({}_{\icd}U(1)\otimes {}_{\icd}SU_{\k_2}(2))$ to a
`complex'-1D subspace, which can be identified with a space 
${}_{\icd}SU_{\k_1}(2)/{}_{\icd}U(1)$. Depending on whether
$\k_1>0, =0, <0$, this space is the elliptic, euclidean or
hyperbolic hermitian `complex' line. Sides are all different from
zero, but angles
$\xI, \xJ, \xK$ must be zero (or  straight), and thus satisfy 
$\ss(\xI)=\ss(\xJ)=\ss(\xK)=0$ (hence $\grammR=\GrammR=0$). For
these values the equations $1J$ reduce to $e^{i \Om}=1$, thus
$\Om=0$ and from  $0iI$ we get:
\be
\Om=0, \qquad \apI -\lpi = \om.
\ee The SR double cosine for sides (\ref{ht:SRcos2}), and double
sine for sides (\ref{ht:SRsin2}) become in this case:
\be
\c(2\xj)=\c(2\xi)\c(2\xk) -\k_1 \s(2\xi)\s(2\xk)\ci(\apJ)
\label{ht:CosRedComplexColl}
\ee
\be
\frac{\s(2\xj)}{\si(\apJ)}=\ctesd
\label{ht:SinRedComplexColl}
\ee so  {\em all} hermitian trigonometric equations reduce in this
case  to the trigonometry of a triangle with sides $2\xi, 2\xj,
2\xk$ and angles $\apI, \apJ, \apK$ in a auxiliar real CK space 
with labels $\k_1$ for sides and $\icd$ for angles, or
equivalently, for a triangle with sides  $\xi, \xj, \xk$ and
angles $\apI, \apJ, \apK$ in a real CK space  with labels $4\k_1$
for sides and $\icd$ for angles, for which
(\ref{ht:CosRedComplexColl}, \ref{ht:SinRedComplexColl}) are the
real cosine and sine theorems. The Lie algebra isomorphism
${}_{\icd}su_{\k_1}(2) \simeq so_{\k_1, \icd}(3)$ lies behind this.

By using (\ref{ht:ExcsOmeg}), and recalling $\Om=0$, the auxiliar
triangle angular excess $\apI+\apJ+\apK$ turns out to be equal to
$2\om$, and thus
$\om$ plays the role of the angular half-excess denoted
$E$ in the previous paper on real type trigonometry
\cite{SpaceTimeTrig}. It is worth remarking that the area
$\cal{A}$ of this auxiliar triangle is related to its angular
excess as
${\cal{A}}=\frac{2\om}{4\k_1}=\frac{\om}{2\k_1}$, thus coinciding 
with the original triangle symplectic area $\symarea$. The lateral
phases
$\lpi$ turn out to coincide with the three auxiliar angles denoted
$E_I$ in
\cite{SpaceTimeTrig}.  In terms of the symmetric invariants, the
`collinear' case corresponds to:
\be
\begin{array}{lllllll}
\Delta = 0  &\quad \Delta_\ap = 2\om   &\quad \om=\om  &\qquad 
\grammR=0 &\quad \symarea =
\symarea  &\quad  \ctesd =
\ctesd
\\
\delta =\delta   &\quad \delta_\lp = -\om  &\quad \Om=0  &\qquad 
\GrammR=0 &\quad
\symcoarea = 0  &\quad  \cteSD = 0.
\end{array}
\quad\ctesS=\infty
\ee

\subsubsection{Concurrent triangles}

The second special case, dual to the previous one, corresponds to a
triangle determined by three different concurrent geodesic sides.
Then sines of sides are equal to zero:
$\s(\xi)=
\s(\xj)=\s(\xk)=0$. Here $\om=0$, and the  the SR dual double
cosine equation (\ref{ht:SRdualcos2}) and SR dual double sine
equation (\ref{ht:CKSRdualsin2}) become the cosine and sine
theorems for a triangle with sides $\xI, \xJ, \xK$ and angles
$\lpi, \lpj,
\lpk$ in a real CK space  with labels $4\k_2$ for sides and $\icd$
for angles. For the angular excess of this auxiliar triangle we
have 
$\lpi+\lpj+\lpk=2\Om$, and thus $\Om$ plays the role of the angular
half-excess ($E$ in \cite{SpaceTimeTrig}).  In terms of the
symmetric invariants, this case correspond to:
\be
\begin{array}{lllllll}
\Delta = \Delta  &\quad \Delta_\ap =-\Om   &\quad \om= 0,
&\qquad   \grammR=0 &\quad
\symarea = 0  &\quad  \ctesd = 0 \\
\delta = 0  &\quad \delta_\lp = 2 \Om  &\quad \Om=\Om  &\qquad 
\GrammR=0 &\quad \symcoarea = \symcoarea  &\quad  \cteSD =
\cteSD
\end{array}
\quad\ctesS=0.
\ee

\subsubsection{Purely Real triangles}

The third special case corresponds to a triangle for which the
lateral and angular phase factors $e^{i\lpi}$ and
$e^{i\apI}$ are {\em real}, and sides and angles are different from
zero; this {\it purely real\/} triangle  is contained in a purely
real totally geodesic submanifold, isometric to $SO_{\k_1,
\k_2}(3)/ (O(1) \otimes SO_{\k_2}(2))$, and locally isometric (as
$O(1)
\equiv Z_2$) to 
$SO_{\k_1, \k_2}(3)/ SO_{\k_2}(2)$; for $\k_1=1, \k_2=1$ this is 
the real projective space $\Re P^2$. Sines of both sets of phases
vanish whenever the other does (see (\ref{ht:Eq3RIDual})); this
corresponds to  the self-dual nature of this case. Angular and
lateral phase excesses, 
$\om$ and $\Om$ and symplectic area and coarea have vanishing
sines. Each individual phase $\lpi$ or $\apI$ can thus have only
two values, either $0$ or twice a quadrant of label $\icd$, which
have opposite cosines $\pm 1$.  In terms of the symmetric
invariants, this case correspond to the values
\be
\begin{array}{lllllll}
\Delta = \Delta  &\quad \Delta_\ap =0   &\quad \om= 0 &\qquad  
\grammR=\grammR &\quad
\symarea = 0  &\quad  \ctesd =
\infty \\
\delta = \delta  &\quad \delta_\lp = 0  &\quad \Om=0  &\qquad
\GrammR=\GrammR &\quad
\symcoarea = 0  &\quad  \cteSD =
\infty
\end{array}
\quad\ctesS=\ctesS.
\ee

This reduction also provides an approach to the trigonometry of
{\em real} projective planes, requiring as triangle elements,
further to sides and angles, a set of {\em discrete} phases,
entering the equations only through their cosines
$\rphi=\ci(\lpi)=\pm 1;\ 
\rphI=\ci(\apI)=\pm 1$. Thus Hermitian trigonometry of the
`complex' spaces simultaneously afford, if we restrict phases to
these two possible discrete values, the trigonometry of the real
`projective' CK spaces family (to which $\Re P^2$ belongs). The
distinction between the trigonometry of the sphere and the real
projective plane  is well known (e.g.\ in Coxeter
\cite{CoxeterNEG}).

\section{Overview and Concluding remarks}

The most direct physical application of Hermitian trigonometry is
to the trigonometry of the Quantum space of states, which is the
elliptic member ($\k_1>0, \k_2>0$) of the family of complex
($\icd>0$) CKD Hermitian spaces; geometric phases appear directly
as trigonometric invariants from this point of view. This will be
discussed in the companion paper \cite{HT2}. 

There are also other possibly interesting applications of an
explicit knowledge of the trigonometry of this family of spaces. 
The real space-time models with zero or constant space-time
curvature (Minkowskian  and de Sitter space-times) are superseded
by a variable curvature pseudoRiemannian space-time; this is the
essence of the Einstenian interpretation of gravitation. The
possibility of a kind  of `Riemannian' Quantum space of states,
whose curvature might be not constant, cannot be precluded a
priori.  A good understanding of the geometry of the Hermitian
constant curvature cases might be helpful to explore and figure
out what consequences might follow from this idea and familiarity
with their trigonometry is a first order tool in this aim.

Another physical problem where the results we have obtained could
apply  lies on the use of pseudo-Hilbert spaces with an indefinite
Hermitian scalar product (Gupta-Bleuler type). These indefinite
Quantum spaces of states are those corresponding to 
$\k_2>0$; and its hermitian trigonometry  should provide the basic
elementary relations in the geometry of these spaces, just as  the
corresponding real relations are the basic space-time relations in
the De Sitter and Anti De Sitter space-times. 

The identification of the Quantum space of states as a member of
this complete CKD family of spaces makes it also natural to
inquire about whether or not the labels $\icd, \k_1, \k_2$ may
have any sensible physical meaning. Within the `kinematical'
($\k_2\leq0$) interpretation of the real CK spaces, $\k_1$ is the
curvature of space-time and
$\k_2=-1/c^2$ is related to the relativistic constant. A natural
query is: are the limits $\icd \to 0, \k_1\to 0$ (and $N\to
\infty$) somehow related to a `classical' limit $\hbar\to 0$ within
some sensible `quantum' interpretation of the `complex hermitian'
spaces? This is worth exploring. 

Real hyperbolic trigonometry, deeply involved in manifold
classification problems, knot theory, etc., is merely a particular
case of real CK trigonometry. It  is not unreasonable to assume
that (some instances at least) of the generic Hermitian
trigonometry may be at least as relevant in the similar
`complexified' problems \cite{ArnoldQBerry}. The intriguing
indications for an  essentially complex nature of space-time at
some deep level makes  also worthy the study of complex spaces in
a way as explicit and visual as possible.  

Aside the physical interest of particular results, another 
potential in the method proposed in \cite{SpaceTimeTrig} and
developed in the present paper lies on the possibility of opening
an avenue for  studying the trigonometry of other symmetric
homogeneous spaces, most of whose trigonometries are still
unknown.  Very few results are known in this area; a general sine
theorem is derived in Leuzinger
\cite{Leuzinger} for non-compact spaces, and the trigonometry of
the rank-two spaces $SU(3)$ and $SL(3,
\Ce)/SU(3)$ is discussed in \cite{Aslaksen1, Aslaksen2} heavily
relying  on the use of the Weyl theorems on invariant theory and
characterization of invariants by means of traces of products of
matrices.  

The trigonometry of the rank-one `quaternionic hyper-hermitian'
spaces  ($Sp(3)/(Sp(1)\otimes Sp(2))$, \ $Sp(2,1)/(Sp(1)\otimes
Sp(2)), \  Sp(2,1)/(Sp(1)\otimes Sp(1,1))$ or  $Sp(6,
\Re)/(SO(2,1)\otimes Sp(4, \Re))$) ---which correspond to further
Cayley-Dickson extensions with a new CD label
$\icd_2$---, and also of the `octonionic type' analogues of the
Cayley plane ---with another CD label
$\icd_3$ altogether---, reduces in some sense to the `complex'
two-dimensional case, since any triangle in these spaces lies on a
`complex' chain; thus in a sense the study of trigonometry in rank
one spaces is essentially complete with the spaces of real
quadratic and  `complex Hermitian' type. This reduction is not
natural however from a purely quaternionic or octonionic
viewpoint. Perhaps quaternionic (and also the exceptional 
octonionic) trigonometry should be understood better. In any case,
this kind of approach in a `$\Re, \Ce, \He$ spirit' fits into the
V.I. Arnold  idea of mathematical trinities; hopefully it may
provide a way to the  quest
\cite{ArnoldQBerry} for the quaternionic analogue of Berry's phase.

A next natural objective along this line is the study of
trigonometry of higher rank grassmannians, either real or complex.
This is still largely unknown (see however \cite{Hangan}).  Should
the method outlined in this paper be able to produce in a direct
form the equations of trigonometry for grassmannians  which are
also very relevant spaces in many physical aplications, this would
make  a further step towards a general approach to trigonometry of
any symmetric homogeneous space. This goal will  require first to
group all symmetric homogeneous spaces into CKD families, and then
to study trigonometry for each family. Work in progress on this
line
\cite{FreudGoslar96, MSBurgos, MSClasLieGroups} opens the
possibility of realizing {\em all} simple Lie algebras (even
$SL(N, \Re), SL(N,
\Ce)$, $SO^*(2n)$,
$SU^*(2n)$ and the exceptional ones) as `unitary' algebras, leaving
invariant an `hermitian' (relative to some antiinvolution)  form
over a tensor product of two pseudo-division algebras. This
realization should allow a test on whether or not some extension
of the ideas outlined here afford the equations of trigonometry
for any homogeneous space in an explicit and simple enough way.


\section*{Acknowledgments}

We want to acknowledge F.\ J.\ Herranz for his collaboration during
the initial stages of this work, as well as for his continued
interest and suggestions. This work was partially supported  by 
DGICYT, Spain  (Project PB98/0370).


\section*{Appendix}

Table Captions: 

\noindent Table 3 $(\icd>0)$.   The Table is arranged after the
values of the pair $\k_1, \k_2$, and the three labels 
$(\icd; \k_1, \k_2)$ are explicitly displayed at each entry.  The
group description $G/H$ of the homogeneous space is shown only
when $G$ has  a standard name; this is not the case for $(\k_1,
\k_2)=(0,0)$. The two spaces of points at the two corners in the 
last row $\k_2=-1$ are the same, but the corresponding geometries
differ by the interchange of first- and second-kind lines
generated by either $P_1$ or $P_2$. Notice the sign difference in
equations involving $a,A$ and involving $b,B; c,C$ and the
relevant  comments in the main text.

\medskip

\noindent Table 4 $(\icd=0)$.  The Table is arranged after the
values of the pair $\k_1,\k_2$, and the three labels  $(\icd;
\k_1, \k_2)$ are explicitly displayed at each entry.  The group
description $G/H$ of the homogeneous space is not shown as when
$\icd=0$ the CKD groups are not simple and have not a standard
name. The fiducial role of the trigonometry for the space $(\icd=0;
\k_1=0, \k_2=0)$ in the center of this Table is clear. All the
trigonometries in Tables 3, 4 and 5 are deformations of this
`purely linear' one.

\medskip

\noindent Table 5 $(\icd<0)$.  The Table is arranged after the
values  $\k_1, \k_2$, and the labels 
$(\icd; \k_1, \k_2)$ are explicitly displayed at each entry.  The
group description $G/H$ of the homogeneous space is shown only
when $G$ has  a standard name. The spaces at the four corners are
equal, but the  trigonometric equations in these geometries are
different as they correspond to  triangles with geodesics sides of
the four not  conjugate different possible types.

\begin{landscape}

\begin{table}[h] {\footnotesize { \noindent
\label{table:htCosineBasicTrigEqns1}
\ \vskip-2.5truecm \caption{Complex Hermitian cosine theorems and
their duals for the nine {\em complex Hermitian} CKD spaces 
($\icd=1$). }
\bigskip
\noindent\hskip 5pt
\begin{tabular}{ccc}
\hline &&\\[-8pt] Complex Hermitian Elliptic\quad $(+1; +1,+1)$&
Complex Hermitian Euclidean\quad $(+1; 0,+1)$& Complex Hermitian
Hyperbolic\quad $(+1; -1,+1)$\\
$SU(3)/U(1){\otimes}SU(2)$&$IU(2)/U(1){\otimes}SU(2)$
    &$SU(2,1)/U(1){\otimes}SU(2)$\\[3pt]
$ \cos a\cos\om=\cos b\cos c- \sin b\sin c\cos A\cos\Psi_A$&
$a^2=b^2+c^2+ 2  b c\cos A\cos\Psi_A$&
$\cosh a\cos\om=\cosh b\cosh c+ \sinh b\sinh c\cos A\cos\Psi_A$\\
$\cos b\cos\om=\cos a\cos c+ \sin a\sin c\cos B\cos\Psi_B$&
$b^2=a^2+c^2- 2  a c\cos B\cos\Psi_B$&
$\cosh b\cos\om=\cosh a\cosh c - \sinh a\sinh c\cos B\cos\Psi_B$\\
$\cos c\cos\om=\cos a\cos b+ \sin a\sin b\cos C\cos\Psi_C$&
$c^2=a^2+b^2- 2  a b\cos C\cos\Psi_C$&
$\cosh c\cos\om=\cosh a\cosh b - \sinh a\sinh b\cos
C\cos\Psi_C$\\[1pt]
$\cos a\sin2\symarea=\sin b\sin c \cos A \sin\Psi_A$&
$2\symarea= b c \cos A \sin\Psi_A$&
$\cosh a\sin2\symarea=\sinh b\sinh c \cos A \sin\Psi_A$\\
$\cos b\sin2\symarea=\sin c\sin a \cos B \sin\Psi_B$&
$2\symarea= c a \cos B \sin\Psi_B$&
$\cosh b\sin2\symarea=\sinh c\sinh a \cos B \sin\Psi_B$\\
$\cos c\sin2\symarea=\sin a\sin b \cos C \sin\Psi_C$&
$2\symarea= a b  \cos C \sin\Psi_C$&
$\cosh c\sin2\symarea=\sinh a\sinh b \cos C \sin\Psi_C$\\[1pt]
$\cos A\sin2\symcoarea = \sin B \sin C \cos a \sin\psi_a$&
$\cos A\sin2\symcoarea = \sin B \sin C \sin\psi_a$&
$\cos A\sin2\symcoarea = \sin B \sin C \cosh a \sin\psi_a$\\
$\cos B\sin2\symcoarea = \sin C \sin A \cos b \sin\psi_b$&
$\cos B\sin2\symcoarea = \sin C \sin A \sin\psi_b$&
$\cos B\sin2\symcoarea = \sin C \sin A \cosh b \sin\psi_b$\\
$\cos C\sin2\symcoarea = \sin A \sin B \cos c \sin\psi_c$&
$\cos C\sin2\symcoarea = \sin A \sin B \sin\psi_c$&
$\cos C\sin2\symcoarea = \sin A \sin B \cosh c \sin\psi_c$\\[1pt]
$\cos A\cos\Om=\cos B\cos C- \sin B\sin C\cos a\cos\psi_a$&
$\cos A\cos\Om=\cos B\cos C- \sin B\sin C\cos\psi_a$&
$\cos A\cos\Om=\cos B\cos C- \sin B\sin C\cosh a\cos\psi_a$\\
$\cos B\cos\Om=\cos A\cos C+ \sin A\sin C\cos b\cos\psi_b$&
$\cos B\cos\Om=\cos A\cos C+ \sin A\sin C\cos\psi_b$&
$\cos B\cos\Om=\cos A\cos C + \sin A\sin C\cosh b\cos\psi_b$\\
$\cos C\cos\Om=\cos A\cos B+ \sin A\sin B\cos c\cos\psi_c$&
$\cos C\cos\Om=\cos A\cos B+ \sin A\sin B\cos\psi_c$&
$\cos C\cos\Om=\cos A\cos B + \sin A\sin B\cosh c\cos\psi_c$\\[3pt]
\hline &&\\[-8pt] Complex Hermitian Co-Euclidean\quad $(+1; +1,0)$&
Complex Hermitian Galilean\quad $(+1; 0,0)$& Complex Hermitian
Co-Minkowskian\quad $(+1; -1,0)$\\ Complex Hermitian Oscillating
Newton-Hooke\quad & 
\ & Complex Hermitian Expanding Newton-Hooke \\
$IU(2)/U(1){\otimes}IU(1)$&$\null$&$IU(1,1)/U(1){\otimes}IU(1)$\\[3pt]
$ \cos a\cos\om=\cos b\cos c- \sin b\sin c\cos\Psi_A$&
$a^2=b^2+c^2+ 2  b c\cos\Psi_A$&
$\cosh a\cos\om=\cosh b\cosh c+ \sinh b\sinh c\cos\Psi_A$\\
$\cos b\cos\om=\cos a\cos c+ \sin a\sin c\cos\Psi_B$&
$b^2=a^2+c^2- 2  a c\cos\Psi_B$&
$\cosh b\cos\om=\cosh a\cosh c - \sinh a\sinh c\cos\Psi_B$\\
$\cos c\cos\om=\cos a\cos b+ \sin a\sin b\cos\Psi_C$&
$c^2=a^2+b^2- 2  a b\cos\Psi_C$&
$\cosh c\cos\om=\cosh a\cosh b - \sinh a\sinh bcos\Psi_C$\\[1pt]
$\cos a\sin2\symarea=\sin b\sin c \sin\Psi_A$&
$2\symarea= b c  \sin\Psi_A$&
$\cosh a\sin2\symarea=\sinh b\sinh c  \sin\Psi_A$\\
$\cos b\sin2\symarea=\sin c\sin a  \sin\Psi_B$&
$2\symarea= c a  \sin\Psi_B$&
$\cosh b\sin2\symarea=\sinh c\sinh a  \sin\Psi_B$\\
$\cos c\sin2\symarea=\sin a\sin b  \sin\Psi_C$&
$2\symarea= a b   \sin\Psi_C$&
$\cosh c\sin2\symarea=\sinh a\sinh b  \sin\Psi_C$\\[1pt]
$2\symcoarea =  B  C \cos a \sin\psi_a$&
$2\symcoarea =  B  C \sin\psi_a$&
$2\symcoarea =  B  C \cosh a \sin\psi_a$\\
$2\symcoarea =  C  A \cos b \sin\psi_b$&
$2\symcoarea =  C  A \sin\psi_b$&
$2\symcoarea =  C  A \cosh b \sin\psi_b$\\
$2\symcoarea =  A  B \cos c \sin\psi_c$&
$2\symcoarea =  A  B \sin\psi_c$&
$2\symcoarea =  A  B \cosh c \sin\psi_c$\\[1pt]
$A^2=B^2+C^2+ 2 B C\cos a\cos\psi_a$&
$A^2=B^2+C^2+ 2 B C\cos\psi_a$&
$A^2=B^2+C^2+ 2 B C\cosh a\cos\psi_a$\\
$B^2=A^2+C^2- 2 A C\cos b\cos\psi_b$&
$B^2=A^2+C^2- 2 A C\cos\psi_b$&
$B^2=A^2+C^2 - 2 A C\cosh b\cos\psi_b$\\
$C^2=A^2+B^2- 2 A B \cos c\cos\psi_c$&
$C^2=A^2+B^2- 2 A B \cos\psi_c$&
$C^2=A^2+B^2- 2 A B \cosh c\cos\psi_c$\\[3pt]
\hline &&\\[-8pt] Complex Hermitian Co-Hyperbolic\quad $(+1;
+1,-1)$& Complex Hermitian Minkowskian\quad $(+1; 0,-1)$& Complex
Hermitian Doubly Hyperbolic\quad $(+1; -1,-1)$\\ Complex Hermitian
Anti-de Sitter&
\ &  Complex Hermitian De Sitter\\
$SU(2,1)/U(1){\otimes}SU(1,1)$&$IU(1,1)/U(1){\otimes}SU(1,1)$&
$SU(2,1)/U(1){\otimes}SU(1,1)$\\[3pt]
$ \cos a\cos\om=\cos b\cos c- \sin b\sin c\cosh A\cos\Psi_A$&
$a^2=b^2+c^2+ 2  b c\cosh A\cos\Psi_A$&
$\cosh a\cos\om=\cosh b\cosh c+ \sinh b\sinh c\cosh A\cos\Psi_A$\\
$\cos b\cos\om=\cos a\cos c+ \sin a\sin c\cosh B\cos\Psi_B$&
$b^2=a^2+c^2- 2  a c\cosh B\cos\Psi_B$&
$\cosh b\cos\om=\cosh a\cosh c - \sinh a\sinh c\cosh B\cos\Psi_B$\\
$\cos c\cos\om=\cos a\cos b+ \sin a\sin b\cosh C\cos\Psi_C$&
$c^2=a^2+b^2- 2  a b\cosh C\cos\Psi_C$&
$\cosh c\cos\om=\cosh a\cosh b - \sinh a\sinh b\cosh
C\cos\Psi_C$\\[1pt]
$\cos a\sin2\symarea=\sin b\sin c \cosh A \sin\Psi_A$&
$2\symarea= b c \cosh A \sin\Psi_A$&
$\cosh a\sin2\symarea=\sinh b\sinh c \cosh A \sin\Psi_A$\\
$\cos b\sin2\symarea=\sin c\sin a \cosh B \sin\Psi_B$&
$2\symarea= c a \cosh B \sin\Psi_B$&
$\cosh b\sin2\symarea=\sinh c\sinh a \cosh B \sin\Psi_B$\\
$\cos c\sin2\symarea=\sin a\sin b \cosh C \sin\Psi_C$&
$2\symarea= a b  \cosh C \sin\Psi_C$&
$\cosh c\sin2\symarea=\sinh a\sinh b \cosh C \sin\Psi_C$\\[1pt]
$\cosh A\sin2\symcoarea = \sinh B \sinh C \cos a \sin\psi_a$&
$\cosh A\sin2\symcoarea = \sinh B \sinh C \sin\psi_a$&
$\cosh A\sin2\symcoarea = \sinh B \sinh C \cosh a \sin\psi_a$\\
$\cosh B\sin2\symcoarea = \sinh C \sinh A \cos b \sin\psi_b$&
$\cosh B\sin2\symcoarea = \sinh C \sinh A \sin\psi_b$&
$\cosh B\sin2\symcoarea = \sinh C \sinh A \cosh b \sin\psi_b$\\
$\cosh C\sin2\symcoarea = \sinh A \sinh B \cos c \sin\psi_c$&
$\cosh C\sin2\symcoarea = \sinh A \sinh B \sin\psi_c$&
$\cosh C\sin2\symcoarea = \sinh A \sinh B \cosh c
\sin\psi_c$\\[1pt]
$\cosh A\cos\Om=\cosh B\cosh C+ \sinh B\sinh C\cos a\cos\psi_a$&
$\cosh A\cos\Om=\cosh B\cosh C+ \sinh B\sinh C\cos\psi_a$&
$\cosh A\cos\Om=\cosh B\cosh C+ \sinh B\sinh C\cosh a\cos\psi_a$\\
$\cosh B\cos\Om=\cosh A\cosh C- \sinh A\sinh C\cos b\cos\psi_b$&
$\cosh B\cos\Om=\cosh A\cosh C- \sinh A\sinh C\cos\psi_b$&
$\cosh B\cos\Om=\cosh A\cosh C- \sinh A\sinh C\cosh b\cos\psi_b$\\
$\cosh C\cos\Om=\cosh A\cosh B- \sinh A\sinh B\cos c\cos\psi_c$&
$\cosh C\cos\Om=\cosh A\cosh B- \sinh A\sinh B\cos\psi_c$&
$\cosh C\cos\Om=\cosh A\cosh B- \sinh A\sinh B\cosh
c\cos\psi_c$\\[3pt]
\hline
\end{tabular} }}
\end{table}
\end{landscape}

\begin{landscape}
\begin{table}[h] {\footnotesize { \noindent
\ \vskip-2.5truecm \caption{`Complex Hermitian' cosine theorems
and their duals for the nine {\em parabolic complex (dual)
`Hermitian'} CKD spaces  ($\icd=0$).  }
\label{table:htCosineBasicTrigEqns2}
\bigskip
\noindent\hskip -30pt
\begin{tabular}{ccc}
\hline &&\\[-8pt] `Parabolic Complex Hermitian' Elliptic\quad $(0;
+1,+1)$& `Parabolic Complex Hermitian' Euclidean\quad $(0; 0,+1)$&
`Parabolic Complex Hermitian' Hyperbolic\quad $(0; -1,+1)$\\
$\null$&$\null$&$\null$\\[3pt]
$ \cos a=\cos b\cos c- \sin b\sin c\cos A$&
$a^2=b^2+c^2+ 2  b c\cos A$& 
$\cosh a =\cosh b\cosh c+ \sinh b\sinh c\cos A$\\
$\cos b =\cos a\cos c+ \sin a\sin c\cos B$&
$b^2=a^2+c^2- 2  a c\cos B$&
$\cosh b =\cosh a\cosh c - \sinh a\sinh c\cos B$\\
$\cos c =\cos a\cos b+ \sin a\sin b\cos C$&
$c^2=a^2+b^2- 2  a b\cos C$&
$\cosh c =\cosh a\cosh b - \sinh a\sinh b\cos C$\\[1pt]
$\cos a\,\,2\symarea=\sin b\sin c \cos A \,\,\Psi_A$&
$2\symarea= b c \cos A \,\,\Psi_A$&
$\cosh a\,\,2\symarea=\sinh b\sinh c \cos A \,\,\Psi_A$\\
$\cos b\,\,2\symarea=\sin c\sin a \cos B \,\,\Psi_B$&
$2\symarea= c a \cos B \,\,\Psi_B$&
$\cosh b\,\,2\symarea=\sinh c\sinh a \cos B \,\,\Psi_B$\\
$\cos c\,\,2\symarea=\sin a\sin b \cos C \,\,\Psi_C$&
$2\symarea= a b  \cos C \,\,\Psi_C$&
$\cosh c\,\,2\symarea=\sinh a\sinh b \cos C \,\,\Psi_C$\\[1pt]
$\cos A\,\,2\symcoarea = \sin B \sin C \cos a \,\,\psi_a$&
$\cos A\,\,2\symcoarea = \sin B \sin C \,\,\psi_a$&
$\cos A\,\,2\symcoarea = \sin B \sin C \cosh a \,\,\psi_a$\\
$\cos B\,\,2\symcoarea = \sin C \sin A \cos b \,\,\psi_b$&
$\cos B\,\,2\symcoarea = \sin C \sin A \,\,\psi_b$&
$\cos B\,\,2\symcoarea = \sin C \sin A \cosh b \,\,\psi_b$\\
$\cos C\,\,2\symcoarea = \sin A \sin B \cos c \,\,\psi_c$&
$\cos C\,\,2\symcoarea = \sin A \sin B \,\,\psi_c$&
$\cos C\,\,2\symcoarea = \sin A \sin B \cosh c \,\,\psi_c$\\[1pt]
$\cos A =\cos B\cos C- \sin B\sin C\cos a$&
$A = B + C $&
$\cos A =\cos B\cos C- \sin B\sin C\cosh a$\\
$\cos B =\cos A\cos C+ \sin A\sin C\cos b$&
$B = A - C$&
$\cos B =\cos A\cos C + \sin A\sin C\cosh b$\\
$\cos C =\cos A\cos B+ \sin A\sin B\cos c$&
$C= A - B$&
$\cos C =\cos A\cos B + \sin A\sin B\cosh c$\\[3pt]
\hline &&\\[-8pt] `Parabolic Complex Hermitian' Co-Euclidean\quad
$(0; +1,0)$& `Parabolic Complex Hermitian' Galilean\quad $(0;
0,0)$& `Parabolic Complex Hermitian' Co-Minkowskian\quad $(0;
-1,0)$\\ `Parabolic Complex Hermitian' Oscillating
Newton-Hooke\quad & 
\ & `Parabolic Complex Hermitian' Expanding Newton-Hooke \\
$\null$&$\null$&$\null$\\[3pt]
$a = b + c$&
$a = b + c$&
$a = b + c$\\
$b = a - c$&
$b = a - c$&
$b = a - c$\\
$c = a - b$&
$c = a - b$&
$c = a - b$\\[1pt]
$\cos a\,\,2\symarea=\sin b\sin c \,\,\Psi_A$&
$2\symarea= b c  \,\,\Psi_A$&
$\cosh a\,\,2\symarea=\sinh b\sinh c \,\, \Psi_A$\\
$\cos b\,\,2\symarea=\sin c\sin a \,\, \Psi_B$&
$2\symarea= c a  \,\,\Psi_B$&
$\cosh b\,\,2\symarea=\sinh c\sinh a \,\, \Psi_B$\\
$\cos c\,\,2\symarea=\sin a\sin b \,\, \Psi_C$&
$2\symarea= a b  \,\, \Psi_C$&
$\cosh c\,\,2\symarea=\sinh a\sinh b \,\, \Psi_C$\\[1pt]
$2\symcoarea =  B  C \cos a \,\,\psi_a$&
$2\symcoarea =  B  C \,\,\psi_a$&
$2\symcoarea =  B  C \cosh a \,\,\psi_a$\\
$2\symcoarea =  C  A \cos b \,\,\psi_b$&
$2\symcoarea =  C  A \,\,\psi_b$&
$2\symcoarea =  C  A \cosh b \,\,\psi_b$\\
$2\symcoarea =  A  B \cos c \,\,\psi_c$&
$2\symcoarea =  A  B \,\,\psi_c$&
$2\symcoarea =  A  B \cosh c \,\,\psi_c$\\[1pt]
$A^2=B^2+C^2+ 2 B C\cos a$&
$A = B + C$&
$A^2=B^2+C^2+ 2 B C\cosh a$\\
$B^2=A^2+C^2- 2 A C\cos b$&
$B = A - C$&
$B^2=A^2+C^2 - 2 A C\cosh b$\\
$C^2=A^2+B^2- 2 A B \cos c$&
$C = A - B$&
$C^2=A^2+B^2- 2 A B \cosh c$\\[3pt]
\hline &&\\[-8pt] `Parabolic Complex Hermitian' Co-Hyperbolic\quad
$(0; +1,-1)$& `Parabolic Complex Hermitian' Minkowskian\quad $(0;
0,-1)$& `Parabolic Complex Hermitian' Doubly Hyperbolic\quad $(0;
-1,-1)$\\ `Parabolic Complex Hermitian' Anti-de Sitter&
\ &  `Parabolic Complex Hermitian' De Sitter\\
$\null$&$\null$&$\null$\\[3pt]
$ \cos a =\cos b\cos c- \sin b\sin c\cosh A$&
$a^2=b^2+c^2+ 2  b c\cosh A$&
$\cosh a =\cosh b\cosh c+ \sinh b\sinh c\cosh A$\\
$\cos b =\cos a\cos c+ \sin a\sin c\cosh B$&
$b^2=a^2+c^2- 2  a c\cosh B$&
$\cosh b =\cosh a\cosh c - \sinh a\sinh c\cosh B$\\
$\cos c =\cos a\cos b+ \sin a\sin b\cosh C$&
$c^2=a^2+b^2- 2  a b\cosh C$&
$\cosh c =\cosh a\cosh b - \sinh a\sinh b\cosh C$\\[1pt]
$\cos a\,\,2\symarea=\sin b\sin c \cosh A \,\,\Psi_A$&
$2\symarea= b c \cosh A \,\,\Psi_A$&
$\cosh a\,\,2\symarea=\sinh b\sinh c \cosh A \,\,\Psi_A$\\
$\cos b\,\,2\symarea=\sin c\sin a \cosh B \,\,\Psi_B$&
$2\symarea= c a \cosh B \,\,\Psi_B$&
$\cosh b\,\,2\symarea=\sinh c\sinh a \cosh B \,\,\Psi_B$\\
$\cos c\,\,2\symarea=\sin a\sin b \cosh C \,\,\Psi_C$&
$2\symarea= a b  \cosh C \,\,\Psi_C$&
$\cosh c\,\,2\symarea=\sinh a\sinh b \cosh C\,\,\Psi_C$\\[1pt]
$\cosh A\,\,2\symcoarea = \sinh B \sinh C \cos a \,\,\psi_a$&
$\cosh A\,\,2\symcoarea = \sinh B \sinh C \,\,\psi_a$&
$\cosh A\,\,2\symcoarea = \sinh B \sinh C \cosh a \,\,\psi_a$\\
$\cosh B\,\,2\symcoarea = \sinh C \sinh A \cos b \,\,\psi_b$&
$\cosh B\,\,2\symcoarea = \sinh C \sinh A \,\,\psi_b$&
$\cosh B\,\,2\symcoarea = \sinh C \sinh A \cosh b \,\,\psi_b$\\
$\cosh C\,\,2\symcoarea = \sinh A \sinh B \cos c \,\,\psi_c$&
$\cosh C\,\,2\symcoarea = \sinh A \sinh B \,\,\psi_c$&
$\cosh C\,\,2\symcoarea = \sinh A \sinh B \cosh c
\,\,\psi_c$\\[1pt]
$\cosh A =\cosh B\cosh C+ \sinh B\sinh C\cos a$&
$A = B + C$&
$\cosh A =\cosh B\cosh C+ \sinh B\sinh C\cosh a$\\
$\cosh B =\cosh A\cosh C- \sinh A\sinh C\cos b$&
$B = A - C$&
$\cosh B =\cosh A\cosh C- \sinh A\sinh C\cosh b$\\
$\cosh C =\cosh A\cosh B- \sinh A\sinh B\cos c$&
$C = A - B$&
$\cosh C =\cosh A\cosh B- \sinh A\sinh B\cosh c$\\[3pt]
\hline
\end{tabular} }}
\end{table}
\end{landscape}

\begin{landscape}
\begin{table}[h] {\footnotesize { \noindent
\ \vskip-2.5truecm \caption{`Complex Hermitian' cosine theorems
and their duals for the nine {\em split complex `Hermitian'} CKD
spaces  ($\icd=-1$).  }
\label{table:htCosineBasicTrigEqns3}
\bigskip
\noindent\hskip -10pt
\begin{tabular}{ccc}
\hline &&\\[-8pt] `Split Complex Hermitian' Elliptic\quad $(-1;
+1,+1)$& `Split Complex Hermitian' Euclidean\quad $(-1; 0,+1)$&
`Split Complex Hermitian' Hyperbolic\quad $(-1; -1,+1)$\\
$SL(3,\Re)/SO(1,1){\otimes}SL(2,\Re)$
    &$IGL(2,\Re)/SO(1,1){\otimes}SL(2,\Re)$
    &$SL(3,\Re)/SO(1,1){\otimes}SL(2,\Re)$\\[3pt]
$ \cos a\cosh\om=\cos b\cos c- \sin b\sin c\cos A\cosh\Psi_A$&
$a^2=b^2+c^2+ 2  b c\cos A\cosh\Psi_A$&
$\cosh a\cosh\om=\cosh b\cosh c+ \sinh b\sinh c\cos A\cosh\Psi_A$\\
$\cos b\cosh\om=\cos a\cos c+ \sin a\sin c\cos B\cosh\Psi_B$&
$b^2=a^2+c^2- 2  a c\cos B\cosh\Psi_B$&
$\cosh b\cosh\om=\cosh a\cosh c - \sinh a\sinh c\cos
B\cosh\Psi_B$\\
$\cos c\cosh\om=\cos a\cos b+ \sin a\sin b\cos C\cosh\Psi_C$&
$c^2=a^2+b^2- 2  a b\cos C\cosh\Psi_C$&
$\cosh c\cosh\om=\cosh a\cosh b - \sinh a\sinh b\cos
C\cos\Psi_C$\\[1pt]
$\cos a\sinh2\symarea=\sin b\sin c \cos A \sinh\Psi_A$&
$2\symarea= b c \cos A \sinh\Psi_A$&
$\cosh a\sinh2\symarea=\sinh b\sinh c \cos A \sinh\Psi_A$\\
$\cos b\sinh2\symarea=\sin c\sin a \cos B \sinh\Psi_B$&
$2\symarea= c a \cos B \sinh\Psi_B$&
$\cosh b\sinh2\symarea=\sinh c\sinh a \cos B \sinh\Psi_B$\\
$\cos c\sinh2\symarea=\sin a\sin b \cos C \sinh\Psi_C$&
$2\symarea= a b  \cos C \sinh\Psi_C$&
$\cosh c\sinh2\symarea=\sinh a\sinh b \cos C \sinh\Psi_C$\\[1pt]
$\cos A\sinh2\symcoarea = \sin B \sin C \cos a \sinh\psi_a$&
$\cos A\sinh2\symcoarea = \sin B \sin C \sinh\psi_a$&
$\cos A\sinh2\symcoarea = \sin B \sin C \cosh a \sinh\psi_a$\\
$\cos B\sinh2\symcoarea = \sin C \sin A \cos b \sinh\psi_b$&
$\cos B\sinh2\symcoarea = \sin C \sin A \sinh\psi_b$&
$\cos B\sinh2\symcoarea = \sin C \sin A \cosh b \sinh\psi_b$\\
$\cos C\sinh2\symcoarea = \sin A \sin B \cos c \sinh\psi_c$&
$\cos C\sinh2\symcoarea = \sin A \sin B \sinh\psi_c$&
$\cos C\sinh2\symcoarea = \sin A \sin B \cosh c \sinh\psi_c$\\[1pt]
$\cos A\cos\Om=\cos B\cos C- \sin B\sin C\cos a\cosh\psi_a$&
$\cos A\cos\Om=\cos B\cos C- \sin B\sin C\cosh\psi_a$&
$\cos A\cos\Om=\cos B\cos C- \sin B\sin C\cosh a\cosh\psi_a$\\
$\cos B\cos\Om=\cos A\cos C+ \sin A\sin C\cos b\cosh\psi_b$&
$\cos B\cos\Om=\cos A\cos C+ \sin A\sin C\cosh\psi_b$&
$\cos B\cos\Om=\cos A\cos C + \sin A\sin C\cosh b\cosh\psi_b$\\
$\cos C\cos\Om=\cos A\cos B+ \sin A\sin B\cos c\cosh\psi_c$&
$\cos C\cos\Om=\cos A\cos B+ \sin A\sin B\cosh\psi_c$&
$\cos C\cos\Om=\cos A\cos B + \sin A\sin B\cosh
c\cosh\psi_c$\\[3pt]
\hline &&\\[-8pt] `Split Complex Hermitian' Co-Euclidean\quad
$(-1; +1,0)$& `Split Complex Hermitian' Galilean\quad $(-1; 0,0)$&
`Split Complex Hermitian' Co-Minkowskian\quad $(-1; -1,0)$\\
`Split Complex Hermitian' Oscillating NH\quad & 
\ & `Split Complex Hermitian' Expanding NH \\
$IGL(2,\Re)/SO(1,1){\otimes}IGL(1,\Re)$
    &$\null$&$IGL(2,\Re)/SO(1,1){\otimes}IGL(1,\Re)$\\[3pt]
$ \cos a\cosh\om=\cos b\cos c- \sin b\sin c\cosh\Psi_A$&
$a^2=b^2+c^2+ 2  b c\cosh\Psi_A$&
$\cosh a\cosh\om=\cosh b\cosh c+ \sinh b\sinh c\cosh\Psi_A$\\
$\cos b\cosh\om=\cos a\cos c+ \sin a\sin c\cosh\Psi_B$&
$b^2=a^2+c^2- 2  a c\cosh\Psi_B$&
$\cosh b\cosh\om=\cosh a\cosh c - \sinh a\sinh c\cosh\Psi_B$\\
$\cos c\cosh\om=\cos a\cos b+ \sin a\sin b\cosh\Psi_C$&
$c^2=a^2+b^2- 2  a b\cosh\Psi_C$&
$\cosh c\cosh\om=\cosh a\cosh b - \sinh a\sinh bcos\Psi_C$\\[1pt]
$\cos a\sinh2\symarea=\sin b\sin c \sinh\Psi_A$&
$2\symarea= b c  \sinh\Psi_A$&
$\cosh a\sinh2\symarea=\sinh b\sinh c  \sinh\Psi_A$\\
$\cos b\sinh2\symarea=\sin c\sin a  \sinh\Psi_B$&
$2\symarea= c a  \sinh\Psi_B$&
$\cosh b\sinh2\symarea=\sinh c\sinh a  \sinh\Psi_B$\\
$\cos c\sinh2\symarea=\sin a\sin b  \sinh\Psi_C$&
$2\symarea= a b   \sinh\Psi_C$&
$\cosh c\sinh2\symarea=\sinh a\sinh b  \sinh\Psi_C$\\[1pt]
$2\symcoarea =  B  C \cos a \sinh\psi_a$&
$2\symcoarea =  B  C \sinh\psi_a$&
$2\symcoarea =  B  C \cosh a \sinh\psi_a$\\
$2\symcoarea =  C  A \cos b \sinh\psi_b$&
$2\symcoarea =  C  A \sinh\psi_b$&
$2\symcoarea =  C  A \cosh b \sinh\psi_b$\\
$2\symcoarea =  A  B \cos c \sinh\psi_c$&
$2\symcoarea =  A  B \sinh\psi_c$&
$2\symcoarea =  A  B \cosh c \sinh\psi_c$\\[1pt]
$A^2=B^2+C^2+ 2 B C\cos a\cosh\psi_a$&
$A^2=B^2+C^2+ 2 B C\cosh\psi_a$&
$A^2=B^2+C^2+ 2 B C\cosh a\cosh\psi_a$\\
$B^2=A^2+C^2- 2 A C\cos b\cosh\psi_b$&
$B^2=A^2+C^2- 2 A C\cosh\psi_b$&
$B^2=A^2+C^2 - 2 A C\cosh b\cosh\psi_b$\\
$C^2=A^2+B^2- 2 A B \cos c\cosh\psi_c$&
$C^2=A^2+B^2- 2 A B \cosh\psi_c$&
$C^2=A^2+B^2- 2 A B \cosh c\cosh\psi_c$\\[3pt]
\hline &&\\[-8pt] `Split Complex Hermitian' Co-Hyperbolic\quad
$(-1; +1,-1)$& `Split Complex Hermitian' Minkowskian\quad $(-1;
0,-1)$& `Split Complex Hermitian' Doubly Hyperbolic\quad $(-1;
-1,-1)$\\ `Split Complex Hermitian' Anti-de Sitter&
\ &  `Split Complex Hermitian' De Sitter\\
$SL(3,\Re)/SO(1,1){\otimes}SL(2,\Re)$
     &$IGL(2,\Re)/SO(1,1){\otimes}SL(2,\Re)$&
$SL(3,\Re)/SO(1,1){\otimes}SL(2,\Re)$\\[3pt]
$ \cos a\cosh\om=\cos b\cos c- \sin b\sin c\cosh A\cosh\Psi_A$&
$a^2=b^2+c^2+ 2  b c\cosh A\cosh\Psi_A$&
$\cosh a\cosh\om=\cosh b\cosh c+ \sinh b\sinh c\cosh
A\cosh\Psi_A$\\
$\cos b\cosh\om=\cos a\cos c+ \sin a\sin c\cosh B\cosh\Psi_B$&
$b^2=a^2+c^2- 2  a c\cosh B\cosh\Psi_B$&
$\cosh b\cosh\om=\cosh a\cosh c - \sinh a\sinh c\cosh
B\cosh\Psi_B$\\
$\cos c\cosh\om=\cos a\cos b+ \sin a\sin b\cosh C\cosh\Psi_C$&
$c^2=a^2+b^2- 2  a b\cosh C\cosh\Psi_C$&
$\cosh c\cosh\om=\cosh a\cosh b - \sinh a\sinh b\cosh
C\cos\Psi_C$\\[1pt]
$\cos a\sinh2\symarea=\sin b\sin c \cosh A \sinh\Psi_A$&
$2\symarea= b c \cosh A \sinh\Psi_A$&
$\cosh a\sinh2\symarea=\sinh b\sinh c \cosh A \sinh\Psi_A$\\
$\cos b\sinh2\symarea=\sin c\sin a \cosh B \sinh\Psi_B$&
$2\symarea= c a \cosh B \sinh\Psi_B$&
$\cosh b\sinh2\symarea=\sinh c\sinh a \cosh B \sinh\Psi_B$\\
$\cos c\sinh2\symarea=\sin a\sin b \cosh C \sinh\Psi_C$&
$2\symarea= a b  \cosh C \sinh\Psi_C$&
$\cosh c\sinh2\symarea=\sinh a\sinh b \cosh C \sinh\Psi_C$\\[1pt]
$\cosh A\sinh2\symcoarea = \sinh B \sinh C \cos a \sinh\psi_a$&
$\cosh A\sinh2\symcoarea = \sinh B \sinh C \sinh\psi_a$&
$\cosh A\sinh2\symcoarea = \sinh B \sinh C \cosh a \sinh\psi_a$\\
$\cosh B\sinh2\symcoarea = \sinh C \sinh A \cos b \sinh\psi_b$&
$\cosh B\sinh2\symcoarea = \sinh C \sinh A \sinh\psi_b$&
$\cosh B\sinh2\symcoarea = \sinh C \sinh A \cosh b \sinh\psi_b$\\
$\cosh C\sinh2\symcoarea = \sinh A \sinh B \cos c \sinh\psi_c$&
$\cosh C\sinh2\symcoarea = \sinh A \sinh B \sinh\psi_c$&
$\cosh C\sinh2\symcoarea = \sinh A \sinh B \cosh c
\sinh\psi_c$\\[1pt]
$\cosh A\cos\Om=\cosh B\cosh C+ \sinh B\sinh C\cos a\cosh\psi_a$&
$\cosh A\cos\Om=\cosh B\cosh C+ \sinh B\sinh C\cosh\psi_a$&
$\cosh A\cos\Om=\cosh B\cosh C+ \sinh B\sinh C\cosh a\cosh\psi_a$\\
$\cosh B\cos\Om=\cosh A\cosh C- \sinh A\sinh C\cos b\cosh\psi_b$&
$\cosh B\cos\Om=\cosh A\cosh C- \sinh A\sinh C\cosh\psi_b$&
$\cosh B\cos\Om=\cosh A\cosh C- \sinh A\sinh C\cosh b\cosh\psi_b$\\
$\cosh C\cos\Om=\cosh A\cosh B- \sinh A\sinh B\cos c\cosh\psi_c$&
$\cosh C\cos\Om=\cosh A\cosh B- \sinh A\sinh B\cosh\psi_c$&
$\cosh C\cos\Om=\cosh A\cosh B- \sinh A\sinh B\cosh
c\cosh\psi_c$\\[3pt]
\hline
\end{tabular} }}
\end{table}
\end{landscape}



\end{document}